\documentclass[twocolumn]{aastex62}
\usepackage[utf8]{inputenc}
\usepackage{float}
\newcommand{\mytilde}{\raise.17ex\hbox{$\scriptstyle\mathtt{\sim}$}}

\begin{document}

\title{Cool, Luminous, and Highly Variable Stars in the Magellanic Clouds from ASAS-SN: Implications for Thorne-\.Zytkow Objects and Super-Asymptotic Giant Branch Stars}

\correspondingauthor{Anna J. G. O'Grady}
\email{ogrady@astro.utoronto.ca}

\author[0000-0002-7296-6547]{Anna J. G. O'Grady}
\affil{David A. Dunlap Department of Astronomy and Astrophysics, University of Toronto,\\ 50 St. George Street, Toronto, Ontario, M5S 3H4 Canada}
\affil{Dunlap Institute for Astronomy and Astrophysics, University of Toronto, 50 St. George Street, Toronto, Ontario, M5S 3H4, Canada}

\author[0000-0001-7081-0082]{Maria R. Drout}
\affil{David A. Dunlap Department of Astronomy and Astrophysics, University of Toronto,\\ 50 St. George Street, Toronto, Ontario, M5S 3H4 Canada}
\affiliation{The Observatories of the Carnegie Institution for Science, 813 Santa Barbara St., Pasadena, CA 91101, USA}

\author{B. J. Shappee}
\affil{Institute for Astronomy, University of Hawaii, 2680 Woodlawn Drive, Honolulu, HI 96822,USA}

\author[0000-0002-4791-6724]{Evan B. Bauer}
\affil{Kavli Institute for Theoretical Physics, University of California, Santa Barbara, CA 93106, USA}

\author{Jim Fuller}
\affil{TAPIR, Mailcode 350-17, California Institute of Technology, Pasadena, CA 91125, USA}

\author{C. S. Kochanek}
\affil{Department of Astronomy, The Ohio State University, 140 West 18th Avenue, Columbus, OH 43210, USA}
\affil{Center for Cosmology and Astroparticle Physics, The Ohio State University, 191 W. Woodruff Avenue, Columbus, OH 43210, USA\\}

\author{T. Jayasinghe}
\affil{Department of Astronomy, The Ohio State University, 140 West 18th Avenue, Columbus, OH 43210, USA}
\affil{Center for Cosmology and Astroparticle Physics, The Ohio State University, 191 W. Woodruff Avenue, Columbus, OH 43210, USA\\}

\author[0000-0002-3382-9558]{B. M. Gaensler}
\affil{David A. Dunlap Department of Astronomy and Astrophysics, University of Toronto,\\ 50 St. George Street, Toronto, Ontario, M5S 3H4 Canada}
\affil{Dunlap Institute for Astronomy and Astrophysics, University of Toronto, 50 St. George Street, Toronto, Ontario, M5S 3H4, Canada}

\author{K. Z. Stanek}
\affil{Department of Astronomy, The Ohio State University, 140 West 18th Avenue, Columbus, OH 43210, USA}
\affil{Center for Cosmology and Astroparticle Physics, The Ohio State University, 191 W. Woodruff Avenue, Columbus, OH 43210, USA\\}

\author[0000-0001-9206-3460]{Thomas~W.-S.~Holoien}
\altaffiliation{Carnegie Fellow}
\affiliation{The Observatories of the Carnegie Institution for Science, 813 Santa Barbara St., Pasadena, CA 91101, USA}

\author{J. L. Prieto}
\affil{N\'ucleo de Astronom\'ia de la Facultad de Ingenier\'ia y Ciencias, Universidad Diego Portales, Av. E\'jercito 441, Santiago, Chile}
\affil{Millennium Institute of Astrophysics, Santiago, Chile}

\author[0000-0003-2377-9574]{Todd A. Thompson}
\affil{Department of Astronomy, The Ohio State University, 140 West 18th Avenue, Columbus, OH 43210, USA}
\affil{Center for Cosmology and Astroparticle Physics, The Ohio State University, 191 W. Woodruff Avenue, Columbus, OH 43210, USA\\}

\begin{abstract}

Stars with unusual properties can provide a wealth of information about rare stages of stellar evolution and exotic physics. However, determining the true nature of peculiar stars is often difficult. In this work, we conduct a systematic search for cool and luminous stars in the Magellanic Clouds with extreme variability, motivated by the properties of the unusual SMC star and Thorne-\.Zytkow Object (T\.ZO) candidate HV2112. Using light curves from ASAS-SN we identify 38 stars with surface temperatures T $<$ 4800K, luminosities $\log$(L/L$_\odot$) $>$ 4.3, variability periods $>$ 400 days, and variability amplitudes $\Delta$V $>$ 2.5 mag. Eleven of these stars possess the distinctive double-peaked light curve morphology of HV2112. We use the pulsation properties and derived occurrence rates for these 12 objects to constrain their nature. From comparisons to stellar populations and models, we find that one star may be a red supergiant with large amplitude pulsations. For the other 11 stars we derive current masses of $\sim$5-10 M$_{\odot}$, below the theoretical minimum mass of $\sim$15 M$_{\odot}$ for T\.ZOs to be stable, casting doubt on this interpretation. Instead, we find that the temperatures, luminosities, mass-loss rates, and periods of these stars are consistent with predictions for super-Asymptotic Giant Branch (s-AGB) stars that have begun carbon burning but have not reached the superwind phase. We infer lifetimes in this phase of $\sim($1$-$7) $\times$ 10$^{4}$ years, also consistent with an s-AGB interpretation. If confirmed, these objects would represent the first identified population of s-AGB stars, illuminating the transition between low- and high-mass stellar evolution.

\end{abstract}


\section{Introduction}

The vast majority of stars fall into well established and understood categories. Smaller populations of stars with unusual properties may be difficult to classify, but provide vital insights into short-lived stages of stellar evolution, exotic physics, and uncertain final fates. HV2112, a red and luminous star in the direction of the Small Magellanic Cloud (SMC), is one such star that has eluded conclusive classification. In this work, we explore the photometric and variability properties of HV2112 and establish that there exists a broader population of stars with similar properties to HV2112 which may provide insight into its stellar identity. 

HV2112 recently gained significant notoriety when identified by \citet{Levesque.E.2014.HV2112disc} as a candidate Thorne-\.Zytkow Object (T\.ZO), but its true nature has been controversial. While briefly proposed to be a foreground S-type star in the Milky Way \citep{Maccarone.T.2016.TZOHighMotion}, new astrometric data from the Gaia satellite indicate that HV2112 is a true member of the SMC \citep{McMillan.P.2018.HV2112Gaia}. At the distance of the SMC, two main proposals for the identity of HV2112 remain: a T\.ZO or a Super Asymptotic Giant Branch (s-AGB) star. We introduce these possibilities below.

\subsection{Thorne-\.Zytkow Objects}\label{sec_intro_tzo}

Thorne-\.Zytkow Objects are a hypothetical class of stars that contain neutron stars at their cores \citep{Thorne.K.1975.TZOprime,Thorne.K.1977.TZOstructure}. Surrounding the neutron star and its thin, hot atmosphere is a fully convective, hydrogen-rich envelope. T\.ZOs can be classified as either giants or supergiants based on the mass of this envelope. The giant and supergiant classes of T\.ZOs have different formation channels, physical properties and observational signatures. While giant T\.ZOs are are powered predominately by accretion onto the central neutron star, the luminosity (and pressure support) in supergiant T\.ZOs comes predominately from nuclear reactions at the base of the convective envelope. Notably, there is a predicted luminosity and mass gap between stable T\.ZO solutions of the giant and supergiant types \citep{Cannon.R.1993.TZOStructure}. In this paper, we focus on the supergiant class of T\.ZOs, as this class would be required to explain the luminosity of HV2112. Henceforth, we refer to supergiant T\.ZOs as ``massive T\.ZOs'' or simply ``T\.ZOs''.

There are two proposed formation channels for massive T\.ZOs. In one, the T\.ZO is formed by an asymmetric supernova kicking a neutron star into its Red Supergiant (RSG) companion \citep{Leonard.P.1994.TZOKick}. In the other, T\.ZOs are hypothesized to be a possible evolutionary outcome for high mass X-ray binaries (HMXBs) with periods less than 100 days \citep{Taam.R.1978.TZOXRBperiods}. Here, the neutron star is dragged into the companion star when the system undergoes common envelope evolution, but fails to eject the envelope \citep{Thorne.K.1975.TZOprime,Thorne.K.1977.TZOstructure,Cannon.R.1992.TZOStrucEvo}. In both cases, the neutron star eventually merges with the core of the RSG.

The existence of T\.ZOs remains an open question. While the initial theory papers presented arguments for T\.ZOs being a stable stellar configuration once formed \citep{Thorne.K.1977.TZOstructure,Cannon.R.1993.TZOStructure}, there are some suggestions that T\.ZOs may be unable to form through the common envelope evolution channel because the inspiralling neutron star would eject the envelope completely \citep{Papish.O.2015.TZOEjectsEnvelope}. In addition, \citet{Fryer.C.1996.NSRapidInfallBH} argue that T\.ZOs would not be stable at all, with the neutron star undergoing enough accretion to collapse into a black hole. 

As a result of these uncertainties, the lifetimes and rates of T\.ZOs are poorly constrained.  A massive T\.ZO will become unstable when nuclear fusion ceases, either due to the depletion of fusible elements or due to the envelope mass falling below the minimum (\mytilde 14 M$_{\odot}$, which combined with a \mytilde 1M$_{\odot}$ core gives a total mass of the T\.ZO of \mytilde 15M$_{\odot}$) required to maintain the requisite physical conditions at the base of the convective region, most likely due to strong mass loss \citep{Cannon.R.1993.TZOStructure,Podsiadlowski.P.1995.TZOEvolution}. Based on these considerations \citet{Cannon.R.1993.TZOStructure} and \citet{Biehle.G.1994.TZOObservational} estimate a T\.ZO lifetime of \mytilde $10^{5}-10^{6}$ yrs.
 
\citet{Podsiadlowski.P.1995.TZOEvolution} estimates a formation rate of \mytilde $2 \times 10^{-4}$ yr$^{-1}$ in the Galaxy, from which an estimated 20 to 200 T\.ZOs could populate the Milky Way. However, no candidates have been identified in the Milky Way to date, possibly implying a shorter lifetime of the T\.ZO phase. Understanding the existence or prevalence of T\.ZOs would clearly inform binary population synthesis models and stability of the T\.ZO phase.

Searching for T\.ZOs has been historically difficult. Photometrically, T\.ZOs should resemble RSGs with a range of luminosities \citep[e.g.\ Fig. 3 of][]{Cannon.R.1992.TZOStrucEvo}, and extremely cool temperatures \citep{Thorne.K.1977.TZOstructure}. The spectra of massive T\.ZOs, however, will be enriched with unusual abundances of isotopes not found in normal RSGs. The extremely high temperatures in the atmosphere of the neutron star facilitate a particular type of nucleosynthesis -- the interrupted rapid proton (irp) process -- that can create elements such as Mo, Rb, Y, and Zr \citep{Biehle.G.1994.TZOObservational}. T\.ZOs should also have enhanced abundances of $^{7}$Li due to the $^{7}$Be-transport mechanism (as in \citealt{Cameron.a.1955.BeTransport}). These elements will be dredged up through the fully convective envelope to the surface of the star \citep{Podsiadlowski.P.1995.TZOEvolution}.

There have been unsuccessful spectroscopic searches for T\.ZOs in the past \citep{Vanture.A.1999.TZOUaqu,Kuncher.M.2002.TZOSpecSearch}, and to date HV2112 is the best candidate. \citet{Levesque.E.2014.HV2112disc} found that HV2112 had a luminosity typical of RSGs and used a comparative line ratio analysis to argue that it displayed enhancements in several of the key elements expected in T\.ZOs. In fact, \citet{Smith.V.1990.MCAGBLienhance} previously identified HV2112 as a luminous AGB star with an excess of lithium. 
However, the full set of abundances of HV2112, and implications for its origin, remain under debate. With their analysis, \citet{Levesque.E.2014.HV2112disc} also argued that an unexpected calcium enhancement was present in HV2112---which has not explicitly been predicted from the irp-process---while \citet{Beasor.E.2018.HV2112AGB} argue that, when compared to the spectra of a different control sample of stars, HV2112 does not show enhancements in either Rb or Ca.

\subsection{Super Asymptotic Giant Branch Stars}\label{sec_intro_sagb}

Whether the properties of HV2112 are more consistent with the late evolutionary stages of single stars has also been explored. In particular, s-AGB stars represent the late evolutionary stages of intermediate mass stars that are massive enough to ignite carbon burning off center, leading to a degenerate O-Ne core. This is in contrast to normal AGB stars, which only progress to helium burning. The typical mass range for s-AGB stars is 6.5 to 12 M$_{\odot}$ \citep{Garcia-Berro.E.1994.SAGBFormation}, though at low metallicties this lower bound can extend to \mytilde 5 M$_{\odot}$ \citep{Girardi.L.2000.MassTracks5SAGB,Doherty.C.2017.SAGBStarsECSNE}. These stars are near the ends of their lives and are undergoing thermal pulses. s-AGB stars are an important connection between low-mass and high-mass stellar evolution, and may be the progenitors of electron capture supernovae \citep{Miyaji.S.1980.OriginalecSN,Doherty.C.2017.SAGBStarsECSNE}. 

Like T\.ZOs, s-AGB stars are also expected to show enhancements of lithium \citep{Cameron.A.1971.SAGBLithium} and heavy elements such as Mo and Rb \citep{Karakas.A.2014.SAGBrpelements}. To date there have been no confirmed s-AGB stars, though one strong candidate in the SMC has been identified \citep{Groenewegen.M.2009.MCRSGAGBStars2112SAGB}. However, significant work has been done on modeling the evolution of s-AGB stars through the carbon burning phase \citep{Siess.L.2010.SuperAGBEarly,Doherty.C.2010.SAGBI,Doherty.C.2014.SAGBMassLossCrit,Doherty.C.2014.SAGBIII,Doherty.C.2015.SAGBIV,Jones.S.2013.SAGBEarlyModel}. s-AGBs are more luminous than typical AGB stars and are expected to sit in a similar area of the Hertzsprung-Russell diagram as RSGs, though they may have colder temperatures. Variability has been posited as a possible avenue for distinguishing s-AGBs from RSGs, as s-AGB stars could have much higher variability amplitudes, similar to the variability observed in some normal AGB stars (Mira variables) \citep{Doherty.C.2017.SAGBStarsECSNE}.

\citet{Tout.C.2014.HV2112SAGB} investigated whether HV2112 could be a s-AGB star. While the photometric observations of HV2112 matched predictions of s-AGB properties, \citet{Tout.C.2014.HV2112SAGB} determined that the calcium enhancement in the spectrum of HV2112 could not be explained by s-AGB nucleosynthesis processes, but could be created during the formation of a T\.ZO. However, a more recent examination of spectroscopy of HV2112 by \citet{Beasor.E.2018.HV2112AGB} suggests that HV2112 does not have enhancements of Ca, Rb, or Mo, showing only an enhancement of Li, and has a luminosity more consistent with an intermediate mass AGB star than with a T\.ZO.

\subsection{This work}

Should HV2112 be either a T\.ZO or s-AGB star, it would be the first confirmed case of either identity. One means to assess its true nature is to determine whether or not it belongs to a larger population. In particular, while large scale spectroscopic surveys capable of detecting abundance anomalies are still on-going, HV2112 is also distinguished by its variability. Its light curve has a period of \mytilde 600 days and a V-band variability amplitude of more than 4 magnitudes \citep{Kochanek.C.2017.ASASSN2}. This level of variability is not standard for its estimated luminosity of log(L/L$_{\odot}$) $\simeq$ 5.0; typical RSG V-band variability is of the order of 1 magnitude \citep{Josselin.E.2000.RSGVariabilityML,Levesque.E.2007.MCRSGTooCold,Soraisam.Monika.2018.RSGVarinM31}.

In this paper, we characterize the optical variability of HV2112 in order to carry out a systematic search in the Magellanic Clouds for more objects at similar luminosities and temperatures that display this type of extreme variability. These objects will be called `HV2112-like-objects' (HLOs). We assess the physical properties of 11 HLOs we identify, as well as 27 other highly variable, luminous, cool stars to determine their possible nature. 

By conducting a systematic survey for a population of these objects, we will be able to discuss rate and lifetime expectations for either a T\.ZO or s-AGB star identity. Additionally, details of the variability of HV2112 and the HLOs can provide important clues to their internal structure. In particular, fundamental-mode pulsations are sensitive to the mean density of stars, and hence offer a means to probe their current mass if information on their current radius and stellar structure are known. As described above, massive T\.ZOs are hypothesized to require total masses of at least 15 M$_\odot$ to sustain the rapid-p process needed to provide the pressure support required for a stable stellar structure, although the precise mass depends on the convective efficiency \citep{Cannon.R.1993.TZOStructure,Podsiadlowski.P.1995.TZOEvolution}. Should details of the stellar pulsations indicate that the current mass of HV2112 or any of the HLOs is below this limit, it would cast doubt on a T\.ZO identity.

In \S\ref{sec_sample_selection} we describe our selection of the HLOs, and in \S\ref{sec_observations} report the observational data available for them. We analyze the observed properties of the 11 newly identified HLOs in \S\ref{sec_properties_observations}, and constrain their physical properties as compared to known populations of stars in \S\ref{sec_properties_physical}. The rates and lifetime expectations for the HLOs, should they be T\.ZOs or s-AGB stars, are explored in \S\ref{sec_lifetimes}. Finally, we discuss our results in \S\ref{sec_discussion}.

\vspace{1em}

\section{Sample Selection}\label{sec_sample_selection}

Our goal is to identify a sample of stars with photometric properties and variability similar to that of HV2112. Here, we describe the optical variability of HV2112, and use this to define a set of criteria to identify HLOs. For this work, we focus on the stellar populations of the Large and Small Magellanic Clouds (LMC and SMC). The distances to these systems are well constrained, extinction in the direction of the Clouds is low, and contamination from foreground dwarfs can be removed using Gaia astrometry \citep{Gaia.Collab.2018.GaiaDR2}.  In addition, the photometric variability of the Clouds has been monitored for more than 30 years through projects such as OGLE \citep{Udalski.A.2003.OGLE}, ASAS \citep{Pojmanski.G.2002.ASAS}, MACHO \citep{Alcock.C.1997.MACHO}, and the All-Sky Automated Survey for Supernovae \citep[ASAS-SN;][]{Shappee.B.2014.ASASSN1,Kochanek.C.2017.ASASSN2}.

\subsection{ASAS-SN Photometry}\label{sec_asassn}

For the purposes of candidate selection, we use V-band light curves from ASAS-SN. ASAS-SN consists of 20 telescopes spread amongst five four-telescope arrays with coverage in both the northern and southern hemispheres. Through normal survey operations, ASAS-SN images the entire night sky to a limiting magnitude of m$_\mathrm{g} \sim$ 18.5 mag with a $\sim$1 day cadence. Each ASAS-SN camera has a 4.5 deg$^{2}$ field-of-view, 8$''$ pixels, and typical point-source full-width half-max of \mytilde 2 pixels. With all-sky coverage, ASAS-SN data has already been used extensively for the analysis of Milky Way variable stars \citep{2018MNRAS.477.3145J,Jayasinghe.T.2018.ASASSNVariables,2019MNRAS.485..961J,Jayasinghe.T.2019.ASASSNVI,Jayasinghe.T.2019.ASASSNVII,Jayasinghe.T.2020.ASASSNV,Shields.J.2018.RCorBorASASSN,Pawlak.M.2019.ASASSNIV,Percy.J.2019.ASASSNRCorBor,Auge.C.2020.ASASSNExtraRef}.

ASAS-SN imaged the LMC and SMC in the V-band for approximately 4.5 years with two arrays between May 2014 and September 2018, with a typical cadence of 1 - 2 days and a limiting magnitude of m$_{\mathrm{V}}$ \mytilde 17.5 mag. (Beginning in Sept. 2017 ASAS-SN added 3 g-band arrays and in Sept. 2018 ASAS-SN switched the southern V-band array to g-band). For our sample selection, we use light curves calculated over this entire time range. Aperture photometry was extracted for each epoch as described by \citet{Kochanek.C.2017.ASASSN2} using the IRAF \verb"apphot" package with a 2-pixel radius aperture. Photometric errors were recalculated as described by \citet{2019MNRAS.485..961J}, and the AAVSO Photometric All-Sky Survey catalog (APASS; \citealt{2015AAS...22533616H}) was used for calibration. On average, there are approximately 910 V-band epochs for each SMC/LMC star.

While ASAS-SN has large pixels and the Magellanic Clouds are crowded, as described below, the stars we select are all very luminous. As a result, most dominate the flux at their location, and the primary effect of blending is to decrease the ASAS-SN limiting magnitude to $\sim$16.5$-$17 mag. This will not impact our results.
We use the aperture photometry light curves for this selection process. Image subtraction light curves (described in \S\ref{sec_asassn_imgsub}) are used for the more substantial analysis of the HLOs.

\subsection{Variability of HV2112}

HV2112 has been identified as optically variable for more than 50 years, appearing in the Harvard Variable catalog \citep{Payne-Gaposchkin.Cecilia.1966.VariablesinSMC} with an amplitude of 4.8 mag and a \mytilde 600 day period. Modern observations of its variability were performed with OGLE and ASAS,  which yield variability amplitudes of $\sim$ 2.2 mag and $>$2.1 mag in the I-band and V-band, respectively (light curves from these surveys are shown in Figure~\ref{hv2112_extra_lightcurves} in the Appendix). Mid-infrared variability was also observed by \citet{Glass.I.1979.MidIRCloudsHV2112}, with HV2112 showing amplitudes of at least 0.87, 1.02, 0.95 mag in the J, H, and K bands, respectively.

\begin{figure}
    \centering
    \includegraphics[width=\linewidth]{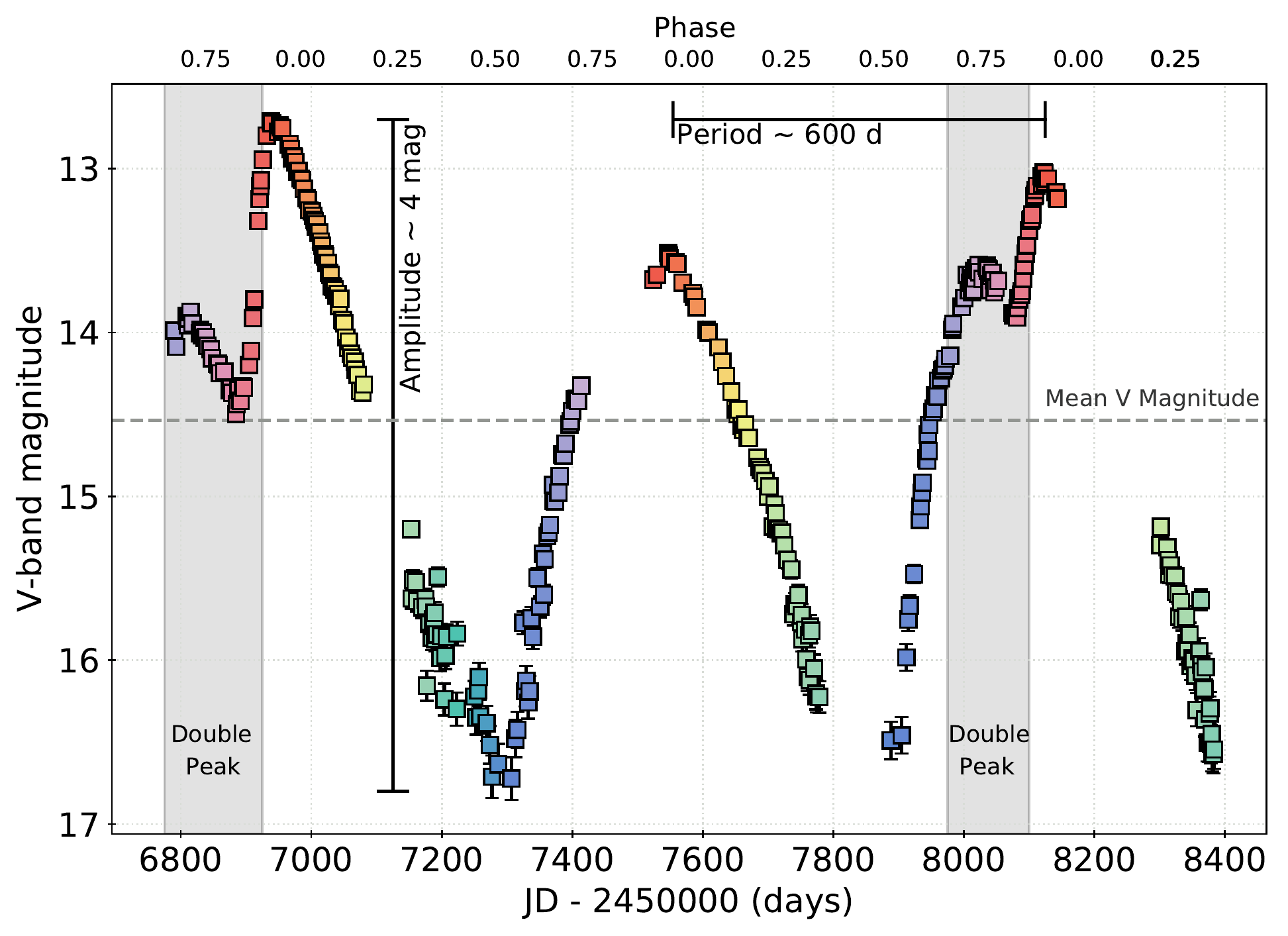}
    \caption{ASAS-SN V-band light curve of HV2112. The color corresponds to the approximate phase of the light curve -- red at the peaks and green/blue in the troughs. We have highlighted the double peak feature mentioned in the text with grey vertical shading, and indicated the mean V-band magnitude.}
    \label{hv2112_lc}
\end{figure}

In Figure~\ref{hv2112_lc} we show the ASAS-SN V-band light curve of HV2112. It has a peak-to-trough variability amplitude of \mytilde 4 mag and a peak-to-peak period of \mytilde 600 days. This level of variability is unusual for stars with luminosities of $\log L/L_{\odot} \sim 5$, as estimated for HV2112 \citep{Levesque.E.2014.HV2112disc,Beasor.E.2018.HV2112AGB,Glass.I.1979.MidIRCloudsHV2112}. Instead, it is more typical of Mira variables, pulsating AGB stars, defined in the General Catalog of Variable Stars (GCVS) \citep{Samus.N.2017.GCVS} as having visual amplitudes greater than 2.5 magnitudes, and periods ranging from 100 to 1000 days (though periods between 200 and 500 days are more typical). However, AGB stars have a maximum luminosity of log$(L/L_{\odot}) = 4.74$ \citep{Paczynski.B.1970.AGBLumLimit}.

The light curve morphology of HV2112 is also unusual; it displays a prominent ``double peak'' feature during the rising phase, as highlighted in Figure~\ref{hv2112_lc}. Unfortunately there was a seasonal gap in the observations around 7500 days, so it is unclear if this feature is present in every pulsation cycle. There is also some cycle-to-cycle variation in the peak V-band magnitude.

This morphology is atypical for both RSGs, whose light curves tend to be complex and only semi-regular, and Mira variables, which tend to be regular and symmetric. However, we note that \citet{Lebzelter.T.2011.MiraLightCurves} find that approximately $\sim$30\% of Mira variables deviate from a strictly sinusoidal morphology. In addition, a double peak feature has been observed in some Mira variables and other large amplitude pulsators \citep{Ludendorff.H.1928.MiraDoublePeak0,Keenan.P.1974.MiraDoubleMaxI,Vardya.M.1988.MiraDoubleMaxII,Marsakova.V.2007.MiraDoubleMaxIII}, and has been attributed to shocks propagating through the stellar atmosphere \citep{Kudashkina.L.1994.MiraDPShocks}.

\subsection{Criteria for Identification as an HLO}

Using the photometric and variability properties of HV2112 as a baseline, we define a set of selection criteria to identify luminous, cool, and highly variable stars in the SMC/LMC. Our goal is to select stars with physical properties as similar to HV2112 as possible.

To be considered an HLO, a star must:
\begin{enumerate}
    \item Be a luminous and red star (\S \ref{sec_phot_select}) with astrometry consistent with membership in the LMC or SMC (\S \ref{sec_gaia_cuts})
    \item Be confirmed as a variable star within the sensitivity limits of ASAS-SN (\S \ref{sec_intrin_var})
    \item Have a variability amplitude $>$ 2.5 magnitudes in the V-band (\S \ref{sec_high_amp_var})
    \item Have a light curve morphology similar to that of HV2112 (\S \ref{sec_double_peak})
\end{enumerate}

In the sections below, we describe each of these criteria in more detail. In Table \ref{tzoc_locs} we summarize the number of star that pass each successive cut.

\subsection{Selection of Luminous and Cool Stars}\label{sec_phot_select}

To construct a sample of cool and luminous stars in the Magellanic Clouds, we first select all sources in the 2 Micron All-Sky Survey \citep[2MASS;][]{2MASS.Skrutskie.2006} in the direction of the Clouds. Sources were taken in a 4.5 degree radius centered at  $\alpha$ = 80.89417, $\delta$ = $-$69.75611 (J2000) for the LMC and a 1.75 degree radius centered at $\alpha$= 13.15833, $\delta$ = $-$72.80028 (J2000) for the SMC. This initial sample contained 1,312,804 sources in the LMC and 207,074 sources in the SMC.

Subsequently, we use the ATLAS9 model atmospheres \citep{Castelli.F.2004.GRIDSmodels} to define a set of color and magnitude cuts. Our goal was to identify stars with T$_\mathrm{eff}$ $<$ 4800 K and luminosities greater than $\log (L/L_\odot) \sim$ 4.2. These criteria are designed to include the T\.ZO models of \citet{Cannon.R.1992.TZOStrucEvo} and the RSG branch at a range of metallicities, while excluding most lower mass AGB stars. In the end, we select LMC stars with $(J-K_{\mathrm{s}}) > 0.9$ mag and $K_{\mathrm{s}} < 10.2$ mag and SMC stars with $(J-K_{\mathrm{s}}) > 0.587$ mag and $K_{\mathrm{s}} < 10.6$ mag. 

We choose to adopt flat K$_{\mathrm{s}}$-band cuts because for stars in our temperature range of interest, the K$_{\mathrm{s}}$-band is near the peak of their spectral energy distribution and is hence a reasonable proxy for luminosity. We estimate that we are complete down to a luminosity of $\log (L/L_{\odot})$ $\sim$ 4.2 and $\sim$ 4.3 in the LMC and SMC, respectively, for stars with 4800 K $>$ T$_\mathrm{eff}$ $>$ 3200 K. These cutoffs also roughly correspond to the ASAS-SN V-band limit of \mytilde 17 mag for stars in this temperature range. The possibility of a population of cooler or heavily dust enshrouded stars will be discussed below. These cuts select 3307 and 917 cool and luminous stars in the direction of the LMC and SMC, respectively.

\subsection{Removal of Foreground Sources with Gaia DR2}\label{sec_gaia_cuts}

In order to minimize contamination from foreground dwarfs, we filtered our sample based on proper motion ($\mu_{\alpha}$, $\mu_{\delta}$) and parallax  ($\pi$) measurements from \emph{Gaia} Data Release 2 \citep[DR2;][]{Gaia.Collab.2018.GaiaDR2}. We follow a procedure modelled closely on that described by \citet{Gaia2018}. We define a three dimensional filter in $\mu_{\alpha}$, $\mu_{\delta}$, and $\pi$ based on a sample of highly-probable SMC/LMC members. This filter is then applied to all 4224 luminous and red stars, except for 86 stars which did not have matches in the Gaia database, in order to assess their consistency with the observed kinematics of the SMC/LMC.

To define the filter, we first select all \emph{Gaia} DR2 sources within 4 degrees and 3.1 degrees of the LMC and SMC centers defined above, respectively, with \emph{Gaia} G $>$ 18 mag. We then exclude all sources with $\pi$/$\sigma_{\pi}$ $>$ 4 and 0.7 mag $<$ (G$_{\mathrm{bp}}$ $-$ G$_{\mathrm{rp}}$) $<$ 1.1 mag, to eliminate likely foreground dwarfs. The latter criterion removes the yellow region of the color magnitude diagram which has been shown to be heavily contaminated by foreground dwarfs \citep[e.g.][]{Neugent.K.2012.RSG.YSG.LMC}. We subsequently determine the median proper motions and parallaxes for the remaining stars and further exclude any sources with parameters that deviate by more than four times the robust scatter estimate in $\mu_{\alpha}$, $\mu_{\delta}$, or $\pi$\footnote{This follows the procedures of \citet{Gaia2018}. The robust scatter estimate (RSE) is defined as RSE $\approx$0.39$\times$(P$_{90}$-P$_{10}$) where P$_{90}$ and P$_{10}$ are the 90th and 10th percentile values of the distribution, respectively. If the distribution is Gaussian, then the RSE is equal to the standard deviation.}.

After applying these cuts, we are left with 906,367 and 190,594 highly-probable members of the LMC and SMC, respectively. These samples have median proper motions of ($\mu_{\alpha}$, $\mu_{\delta}$) = (1.82, 0.29) and (0.71, $-$1.22) mas/yr, respectively, which agree well with the center-of-mass proper motions of ($\mu_{\alpha}$, $\mu_{\delta}$) = (1.89,0.31) and (0.69,$-$1.23) mas/yr, respectively, from \citet{Gaia2018}. We use these samples to define a covariance matrix {\bf $\sigma$} for the variables $\vec{\mu} =$ ($\mu_{\alpha}$, $\mu_{\delta}$, $\pi$) for each galaxy. We remove objects with {\bf $\mu^{T} \sigma^{-1} \mu$} $>$ 12.8 as probable foreground dwarfs. This filter identifies stars that fall outside the region that contains 99.5\% of likely SMC/LMC members. The effect of choosing this particular threshold on the purity and completeness of our final sample is discussed in \S\ref{sec_sample_complete}. After removing all likely foreground dwarfs from our sample, we are left with 2897 and 633 luminous and cool stars in the LMC and SMC, respectively, for a combined total of 3530 stars.

\subsection{Determination of Basic Variability Properties}\label{sec_basic_var}

In order to assess which luminous and red stars in the Clouds have variability properties similar to HV2112, we begin by using the ASAS-SN aperture photometry light curves to determine basic variability properties for the entire sample. We calculate the mean and median V-band magnitude, RMS light curve variation, and peak-to-peak amplitude $\Delta$V over the \mytilde 4.5 years of ASAS-SN V-band coverage. All properties were calculated after removing points with magnitude errors $>$ 0.3 mag. Variability amplitudes were calculated using only ASAS-SN detections, not upper limits, and we performed sigma clipping -- removing points more than 4 standard deviations away from the mean -- to mitigate incorrect amplitudes due to outliers. The top panel of Figure~\ref{chi2_amp_plots} shows the mean magnitude versus the V-band variability amplitude $\Delta$V for the complete sample. HV2112 is highlighted in blue, and is clearly separated from almost all other sources.

For 242 sources in our sample, $>$90\% of the ASAS-SN light curve points yield non-detections, precluding a detailed assessment of their variability. These stars still passed our initial color and magnitude cuts, indicating they are likely very red, possibly self-extincted. We remove these sources from further consideration, leaving 2670 in the LMC and 618 in the SMC for a total of 3288 sources. The impact of this on sample completeness is discussed in \S\ref{sec_sample_complete}.

\subsection{ASAS-SN Sensitivity to Intrinsic Variability}\label{sec_intrin_var}

\begin{figure}
    \centering
    \includegraphics[width=\linewidth]{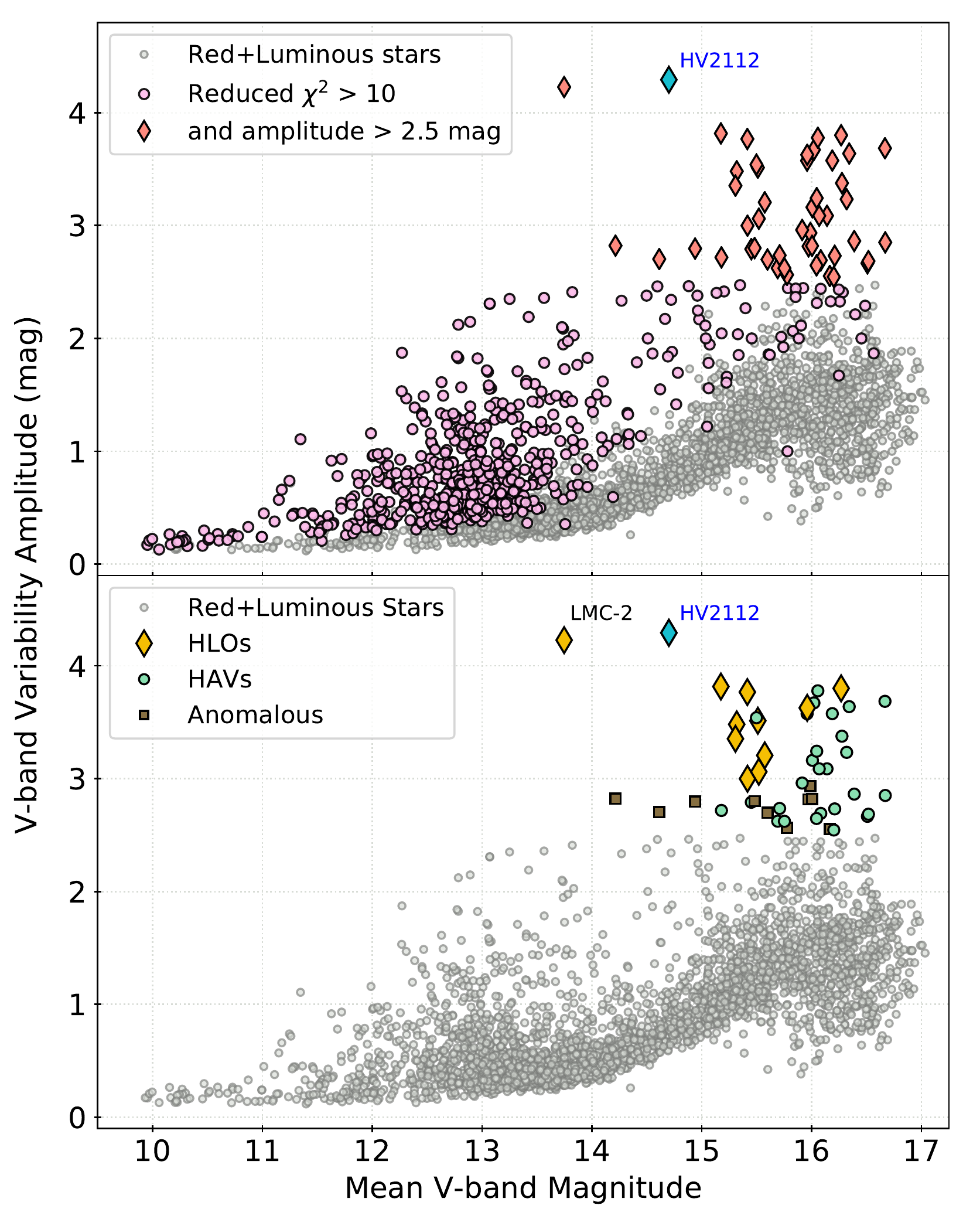}
    \caption{Mean V-band magnitude vs variability amplitude $\Delta$V. Red and luminous stars are shown as grey circles, and HV2112 is highlighted as a blue diamond. \textit{Top:} The top panel illustrates which stars have reduced $\chi^{2} > $10 as pink circles and those which also have amplitudes $\Delta$V $> $ 2.5 mag as red diamonds. \textit{Bottom:} The final division between HAVs (teal circles) and HLOs (gold diamonds). Anomalous sources are shown as brown squares. The only other star with a variability amplitude $>$ 4 mag is the star we refer to as LMC-2.}
    \label{chi2_amp_plots}
\end{figure}

The sensitivity of ASAS-SN photometry to intrinsic variability as a function of magnitude for point sources within crowded regions---such as in the Magellanic Clouds---has not previously been explored.  The impact of this sensitivity is evident in Figure~\ref{chi2_amp_plots} where some bright stars ($\langle \mathrm{m}_{\mathrm{V}} \rangle$ $\sim$ 10 mag) have measured variability amplitudes of $\lesssim$0.2 mag, while no star with $\langle \mathrm{m}_{\mathrm{V}} \rangle$ $\sim$ 15 mag has a measured amplitude less than 0.65 mag. This effect may partially be caused by selection effects, but it is important to ascertain whether the measured amplitudes are real. In order to determine whether the variability observed in our sample represents intrinsic source variability, we calculate the reduced $\chi^{2}$ that results from fitting a flat line to each ASAS-SN light curve at the mean magnitude of the star. A low $\chi^{2}$ will result either from an ASAS-SN light curve with low RMS variability, or with higher measured scatter, but accompanied by larger error bars. 
In order to select only stars with intrinsic variability we restrict our sample to those with reduced $\chi^{2}$ values of 10 or greater. These sources are pink circles in the top panel of Figure~\ref{chi2_amp_plots}.

\subsection{Selection of High Amplitude Variables}\label{sec_high_amp_var}

As can be seen in Figure~\ref{chi2_amp_plots}, HV2112 is an extreme outlier in having a large variability amplitude for its mean magnitude. In fact, HV2112 possess \emph{the largest} variability amplitude of all 3288 luminous and red stars in the SMC/LMC. Therefore, in order to select HLOs, we restrict our search to sources with variability amplitudes $>$2.5 mag in V-band. This threshold was selected because it is the canonical dividing line between semi-regular and high-amplitude Mira variables in the General Catalog of Variable Stars \citep{Samus.N.2017.GCVS}. Sources that pass both this amplitude cut and the reduced $\chi^{2}$ cut are shown as red diamonds in the top panel of Figure~\ref{chi2_amp_plots}. These cuts leave 49 objects, with 12 in the SMC and 37 in the LMC.

\subsection{Selection Based on Light Curve Morphology}\label{sec_double_peak}

\begin{figure*}
    \centering
    \plottwo{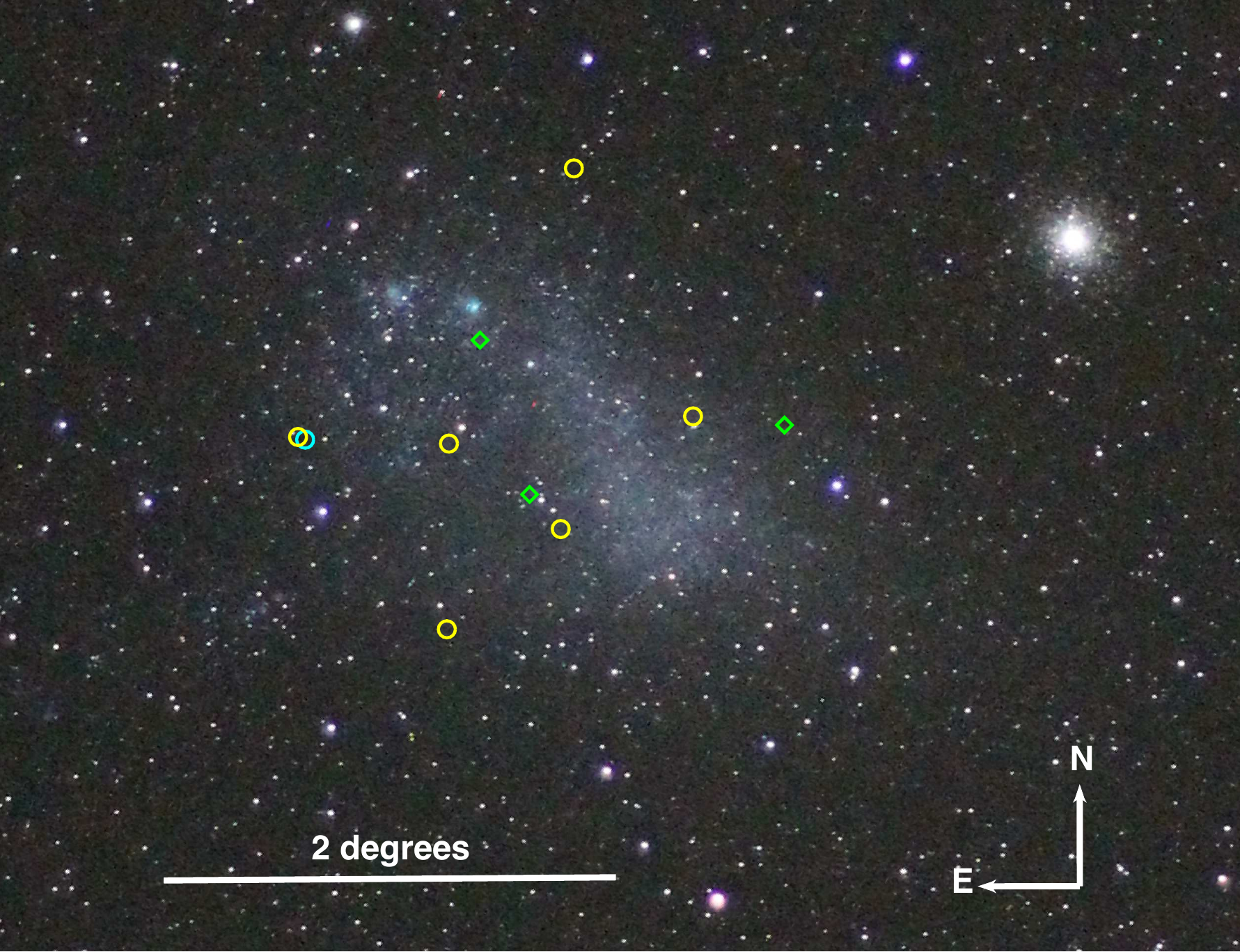}{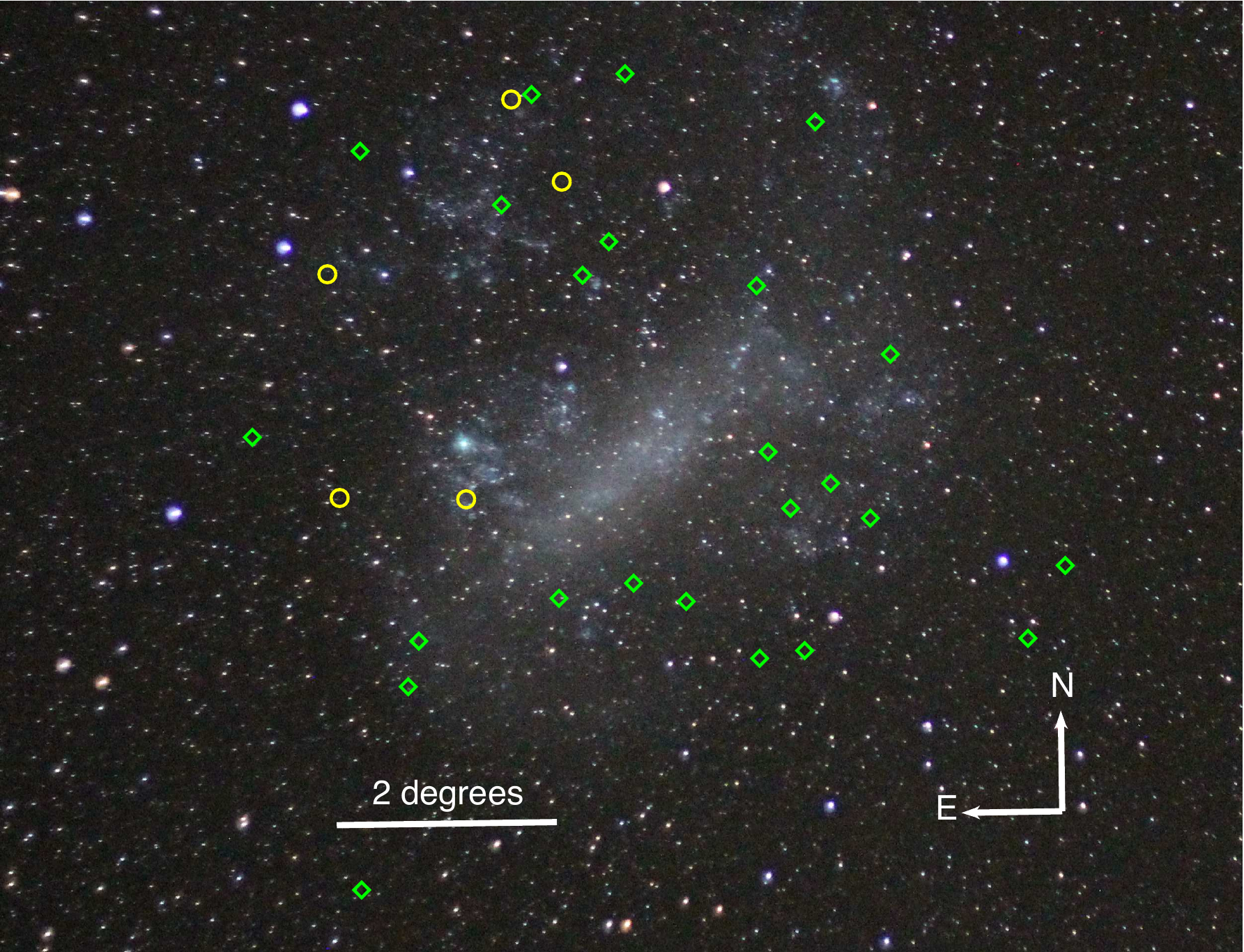}
    \caption{Locations of the HLOs (yellow circles) and HAVs (green diamonds) in the SMC (left) and LMC (right). HV2112 is a cyan circle. North is up and east is left in both photos, and a scale is provided. Photographs taken by Anna O'Grady at Las Campanas Observatory in December 2019. Coordinates provided by Astrometry.net \citep{Lang.D.2012.Astrometry}.}
    \label{locations}
\end{figure*}

Finally, we visually examined the ASAS-SN light curves of these 49 stars to determine their morphological similarity to that of HV2112. We separate the sources into three categories. We classify stars that show smooth and asymmetric light curves with secondary maxima features during the rising phase as HLOs. Any additional stars that strongly resemble the overall asymmetric morphology of HV2112, but lack an observed double-peak feature, we call High Amplitude Variables (HAVs). Finally, any remaining stars with morphologies extremely unlike that of HV2112 are classified as ``anomalous''.

In total, we identify 11 additional stars (5 in the LMC and 6 in the SMC) with light curve morphologies similar to that of HV2112---including the double-peak feature. Thus, including HV2112, there are 12 HLOs in the Magellanic Clouds. Ascertaining the nature of these stars, highlighted as gold diamonds in the lower panel of Figure~\ref{chi2_amp_plots}, will be the primary focus of the rest of this paper. In addition, we classify 27 stars (24 in the LMC and 3 in the SMC) as HAVs---these stars appear similar to HV2112 but lack a double peak feature in their light curves. In the sections below, the possibility that some HAVs have the same physical origin as the HLOs will be discussed. The HAVs are shown as teal circles in the lower panel of Figure~\ref{chi2_amp_plots}. 

Finally, we classify 10 stars (8 in the LMC and 2 in the SMC) as ``anomalous''. In general these stars barely pass the 2.5 mag variability threshold and possess either complex light curves with variability on short timescales or very symmetric or triangular light curves unlike that of HV2112. Some examples of these light curves are shown in Appendix~\ref{anom_app} (Figure~\ref{anom_lcs}). These stars are discarded for the rest of this paper, and are shown as brown squares in the lower panel of Figure~\ref{chi2_amp_plots}.

\begin{deluxetable}{l|rr}
\centering
\tablecaption{Sample Construction\label{tzoc_locs}}
\tablehead{\multicolumn{1}{c|}{Criterion} & \colhead{SMC} & \colhead{LMC}}
\startdata
2MASS Sources & 207,074 & 1,312,804 \\ 
Color-Magnitude Cuts & 917 & 3307 \\ 
Gaia Astrometry Cut  & 633  & 2897 \\ 
High Variability & 12  & 37 \\ 
\hline
HAVs & 3 & 24  \\ 
HLOs & 7 & 5  \\
\hline
\hline
\enddata
\tablecomments{Each row displays the number of stars remaining after applying each selection criteria for defining HLOs. Note that the number of HAVs does not include the HLOs.}
\end{deluxetable}

\subsection{Final Sample Summary}

In Table \ref{tzoc_locs} we show the progression of our selection criteria to identify luminous ($\log (L/L_\odot) \gtrsim 4.2$) and cool (T$_{\mathrm{eff}}$ $<$ 4800 K) stars in the Magellanic Clouds with variability amplitudes and light curve morphologies similar to HV2112. The locations of the 12 HLOs and 27 HAVs within the SMC and LMC are shown in Figure~\ref{locations}. The ASAS-SN light curves of the 12 HLOs are shown in Figure~\ref{all_hlos}. Light curves for the HAVs are included in Appendix \ref{hav_appendix} as Figures \ref{hav_lcs1}-\ref{hav_lcs2}, and HAV photometric information is included in Table \ref{hav_info}.

Throughout the paper we refer to individual HLOs with sequential numbers according to the Magellanic Cloud in which they are located. In Table \ref{hlo_info} we list the coordinates of each star alongside these names. HV2112 will continue to be referred to as such. Table \ref{hlo_info} also provides the variability amplitudes and periods of the HLOs, as well as previous classifications retrieved from SIMBAD \citep{SIMBAD.Wenger.M.2000}. 

Essentially all of the HLOs have been previously classified as either Mira variables or AGB stars. Most are photometric classifications from the OGLE \citep{Soszynski.I.2011.OGLELPVsinSMC,Ulaczyk.K.2013.OGLELMCVars} or SAGE \citep{Vijh.U.2009.SAGEVariableStarsLMC} surveys, based primarily on their red colors and variability. HV2112, SMC-1, and SMC-4 were also identified as likely AGB stars by \citet{Wood.P.1983.LPVinMC} due to a combination of their absolute magnitudes and periods. We note that \citet{Frith.J.2013.HPMBrightMDwrfs} classified LMC-4 as a foreground M-dwarf, based on a high proper motion measurement of ($\mu_{\alpha}$, $\mu_{\delta}$) $=$ (28.9$\pm$14,$-$74.4$\pm$14) mas/yr. However, after querying \emph{Gaia} DR2, we find that no star within 30\arcsec of the position of LMC-4 has a proper motion greater than $\sim$6 mas/yr. Hence, we suspect that this association is spurious and LMC-4 is a true member of the LMC.

We note that the second T\.ZO candidate identified in \citet{Beasor.E.2018.HV2112AGB}, HV11417, is not contained in our final sample. While it was identified as one of the original luminous and red stars overlapping with the SMC (\S\ref{sec_phot_select}), its \emph{Gaia} DR2 proper motions are inconsistent with other likely SMC sources based on our analysis in \S\ref{sec_gaia_cuts}, indicating it may be a foreground halo star. However, even if it is a true member of the SMC (the Gaia proper motion errors are large), we note that it likely would have been filtered from our sample in \S\ref{sec_high_amp_var} -- it's V-band variability amplitude during the time period observed by ASAS-SN is only $\sim$1.25 mag. It also does not display a double peak structure in its light curve.

In the sections below, we reassess the nature of the HLOs. We focus on a comparison of their physical and pulsation properties to modern models of stellar structure and the implications of the total number of HLOs for their rates and lifetimes.

\begin{deluxetable*}{ccccrrrrrrr}
\tabletypesize{\footnotesize}

\tablecaption{Basic Properties of HLOs and HV2112. \label{hlo_info}}
\tablehead{\colhead{RA} & \colhead{DEC} & \colhead{2MASS} & \colhead{Name} & \colhead{Period\tablenotemark{a}} & \colhead{$\Delta$V\tablenotemark{b}} & \colhead{Mean V\tablenotemark{c}} & \colhead{2MASS} & \colhead{Gaia} & \colhead{SIMBAD} & \colhead{Ref.\tablenotemark{e}} \vspace{-0.275cm}\\
\colhead{J2000} & \colhead{J2000} & \colhead{Name} & \colhead{in paper} & \colhead{} & \colhead{} & \colhead{} & \colhead{K$_{\mathrm{s}}$-band\tablenotemark{c}} & \colhead{$\chi^2$ value\tablenotemark{d}} & \colhead{Class} & \colhead{} \vspace{-0.2cm}\\
\colhead{(deg)} & \colhead{(deg)} & & & \colhead{(days)} & \colhead{(mag)} & \colhead{(mag)} & \colhead{(mag)}  & \colhead{} &
}
\startdata
   17.515856 & $-$72.614603 & J01100385$-$7236526  & HV2112 & 600 & 4.0$\pm$0.13 & $-$4.80$\pm$1.02 & $-$10.33$\pm$0.02 & 4.23 & Mira & \tablenotemark{[1,2]} \\ 
   11.703220 & $-$72.763824 & J00464877$-$7245497  & SMC-1 & 510 & 3.7$\pm$0.17 & $-$3.71$\pm$0.76 & $-$9.13$\pm$0.03 & 2.34 & Mira & \tablenotemark{[1]} \\
   13.036803 & $-$71.606606 & J00520884$-$7136240 & SMC-2 & 530 & $>$ 3.3 & $-$3.96$\pm$0.71 & $-$9.52$\pm$0.02 & 0.60 & Mira & \tablenotemark{[2,3]} \\ 
   13.909812 & $-$73.194845 & J00553821$-$7311410 & SMC-3 & 660 & $>$ 4.1 & $-$3.73$\pm$0.94 & $-$9.62$\pm$0.02 & 4.64 & Mira & \tablenotemark{[1,4]} \\ 
   15.402500 & $-$72.744762 & J01013681$-$7244411 & SMC-4 & 590 & $>$ 3.8 & $-$3.30$\pm$0.89 & $-$9.85$\pm$0.02 & 10.55 & Mira & \tablenotemark{[1]} \\ 
   15.903689 & $-$73.560525 & J01033691$-$7333377 & SMC-5 & 520 & 2.9$\pm$0.15 & $-$3.79$\pm$0.72 & $-$9.36$\pm$0.02 & 2.22 & Mira & \tablenotemark{[1,2]} \\ 
   17.612562 & $-$72.596670 & J01102693$-$7235486 & SMC-6 & 570 & $>$ 3.7 & $-$4.24$\pm$0.83 & $-$9.97$\pm$0.02 & 2.89 & Mira & \tablenotemark{[1]} \\ \hline
   80.824095 & $-$66.952095 & J05231778$-$6657073 & LMC-1 & 690 & 3.2$\pm$0.14 & $-$3.82$\pm$0.96 & $-$10.30$\pm$0.02 & 1.93 & AGB & \tablenotemark{[5]} \\ 
   81.567365 & $-$66.116348 & J05261606$-$6606589 & LMC-2 & 550 & 4.3$\pm$0.14 & $-$5.65$\pm$0.96 & $-$10.50$\pm$0.03 & 1.29 & AGB & \tablenotemark{[5]} \\ 
   84.986223 & $-$69.589014 & J05395683$-$6935210 & LMC-3 & 400 & $>$ 3.7 & $-$2.56$\pm$0.82 & $-$8.92$\pm$0.02 & 0.37 & LPV & \tablenotemark{[6]} \\
   86.709478 & $-$67.246312 & J05465030$-$6714468 & LMC-4 & 560 & 3.7$\pm$0.15 & $-$3.31$\pm$0.91 & $-$9.52$\pm$0.02 & 1.48 & HPM\tablenotemark{f} & \tablenotemark{[7]} \\
   88.116079 & $-$69.236122 & J05522785$-$6914100 & LMC-5 & 590 & 3.8$\pm$0.16 & $-$2.59$\pm$0.91 & $-$9.33$\pm$0.03 & 11.85 & AGB & \tablenotemark{[5]} \\
\enddata
\tablenotetext{a}{Determination of periods is detailed in \S\ref{sec_lpvs}. Values in this table have been rounded to the nearest 10 days because we see cycle-to-cycle variations in the period on the order of $\pm$ 30 days.}
\tablenotetext{b}{Determination of amplitudes is detailed in \S\ref{sec_high_amp_var}. ``$>$'' designates a lower limit. Errors are statistical, not systematic.}
\tablenotetext{c}{Absolute 2MASS M$_{K_{\mathrm{s}}}$ magnitudes, corrected for extinction (see \S\ref{sec_dist_met_red})}
\tablenotetext{d}{Gaia $\chi^2$ is a measure of how consistent proper motion and parallax of the source is with the distribution of parameters found for likely SMC/LMC members (see Section~\ref{sec_gaia_cuts}. A value of $\chi^{2}$ $<$ 4.11, 7.81, 12.8 indicates that a star falls within the region that encompasses 75\%, 95\%, and 99.5\% or the ``highly likely'' SMC/LMC members based on their Gaia parallaxes and proper motions.}
\tablenotetext{e}{References for SIMBAD classifications: [1] \citet{Soszynski.I.2011.OGLELPVsinSMC}, [2] \citet{Wood.P.1983.LPVinMC}, [3] \citet{Samus.N.2017.GCVS}, [4] \citet{Ruffle.P.2015.SpitzerSMC},[5] \citet{Vijh.U.2009.SAGEVariableStarsLMC}, [6] \citet{Ulaczyk.K.2013.OGLELMCVars}, [7] \citet{Frith.J.2013.HPMBrightMDwrfs}} 
\tablenotetext{f}{While \citet{Frith.J.2013.HPMBrightMDwrfs} classify LMC-4 as a high proper motion star, no sources with proper motions $>$6 mas/yr are found within a 30 arcsec radius of these coordinates in \emph{Gaia} DR2.  Hence, we regard this classification as spurious.}

\end{deluxetable*}

\begin{figure*}
    \centering
    \includegraphics[width=0.95\textwidth]{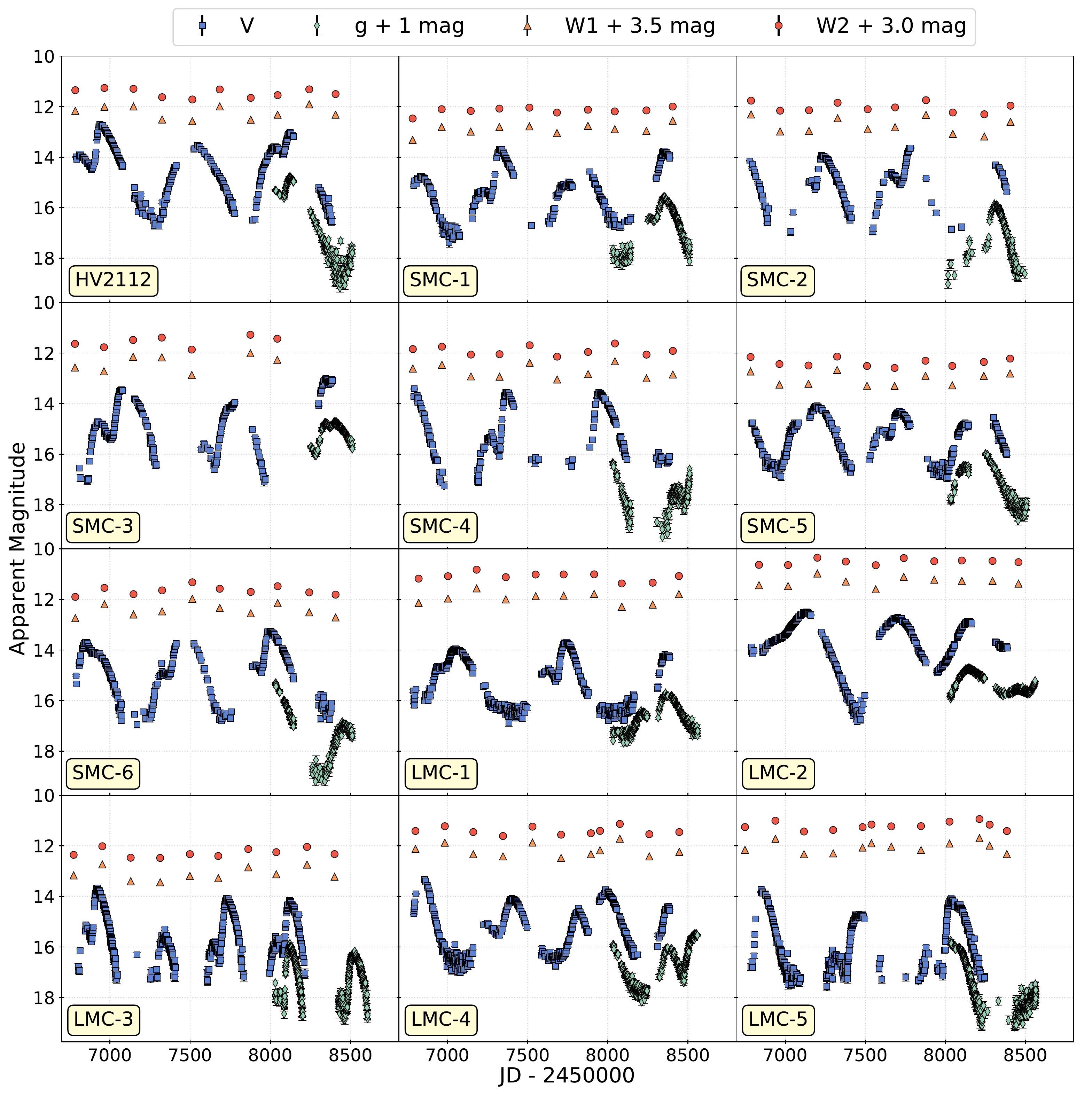}
    \caption{ASAS-SN and NEO-WISE light curves of the HLOs and HV2112. Blue squares are ASAS-SN V-band, green diamonds are ASAS-SN g-band, orange triangles are NEO-WISE W1-band, and red circles are NEO-WISE W2-band photometry. The g-, W1-, and W2-band photometry are offset by 1.0, 3.5, and 3.0 magnitudes respectively, so as to aid in viewing. The panels are labeled with the names of the sources (see Table \ref{hlo_info}).}
    \label{all_hlos}
\end{figure*}

\subsection{Sample Completeness and Purity}\label{sec_sample_complete}

In order to use the number of identified HLOs to estimate the intrinsic rate/lifetime of their evolutionary state, it is critical that the completeness and purity of the sample is understood. The initial color cuts described in Section~\ref{sec_phot_select}, were designed to be complete for stars with temperatures between 4800 K and 3200 K and $\log (L/L_\odot)$ $>$ 4.2 (in the LMC) or $\log (L/L_\odot)$ $>$ 4.3 (in the SMC). However, there are a number of factors that may impact this.

First, our initial color cuts were performed using single epoch data from the 2MASS survey. Yet, HV2112 has historical variability of $\sim$1 mag in the near-infrared (NIR) bands \citep{Glass.I.1979.MidIRCloudsHV2112}. Additionally, the HLOs have variability in the NEO-WISE bands of  $\Delta \mathrm{W1} \mytilde 0.7$ mag and $\Delta \mathrm{W2} \mytilde 0.5$ mag (see Section~\ref{sec_neowise}). Thus, if a star with a \emph{mean} K$_{\mathrm{s}}$-band magnitude (and hence luminosity) only slightly above our cutoff were to exhibit similar levels of NIR variability, it could have be excluded from our sample if it was in the low point of its light curve when the 2MASS data was taken. To quantify this, we examine the number of highly variable stars in our sample with K$_{\mathrm{s}}$-band magnitudes within 0.5 mag of our adopted cutoffs. While all of the identified HLOs are brighter than this cutoff in the NIR, 7 out of 27 HAVs (all in the LMC) have K$_{\mathrm{s}}$-band magnitudes within 0.5 mag of our limit. If we assume that roughly half these stars were observed above their mean K$_{\mathrm{s}}$-band magnitudes, then we estimate that $\lesssim$3 highly variable stars with mean K$_{\mathrm{s}}$-band magnitudes above our threshold may have been excluded due to the timing of the 2MASS observations. However, we emphasize that these would be among the lowest luminosity stars in our sample, and may not exhibit the double-peaked morphology of HV2112. 

Second, because we assessed variability using ASAS-SN V-band light curves---with a limiting magnitude of $\sim$17 mag---our sample is not sensitive to some stars with (V$-$K$_{\mathrm{s}}$) $>$ 6.5 mag. While a star with a similar V$-$K$_{\mathrm{s}}$ color to HV2112 (V$-$K$_{\mathrm{s}}$ $\sim$ 5.5 mag) or a temperature of $\sim$3200 K would be easily detected in ASAS-SN over the full range of K$_{\mathrm{s}}$-band magnitudes considered, some cool stars are heavily dust enshrouded, leading to extreme V$-$K$_{\mathrm{s}}$ colors \citep[e.g.][]{vanLoon.J.2005.MLRformulaRSGAGB}. Indeed, 242 out of the 3530 red and luminous stars were not recovered in ASAS-SN -- 90\% of their photometric points were upper limits -- despite passing the 2MASS color cuts. The impact of our insensitivity to heavily dust enshrouded stars on our physical interpretation of the HLOs will be discussed in \S\ref{sec_discussion}. 

Finally, our sample composition is influenced by our threshold for classifying stars as likely foreground stars based on their \emph{Gaia} astrometry in \S\ref{sec_gaia_cuts}. To err on the side of sample completeness, we chose a generous threshold, only eliminating stars with kinematics outside the region occupied by 99.5\% of likely SMC/LMC members. However, to assess whether this threshold could have eliminated any additional HLOs with unusual kinematics, we also examine the variability of all stars with 12.8 $<$ $\mu^{T} \sigma^{-1} \mu$ $<$ 50. Two additional stars exhibit variability amplitudes $>$2.5 mag. One shows a symmetric light curve characteristic of a Mira variable, while the other may have been classified as an HAV if not eliminated based on kinematics.

Conversely, we also investigate whether this generous threshold may have impacted our sample purity by failing to remove some bona fide foreground stars. To do so, we select a control field with a radius of 10 degrees centered on ($\alpha$,$\delta$) (J2000) $=$ (22:13:28.17, $-$63:06:55.92). This field was chosen to be at a similar Galactic latitude as the Clouds. We identify 1168 stars in this field that pass the color and magnitude cuts of \S\ref{sec_phot_select} and cross match them with \emph{Gaia} DR2. Only 5 stars would have passed the kinematic filters applied in \S\ref{sec_gaia_cuts} (3 for the LMC, 2 for the SMC). Given that Galactic kinematics may be slightly different in the direction of the Clouds than in the control field, we examine the light curves for all 73 control field stars that have $\mu^{T} \sigma^{-1} \mu$ $<$ 50 when compared to the kimematics of either the LMC or SMC. No high amplitude variables are identified.

Thus, we conclude that our sample should be complete to within $\sim$ a few stars, and while it is not impossible for an individual star in our sample to be a foreground dwarf, contamination should be minimal. The impact of these small uncertainties on our implied rates and lifetimes will be addressed in Section~\ref{sec_lifetimes}.

\section{Photometric Observations of HLO\lowercase{s}}\label{sec_observations}

To assess the nature of the 12 HLOs, we gathered additional multi-wavelength photometric data and light curves from a number of surveys.

\subsection{ASAS-SN Image Subtraction Light Curves}\label{sec_asassn_imgsub}

We extract image subtraction light curves of the 12 HLOs from ASAS-SN. Light curves are produced for both the V-band (May 2014 to Sept.\ 2018) and g-band (Sept.\ 2017 onwards). There is one year of overlap where both V$-$ and g$-$band are available. Light curves were produced as described by \citet{2018MNRAS.477.3145J} using the ISIS image subtraction software \citep{1998ApJ...503..325A,2000A&AS..144..363A} and subsequently preforming aperture photometry on the subtracted images with a 2 pixel radius aperture. Image subtraction is performed on a co-add of the  $\sim$2$-$3 images taken on each night, thus achieving a deeper limiting magnitude than the aperture photometry light curves used for sample selection. This allows for a more accurate measurement of the full variability amplitude of the HLOs. Photometric errors were recalculated as described by \citet{2019MNRAS.485..961J} and the zero point offsets between the different cameras were corrected as described by \citet{2018MNRAS.477.3145J}. Calibration was performed using stars from APASS \citep{2015AAS...22533616H}. The resulting light curves are shown in Figure~\ref{all_hlos}.

\subsection{NEO-WISE Light Curves}\label{sec_neowise}

We utilize infrared light curves for the 12 HLOs from the 2019 data release of the NEOWISE mission \citep{Mainzer.A.2011.NEOWISE}, retrieved from the Infrared Science Archive (IRSA). Since 2013, the WISE satellite has repeatedly surveyed the entire sky in the W1 (3.4$\mu$m) and W2 (4.6$\mu$m) bands, primarily searching for near-Earth objects. NEOWISE takes a new exposure every 11 seconds and, due to its survey strategy, the Magellanic Clouds are imaged repeatedly over a $\sim$2 day period every $\sim$180 days. We consider each of these $\sim$2 day observation periods as an individual epoch. 

For each epoch we average all the observations, after clipping at 4-sigma around the mean and removing any points with a quality flag less than 10. The standard deviation of the sigma-clipped magnitudes is included in our final photometric errors for each epoch. The resulting light curves are shown in Figure~\ref{all_hlos}. All HLOs show variability of $\lesssim$1 mag in these mid-IR bands. It is unclear if the IR light curves exhibit a similar double-peaked morphology to the visible due to the low cadence of the NEOWISE observations. For each star, there are between 6 and 9 epochs with contemporaneous NEOWISE and ASAS-SN data.

\subsection{Additional Archival Photometry}\label{sec_photometry}

In addition to the contemporaneous light curves from ASAS-SN and NEOWISE, we make use of single-epoch photometry from a variety of surveys to compare the HLOs to known classes of stars.  We use photometry from 2MASS (J-, H-, and K$_{\mathrm{s}}$-bands; \citealt{2MASS.Skrutskie.2006}), the WISE All-Sky Survey (3.4-, 4.6-, 12-, and 22-$\mu$m; \citealt{WISE.Wright.E.2010}), and the Spizter SAGE survey (3.6-, 4.5-, 5.8-, 8.0-, and 70$\mu$m; \citealt{Meixner.M.2006.SAGE}).

\subsection{Distance and Reddening}\label{sec_dist_met_red}

Throughout this paper, we adopt distances to the SMC and LMC of 61 kpc \citep{Hilditch.R.2005.SMCDistance} and 50 kpc \citep{Pietrzynski.G.2013.LMCDistance}, respectively. We correct the photometry for reddening in both the Galaxy and the Magellanic Clouds. Extinction curves for the Magellanic Clouds were obtained from \citet{Gordon.Karl.2003.MCExtinction}, and the reddening was estimated using the Zaritsky MCPS extinction maps for cool stars \citep{Zaritsky.D.2002.MCPSSMC,Zaritsky.D.2004.MCPSLMC}. For comparison samples of Galactic sources, extinctions were estimated using the \citet{Schlafly.Edward.2011.GalacticReddening} reddening map, obtained through the IRSA database.

\section{Observed Properties}\label{sec_properties_observations}

Here we outline the observed properties of our HLOs, such as variability, color, and magnitude, and compare them to other classes of red and luminous stars to determine whether the HLOs appear to be a unique class.

\begin{figure*}[ht!]
    \centering
    \includegraphics[width=0.85\textwidth]{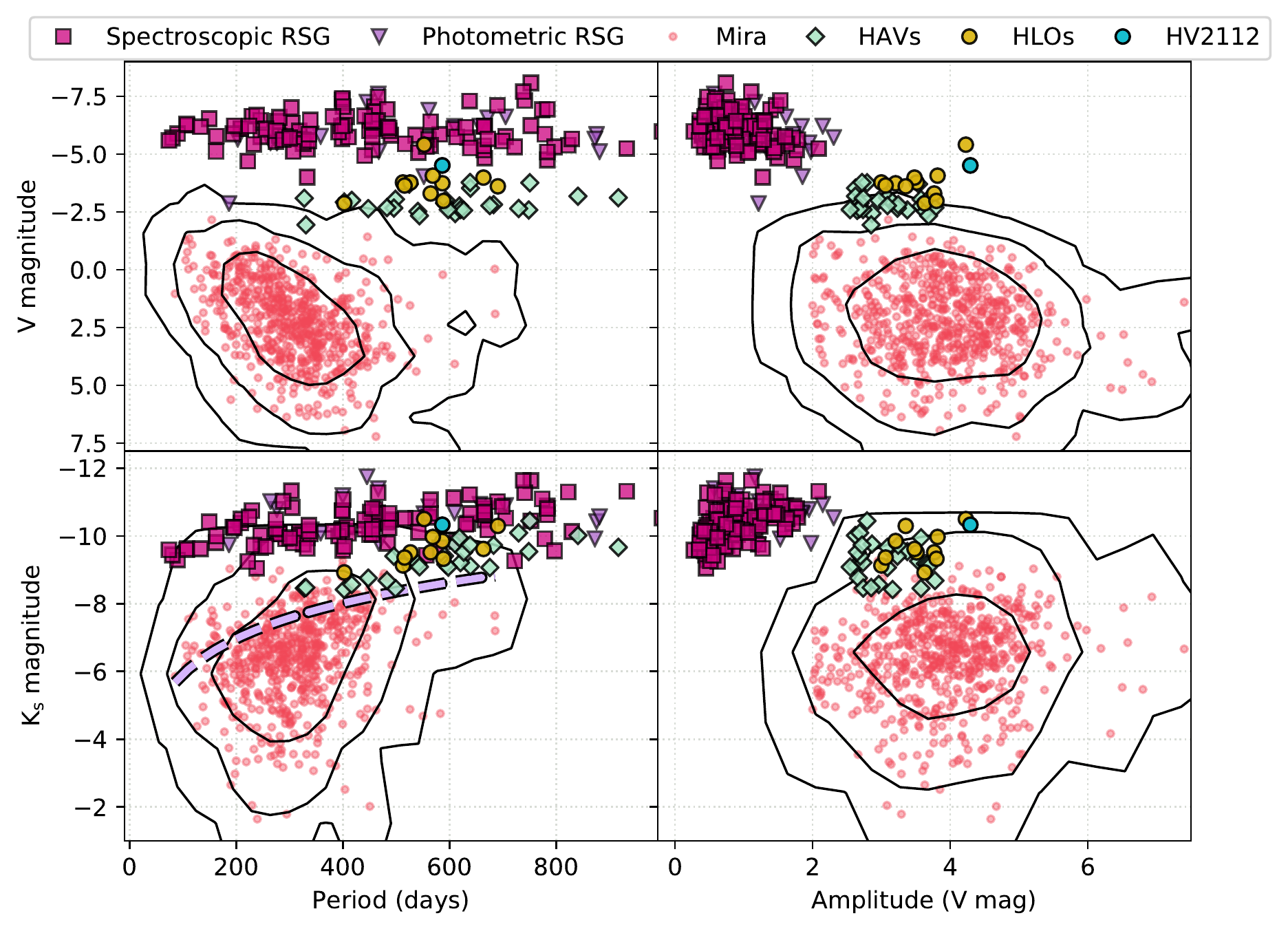}
    \caption{Magnitudes, periods, and amplitudes of the HLOs, HAVs, and other luminous red variable stars. HLOs are shown as gold circles with HV2112 in blue, other High Amplitude Variables as light green diamonds, spectroscopically confirmed RSGs as dark magenta squares, photometrically selected RSGs as violet triangles, and Mira variables as smaller red circles. The contours surrounding the Mira distribution represent the 1-, 2-, and 3-$\sigma$ levels of the density distribution of Mira sources. \textit{Top left:} Absolute mean V magnitude vs Period. \textit{Top right:} Mean V-band magnitude vs V-band amplitude. \textit{Bottom left:} Absolute 2MASS K$_{\mathrm{s}}$-band magnitude vs period. The PLR for Mira variables \citep{Feast.M.1989.MiraPLR} is included as a dashed violet line. \textit{Bottom right:} Absolute 2MASS K$_{\mathrm{s}}$-band magnitude vs V-band Amplitude.}
    \label{phase_space_plots}
\end{figure*}

\subsection{Period Determination}\label{sec_lpvs}

We estimate the periods of the 39 HLOs and HAVs using Lomb-Scargle Periodograms \citep{Lomb.N.1976.LombScargle,Scargle.J.1982.LombScargle}, calculated with the \texttt{LombScargle} feature in the \texttt{astropy} Python package \citep{Astropy.Collab.2018.Astropy}. Periods were estimated for each HAV by selecting the frequency of maximum power, after excluding frequency peaks caused by aliasing. Alias peaks were identified by running the periodogram on a flat light curve with the cadence of ASAS-SN observations. Periods for all HLOs are listed in Table~\ref{hlo_info}. While period was not included as a selection criteria, no HLO has an observed period less than 400 days.

We emphasize that these periods are estimates. For stars with periods on the order of HV2112's \mytilde 600 days, the ASAS-SN light curves cover only 2 or 3 cycles, and the cycle-to-cycle periodicity of these stars can vary slightly. This behaviour has also been observed in other long period variables such as Miras \citep{Zijlstra.A.2002.MiraPeriodChanges,Neilson.H.2016.PeriodChanges}. All light curves were visually inspected to ensure that the estimates from the Lomb-Scargle Periodograms were reasonable, and we apply a systematic error of $\pm$ 30 days to our final values.

\subsection{Comparison to Known Stellar Classes: Variability}\label{sec_phase_space}

\begin{figure*}[ht!]
    \centering
    \includegraphics[width=0.85\textwidth]{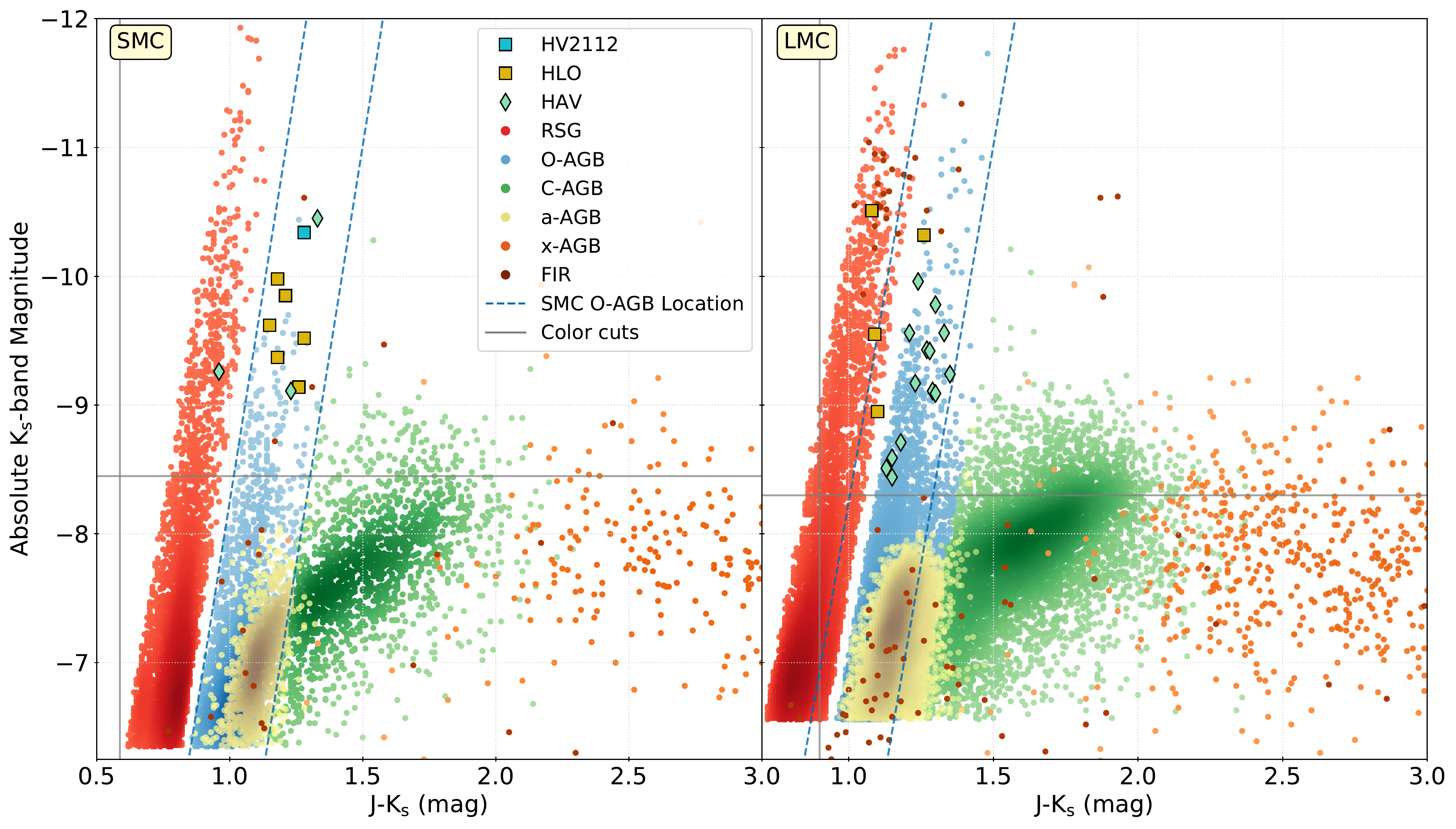}
    \caption{J $-$ K$_{\mathrm{s}}$ vs K$_{\mathrm{s}}$ color-magnitude diagrams for the SMC (left) and LMC (right). HLOs are highlighted as gold, and HV2112 in cyan. HAVs are green diamonds. The circles show RSGs (red), O-rich AGB (blue), C-rich AGB (green), `anomalous'-AGB (yellow), extreme IR AGB (orange), and far-IR objects (dark brown), based on the classification of \citet{Boyer.M.2011.SAGE.MC.Photom}. The gap of 0.05 mag between the RSG and O-rich AGB branch was included by \citet{Boyer.M.2011.SAGE.MC.Photom} to avoid contamination between the classes. The solid grey lines indicate the photometry color cuts described in Section \ref{sec_phot_select}, while the dashed blue lines indicate the location of the O-rich AGB branch in the SMC (on both plots). These magnitudes are not extinction corrected, but the corrections will be small.}
    \label{boyer_cmd}
\end{figure*}

In Figure~\ref{phase_space_plots} we show the absolute V-band and K$_{\mathrm{s}}$-band magnitudes versus variability amplitude and period for the HLOs (gold circles) and HAVs (green diamonds) in comparison to a sample of 132 RSGs (magenta squares and violet triangles) and 593 Mira variables (red circles).

The comparison sample of RSGs plotted in Figure~\ref{phase_space_plots} are all members of the Magellanic Clouds, and are pulled from multiple sources to ensure both purity and completeness. First, we take the spectroscopically confirmed RSGs from \citet{Massey.P.2003.RSGinMC} and \citet{Davies.B.2018.RSGHDL} that were in our original sample of luminous and cool stars and have measured periods from the ASAS-SN catalog of variable stars \citep{Jayasinghe.T.2018.ASASSNVariables}. These 112 stars are plotted as magenta squares. Second, we identify an additional 20 stars (violet triangles) which were photometrically selected as RSGs based on their IR colors in 
\citet{Yang.M.2011.RSGsinLMC,Yang.M.2012.PLRSGSMC} and are not contained in the other samples. While this photometric sample avoids potential bias in the spectral types selected for spectroscopic follow-up, we note that some contamination is also possible. Indeed, 9 photometric RSGs from \citet{Yang.M.2011.RSGsinLMC,Yang.M.2012.PLRSGSMC} were identified as HLOs or HAVs in our sample. Other than these cases, the photometric RSG sample occupies a similar portion of phase space as the spectroscopically confirmed RSGs.

The comparison sample of Mira variables plotted in Figure~\ref{phase_space_plots} are Galactic stars with variability properties retrieved from the ASAS-SN Variable Star Database \citep{Jayasinghe.T.2018.ASASSNVariables}. The full ASAS-SN sample of Galactic Mira variables was restricted to stars with a classification probability greater than 99.7\%, $\emph{Gaia}$ parallax uncertainty $<$ 20\%, and in a direction with total Galactic V-band extinction less than 1.0 mag.

Some highly variable sources drop below the ASAS-SN detection limit for part of their variability cycle, and hence the V-band amplitudes will be underestimated and mean magnitudes overestimated. To characterize the effect of the latter, we compare the median magnitudes for all stars, including upper limits from non-detections, to the mean magnitudes. For stars with no upper limits, the difference between median and mean magnitudes are $\lesssim$0.2 mag. For the HLOs, the average difference between median and mean V-band magnitudes is 0.6 mag and the maximum is 1.4 mag.

The K$_{\mathrm{s}}$-band magnitudes are all single epoch 2MASS observations. While the HLOs, HAVs, and Miras may show K$_{\mathrm{s}}$-band variability of up to \mytilde1 mag, which may impact the location of individual stars on these plots, we do not expect a systematic shift between classes of objects. All magnitudes are extinction corrected.

From Figure~\ref{phase_space_plots}, we see that most HLOs and HAVs have properties that are inconsistent with the bulk of the RSG and Mira populations. While their V-band variability amplitudes are similar to those of Mira variables, they are completely disjoint from the RSGs (which all have $\Delta$V $<$ 2.5 mag). Conversely, the HLOs and HAVs have periods similar to the RSGs (300 days $<$ P $<$ 900 days), but significantly longer periods than typical Mira variables: 92\% (74\%) of the HLOs (HAVs) have periods greater than 500 days, whereas only 2\% of Mira variables have periods this long.

In mean V-band magnitude, the HLOs and HAVs lie between the Mira variables and the RSGs, with only LMC-2 overlapping with the bulk of the RSG population. In K$_{\mathrm{s}}$-band, the HLOs and the HAVs lie on the extreme bright end of Mira variables and show a distinct period-luminosity relationship (PLR). The longest period HLOs and HAVs begin to overlap with the distribution of RSGs. In the lower left panel of Figure~\ref{phase_space_plots}, we also plot the PLR for O-type Mira variables in the LMC from \citet{Feast.M.1989.MiraPLR}; the slope and zero point of this PLR is invariant with metallicity within uncertainties, as determined by \citet{Feast.M.2004.AGBMiraPLR}. Once again, LMC-2 is an outlier, showing a K$_{\mathrm{s}}$-band magnitude more consistent with RSGs of a similar period rather than the HLOs/HAVs. The HLOs/HAVs follow a similar slope, but are offset to higher luminosities. Notably, \citet{Whitelock.P.2003.AGBLumLi} find that high amplitude pulsating AGB stars with strong lithium lines (HV2112 also has strong lithium features) often have luminosities higher than predicted by the Mira PLR.  

Overall, the HAVs show broadly similar properties to the HLOs. However, on average they possess lower magnitudes for a given period and smaller variability amplitudes. After constructing a 4-dimensional distribution based on the parameters plotted in Figure~\ref{phase_space_plots}, we find that all HLOs and all but 2 HAVs are $>$ 3$\sigma$ outliers compared to the population of Galactic Mira variables. This extreme inconsistency is driven in large part by the high V-band magnitudes of the HLOs/HAVs. When constructing either a 3-dimensional distribution (M$_\mathrm{K_{s}}$-Period-Amplitude) or 2-dimensional distribution (M$_\mathrm{K_{s}}$-Period) we find that 92\% of HLOs and 74\% of HAVs are still $>$ 2$\sigma$ outliers (i.e. lying outside the parameter space occupied by 95\% of Mira variables). A single HLO (LMC-3) and 7 HAVs, those with the lowest luminosities and shortest periods, more closely overlap with the luminous end of the Mira distribution. Below we consider whether the physical properties of these objects are consistent with an intrinsically rare, high-luminosity extension of Mira variables and the implications if the HAVs are members of the same physical class as the HLOs.

\subsection{Comparison to Known Stellar Classes: Color}\label{sec-cmag-jk}

\begin{figure*}
    \centering
    \includegraphics[width=0.9\textwidth]{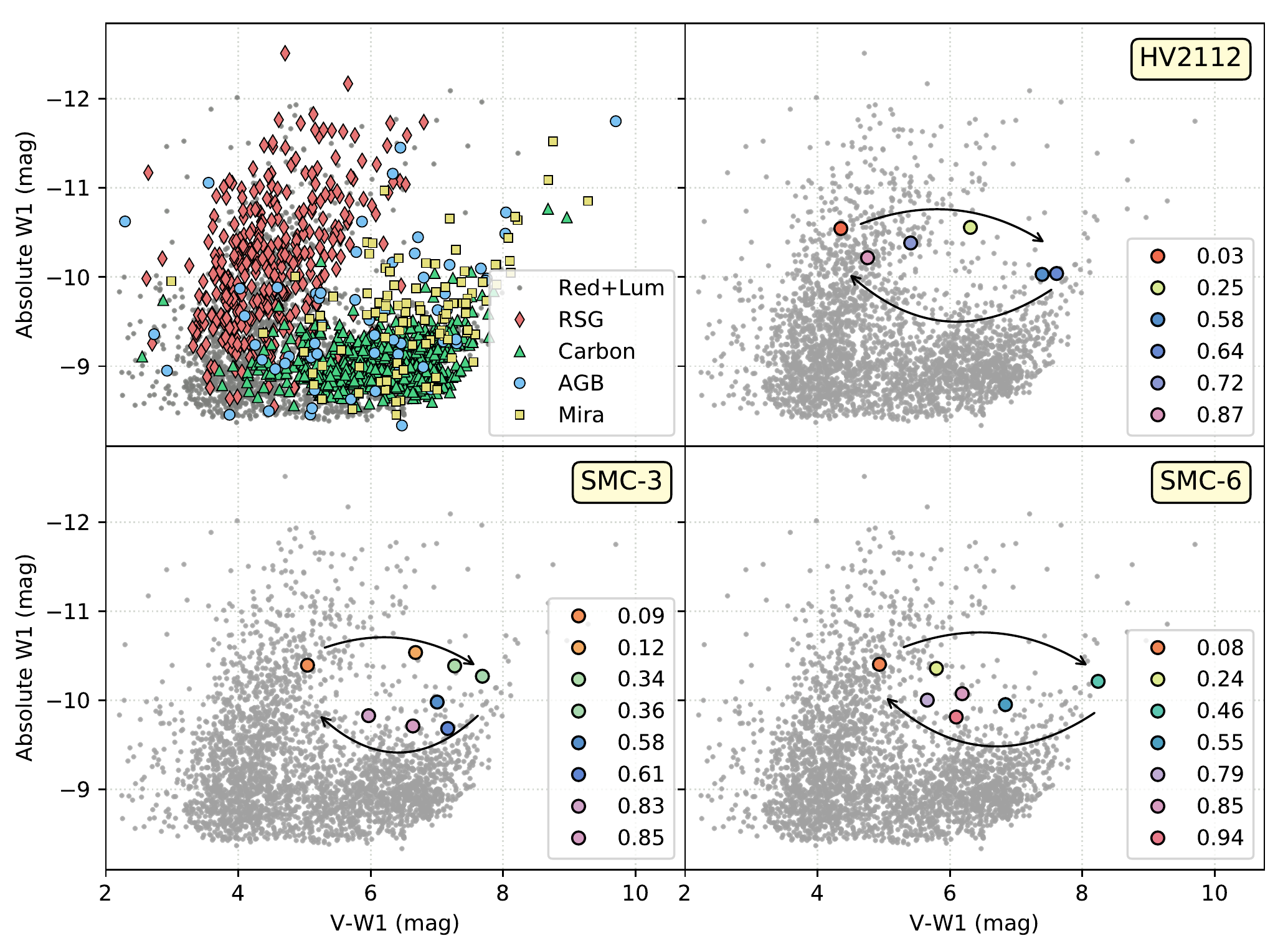}
    \caption{V$-$W1 vs W1 (WISE 3.4 micron) color-magnitude diagrams. The V-magnitudes are the mean values from the ASAS-SN light curves, and the W1 magnitude comes from the WISE survey. The magnitudes in these plots are not extinction corrected. \textit{Top left:} The grey points are the \mytilde 3500 luminous red stars towards the LMC and SMC from our initial selection. These were cross-matched to SIMBAD, and some relevant classifications are shown. Spectroscopically verified RSGs are red diamonds, AGB stars are blue circles, carbon stars are green upward-pointing triangles, and Mira variables are yellow squares. The RSGs largely occupy the blue ``arm'' of the diagram, with the Mira, carbon, and AGB stars occupying the red ``arm''. \textit{Top right:} V$-$W1 vs W1 color-magnitude diagram, with the same red and luminous stars from the top left panel in grey. The colored points correspond to times when we had photometry from both ASAS-SN in V and NEO-WISE in W1 for the star HV2112. The points are colored by their phase, with the same coloring as in Figure~\ref{hv2112_lc}. The peak of the light curve (phase $=$ 0) appears in the left arm of the diagram, while the phase corresponding to the trough (phase $=$ 1/2) lies in the right arm. Arrows show the overall motion. \textit{Bottom left:} The same, but for SMC-3. \textit{Bottom right:} The same, but for SMC-6.}
    \label{phase_motion_diagrams}
\end{figure*}

In Figure~\ref{boyer_cmd} we show a \emph{2MASS} color magnitude diagram to compare HLOs and HAVs to the sample of nearly 150,000 (440,000) red stars in the SMC (LMC) from \citet{Boyer.M.2011.SAGE.MC.Photom}. Using data from Spitzer SAGE survey and color cuts from \citet{Cioni.M.2006.ColorcutSAGBs}, \citet{Boyer.M.2011.SAGE.MC.Photom} differentiate between RSGs, Red Giant Branch stars, and various types of AGB stars (see \citealt{Boyer.M.2011.SAGE.MC.Photom,Boyer.M.2015.SAGE.MC.Spectra} for more information). All these data are based on single-epoch surveys.

In the SMC, all the HLOs lie within the region of color-magnitude space occupied by the most luminous oxygen-rich AGB stars.  In the LMC, the situation is more complex. While LMC-1 also falls in the high luminosity end of the O-rich AGB branch, LMC-2, -3, and -4 all lie in the region of color-magnitude space occupied by RSGs. LMC-5 did not have a counterpart in the SAGE survey. However, we note that the location of the RSG and AGB branches are metallicity dependent and all LMC HLOs other than LMC-3 are located in the outskirts of the LMC (Figure~\ref{locations}). The outer portions of the LMC tend to have lower metallicity, with some regions possessing [Fe/H] $\leq$ $-$0.67 \citep{Choudhury.S.2016.LMCMetallicity}, more typical of the SMC.
Dashed blue lines designating the edges of the SMC O-rich AGB branch are also shown in the right panel of Figure~\ref{boyer_cmd}. Based on these cuts LMC-3 would also be classified as an O-rich AGB star in the SMC, and LMC-4 is extremely close to this boundary. LMC-2 occupies the RSG portion of color-magnitude space regardless of metallicity.

All of the HAVs lie in the O-rich AGB branch, with the exception of one SMC HAV.

\subsection{Color and Magnitude Evolution}\label{sec-cmag-vw1}

In order to assess how the colors of the HLOs vary throughout their pulsation periods, we have constructed ``phase motion'' color-magnitude diagram using contemporaneous ASAS-SN V-band and NEOWISE W1-band observations. In Figure~\ref{phase_motion_diagrams} we show the V$-$W1 vs.\ W1 color-magnitude diagrams for the 3288 luminous and red stars identified in \S\ref{sec_sample_selection}. In the upper left panel we color-code these sources based on their SIMBAD classifications. We distinguish between RSGs, AGB stars, carbon stars, and Mira variables. As above, RSGs are well separated from the various types of AGB stars in color-magnitude space.

In the remaining panels of Figure~\ref{phase_motion_diagrams}, we again show this full sample---whose magnitudes were derived from single-epoch surveys---in grey, and overlay the contemporaneous ASAS-SN and NEO-WISE observations of HV2112, SMC-3, and SMC-6. Points are color coded as in Figure~\ref{hv2112_lc} such that the peak of the V-band light curve is colored red and the trough is green-blue. We include arrows to illustrate the movement of the stars in color-magnitude space as they vary.  During their light curve peaks, HLOs occupy the same region of color-magnitude space as RSGs, while at minimum they occupy a similar region to the cool AGB stars, changing by $\gtrsim$3 mag in V$-$W1.  This indicates that the HLOs are likely undergoing substantial temperature and radius variations throughout their pulsation cycle. We quantify this in Section~\ref{sec_properties_physical}.

\section{Derived Physical Properties}\label{sec_properties_physical}

\subsection{Temperatures and Luminosities}

In order to constrain how the temperatures and luminosities of the HLOs vary, we model their spectral energy distributions (SEDs) at multiple epochs through their pulsation cycles. These results are then compared to the temperatures and luminosities of control samples of stars and theoretical models.

\subsubsection{Spectral Energy Distribution Construction}

We construct SEDs for the HLOs at multiple points through their pulsation cycles using the contemporaneous ASAS-SN and NEOWISE data described above. We consider only epochs with both ASAS-SN and NEOWISE data, as together they sample both sides of the SED peak, allowing for better temperature constraints.

In addition, we supplement the ASAS-SN/NEOWISE g$-$, V$-$, W1$-$, and W2$-$band observations with the W3$-$band (12 $\mu$m) data from the WISE All-Sky Survey. Observations at these wavelengths provide stronger constraints on the quantity of dust/mass loss surrounding cool and massive stars. In order to account for possible low-level variability in the mid-IR and the non-simultaneous nature of the W3 observations, we apply a systematic error of 20\% to the W3 fluxes. Following \citet{Adams.Scott.2017.RSGMCMCFitter} we also adopt a minimum flux error of 10\% in all other observed bands. In total, we construct SEDs at 4$-$8 phases for each HLO, with 4$-$5 data points per phase.

\subsubsection{SED Modeling with MARCS \& DUSTY}\label{sec_sed_marcs}

\begin{figure}
    \centering
    \plotone{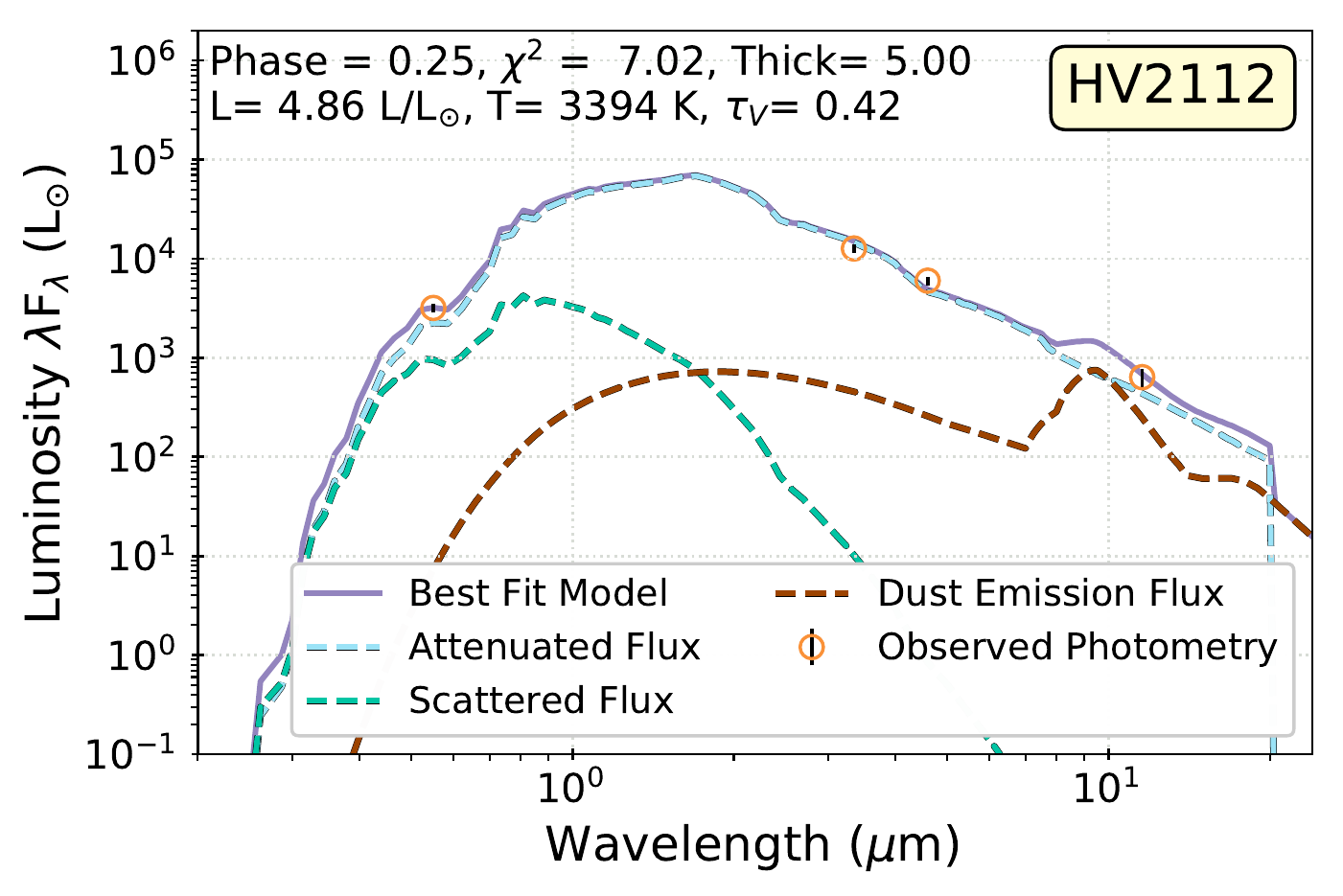}
    \plotone{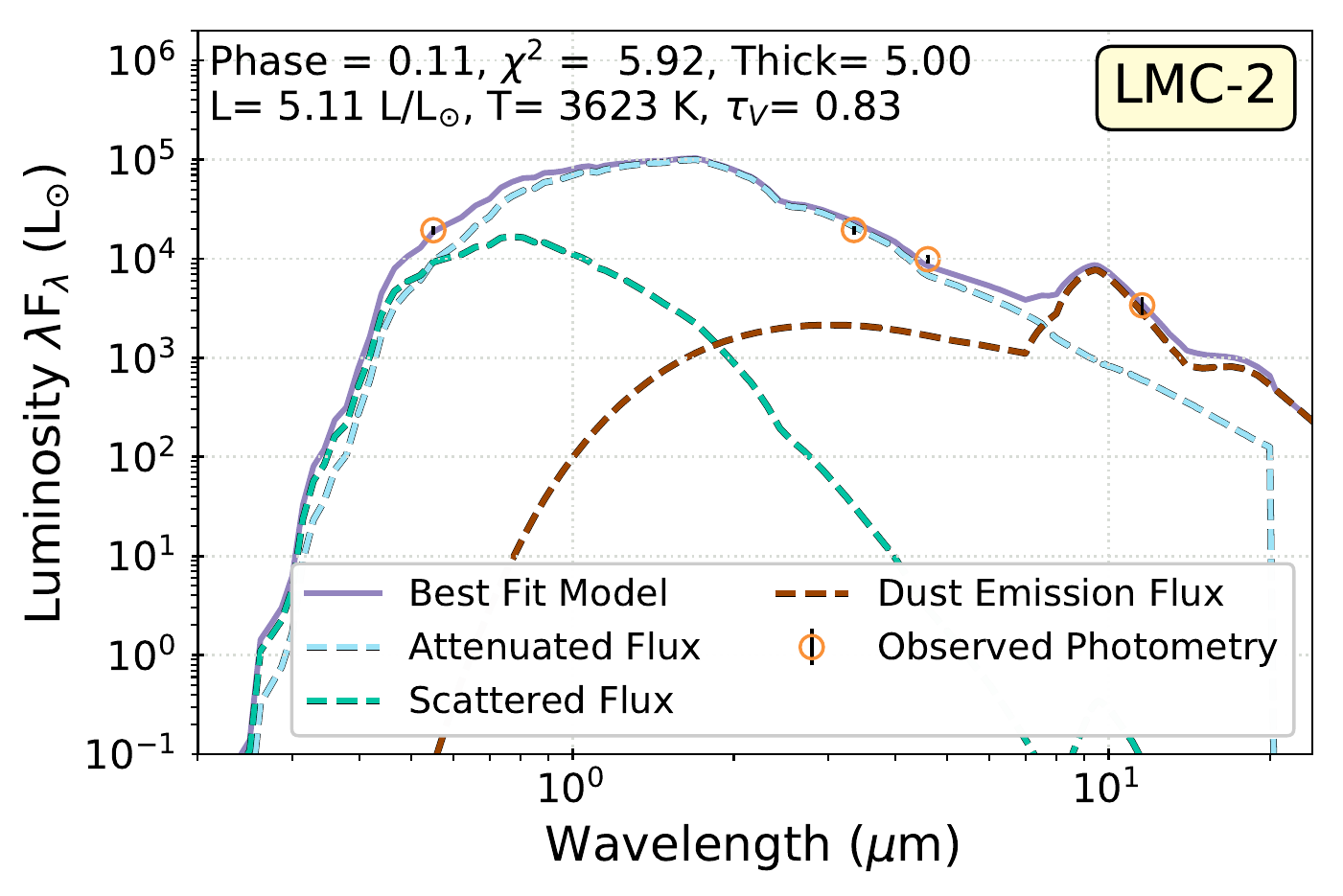}
    \caption{Model spectral energy distributions of one phase of the light curve for HV2112 (top) and LMC-2 (bottom). The best-fit model SED obtained from DUSTY is shown as a violet solid line; the observed photometry for that phase are shown as orange circles. The phase, reduced $\chi^{2}$ of the fit, the model temperature, luminosity, and $\tau_{\text{V}}$ values are in the upper left corner each panel.}
    \label{hv2112_model_ex}
\end{figure}

To estimate temperatures and luminosities for the HLOs, we use a Markov chain Monte Carlo (MCMC) wrapper \citep{Adams.Scott.2017.RSGMCMCFitter} that combines MARCS stellar atmosphere models \citep{Gustafsson.B.2008.MARCSModels} with the DUSTY radiative transfer code \citep{Nenkova.M.2000.DUSTY}. We reprocess the MARCS model spectra through DUSTY because luminous and cool stars---such as RSGs and thermally pulsing AGBs---can be dust enshrouded due to mass loss \citep[e.g.][]{vanLoon.J.2005.MLRformulaRSGAGB}.

We use MARCS stellar atmosphere models with temperatures that range from 2600 K to 4600K, have solar composition, a surface gravity $\log$(g) = $-$0.5, and a mictroturbulent velocity of 5 km/s. Additionally, we assume a density distribution approximating a steady-state wind, and set the thickness of the dust shell (the ratio of outer and inner boundary) to 5 (varying the thickness parameter has been shown not to have a large affect on other model properties; \citealt{Adams.Scott.2017.RSGMCMCFitter}). We assume a silicate dust composition \citep{Draine.B.1984.PropertiesSilicateGrains} because the HLOs are most consistent with O-rich AGB stars (Figure~\ref{boyer_cmd}), and a standard MRN grain size distribution \citep{Mathis.J.1977.MRNDist}. The remaining free parameters within the MCMC wrapper are the effective temperature of the star, the dust temperature at the inner boundary of the dust shell, and dust opacity at V-band, $\tau_{\text{V}}$.

The MCMC wrapper requires an initial guess of these free parameters. We perform the fitting in a two-step process. We first execute a short MCMC run with large step sizes. The best-fit stellar temperature from this run is then the input as the initial guess for a longer MCMC run. At each MCMC step, the $\chi^{2}$ value between the observed photometry and synthetic photometry on the model spectrum is computed. Final best-fit values and errors for stellar T$_{\rm{eff}}$, $\log$(L/L$_\odot$), and $\tau_{\text{V}}$ are based on 4000 accepted MCMC trials. The luminosity $\log$(L/L$_{\odot}$) is calculated by integrating under the output model SED from DUSTY. Best-fit models for one phase of HV2112 and LMC-2 are shown in Figure~\ref{hv2112_model_ex} to illustrate the contribution to the SED from absorbed emission, scattered emission, and dust. The slopes of the SEDs going into the mid-IR are not consistent with Rayleigh Jeans tails, and thus require dust.

\subsubsection{Model Fitting Results}\label{sec_model_results}

\begin{figure}
    \centering
    \plotone{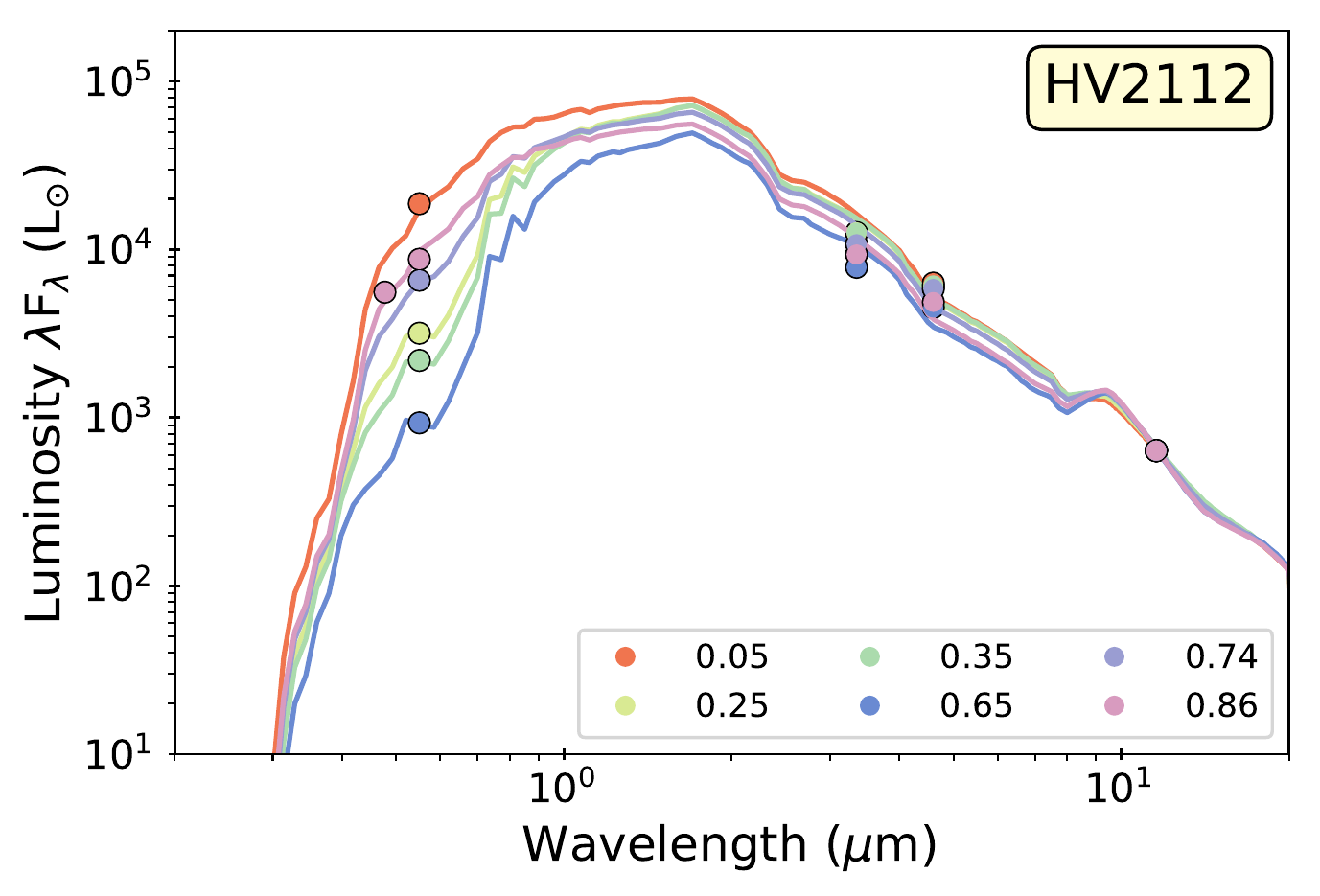}
    \plotone{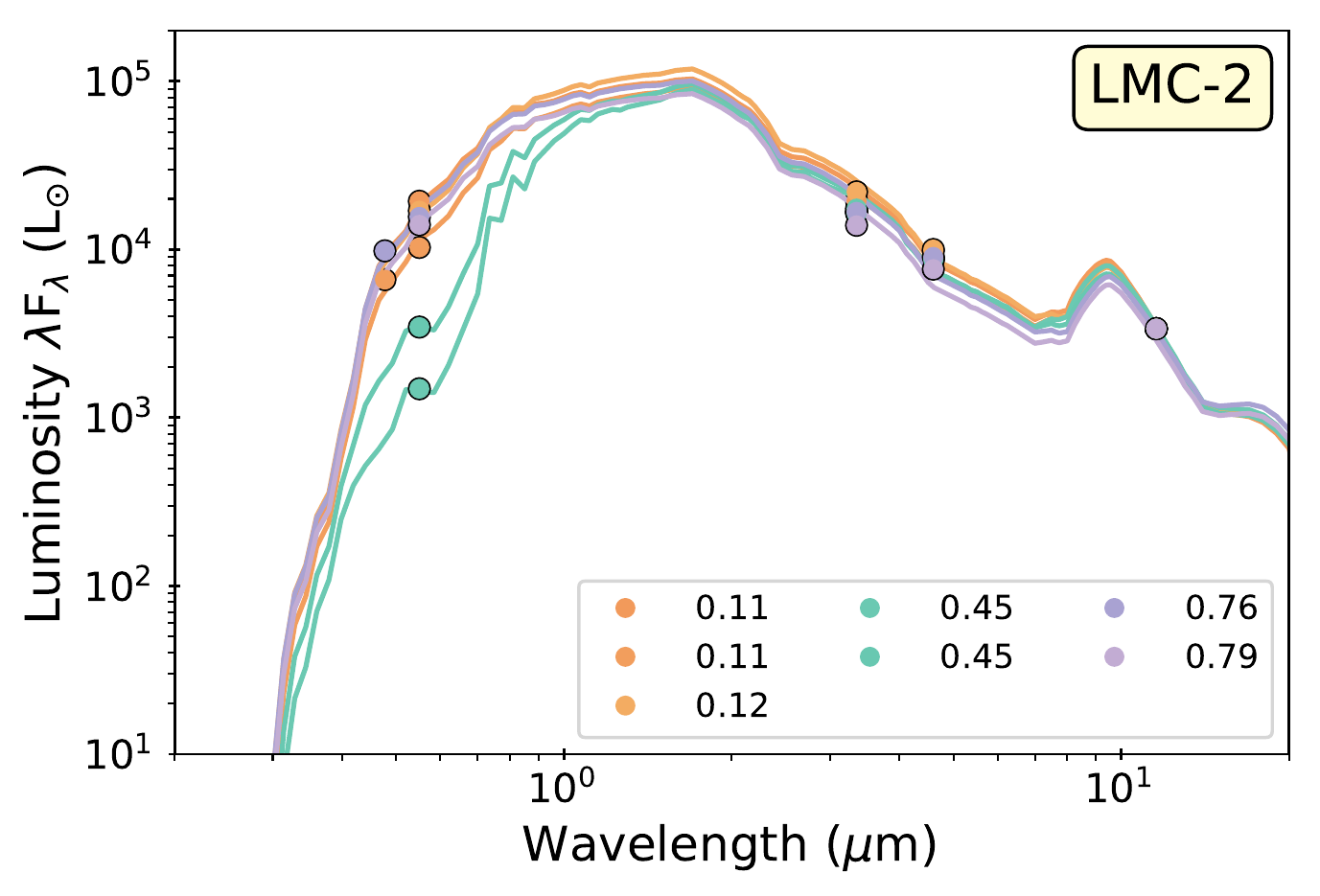}
    \caption{Multi-epoch spectral energy distribution of two HLOs. The best-fit model is plotted as a line, and the observed photometry as circles. Each model is fit to a different phase, and the colors indicate the phase with the same colormap as Figure~\ref{hv2112_lc}. A phase of 0 corresponds to the peak of the light curve and is red. The photometric point at 12 microns (WISE W3) is not contemporaneous with the rest of the ASAS-SN and NEO-WISE data.}
    \label{multi_epoch_seds}
\end{figure}

In Appendix \ref{mcmc_appendix} (Table~\ref{mcmc_results}) we present the full results of the MCMC fitting with DUSTY and MARCS, as well as an example of the posterior distributions for the fits to one of the phases of HV2112. Overall, we find temperatures that range from 3250 K to 3600 K and luminosities that range from $\log$(L/L$_\odot$) $\sim$ 4.15 to 5.15. Throughout their pulsation cycles, HLOs exhibit larger fractional variations in their luminosities than their temperatures. We find temperature variations of $\sim$6$-$9\% and luminosity variations of $\sim$60$-$95\%. This is highlighted in Figure~\ref{multi_epoch_seds} where we show the multi-epoch SEDs and best-fit models for HV2112 and LMC-2.

\subsection{Hertzsprung-Russell Diagrams}\label{sec_hrd}

\begin{figure}
    \centering
    \includegraphics[width=\linewidth]{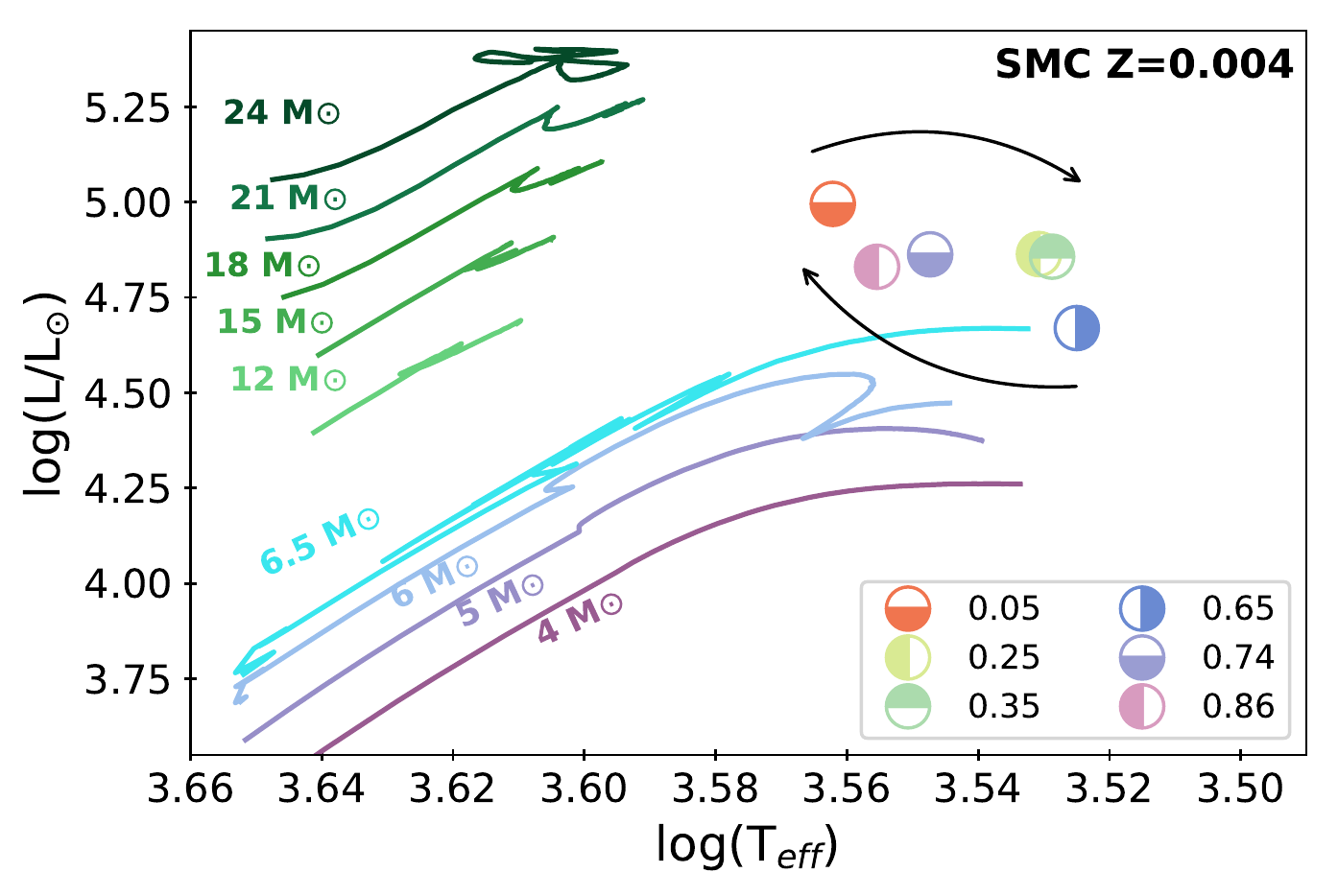}
    \caption{Hertzsprung-Russell Diagram with HV2112 and stellar evolution tracks. The colored tracks represent different masses of MESA evolutionary tracks at the metallicity of the host galaxy, with their masses printed beside the tracks and the metallicity indicated. The 12-24 M$_{\odot}$ tracks (in green) are RSG models, while the 4-6.5 M$_{\odot}$ tracks (violet and blue) are AGB models. The highest mass AGB model (6.5 M$_{\odot}$) has evolved through the carbon burning phase. The large, half-colored circles indicate the phases, with the colors of each point corresponding to the phase as in Figure~\ref{multi_epoch_seds}. Arrows indicate overall motion.}
    \label{hv2112_hrd_phase}
\end{figure}

Next we examine the relationship between the temperature and luminosity estimates of the HLOs in Hertzsprung-Russell Diagrams (HRDs). First we examine how these properties change as the stars vary, as we did for their colors in \S\ref{sec-cmag-vw1}. In Figure~\ref{hv2112_hrd_phase} we show the best-fit model temperature and luminosity for each available phase of HV2112, and we see a cyclical motion through the luminosity-temperature space, with the peak of the V-band light curve on the left and the trough on the right. This displays the same behavior as the V$-$W1 color in Figure~\ref{phase_motion_diagrams} (upper right).

\subsubsection{Comparison to Stellar Evolution Models}\label{sec_comp_models}

Figure~\ref{hv2112_hrd_phase} also displays theoretical predictions for RSGs and AGBs of various masses based on stellar evolution models from MESA \citep[version 10398,][]{Paxton.B.2011.MESAPaperI,Paxton.B.2015.MESAPaperII,Paxton.B.2018.MESAPaperIII,Paxton.B.2019.MESAPaperIV}. Our MESA models for pulsating RSGs were constructed following the methods described in \citet{Soraisam.Monika.2018.RSGVarinM31}. These models employ the calibrations of \citet{Chun.S.2018.MESAEvoRSGSNe} to reproduce average HRD positions of RSGs in the LMC and SMC. In particular, we adopt a metallicity-dependent mixing length parameter of $\alpha_{\rm MLT} = 2.0$ with metallicity $Z = 0.007$ in the LMC, and $\alpha_{\rm MLT} = 2.2$ with metallicity $Z = 0.004$ in the SMC. For more details on MESA evolutionary models for RSGs, see \citet{Soraisam.Monika.2018.RSGVarinM31} and \citet{Chun.S.2018.MESAEvoRSGSNe}. Our RSG models include masses ranging from 12-24~$M_\odot$ for both the LMC and SMC.

We also employ similar MESA settings to evolve lower mass stars to the AGB. Our MESA AGB models follow the settings of \citet{Fuller.J.2019.MESASpinsofStellarCores}, except that rotation is turned off because we are not concerned with angular momentum transport in the core, and we adopt the calibrated mixing length parameters described above for consistency with RSGs in the LMC and SMC. We run these models either until AGB winds have removed half of the initial mass of the star or until the carbon burning phase for the higher mass models where degenerate carbon ignition occurs. The carbon burning phase requires small timesteps that become computationally expensive, so we only evolve our highest mass models all the way through this phase (6.5 M$_{\odot}$ for the SMC and 7 M$_{\odot}$ for the LMC). The difficulty in running higher mass model tracks is also why we do not include tracks for AGB stars with initial masses of 8-10 M$_{\odot}$.

\begin{figure*}
    \centering
    \plottwo{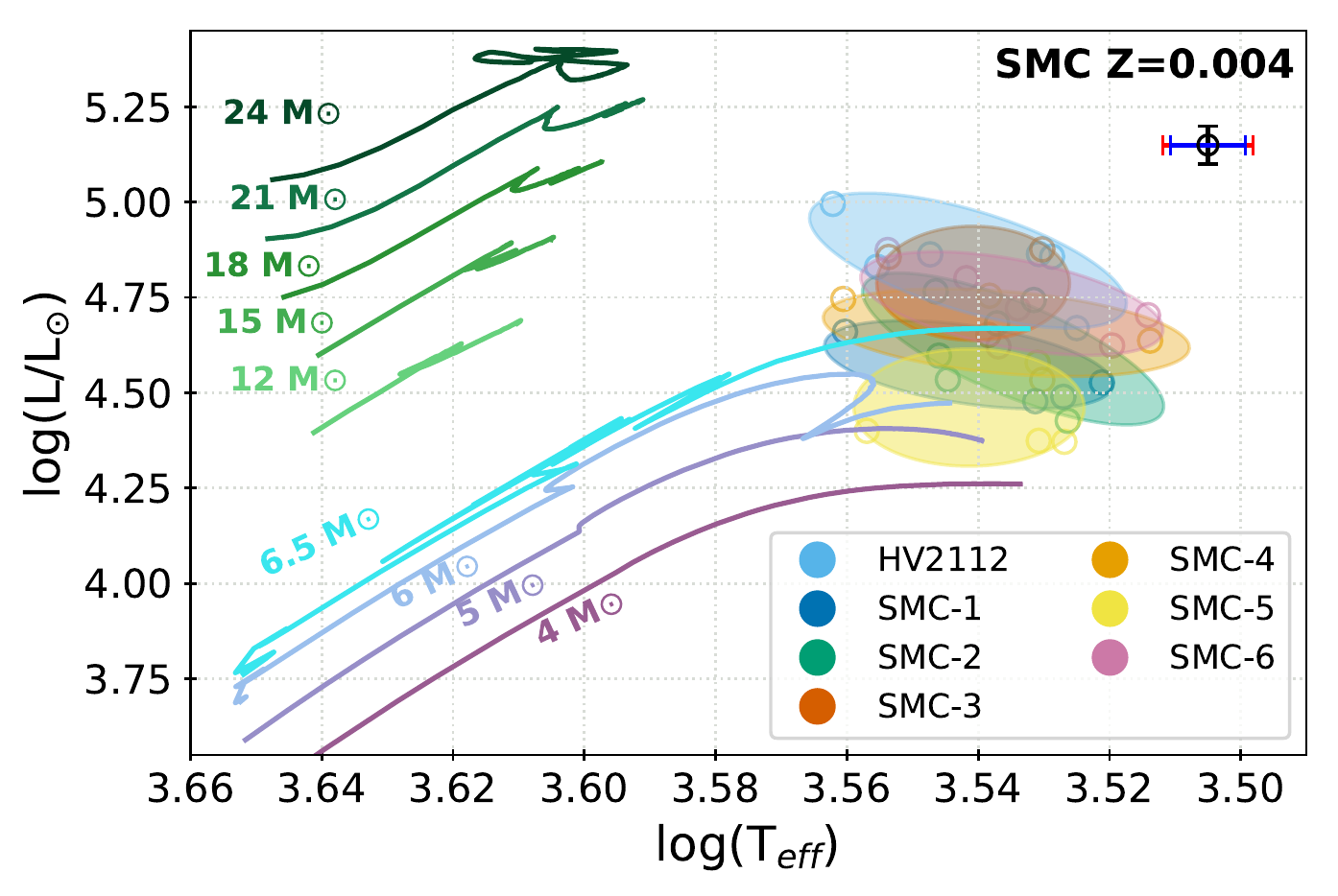}{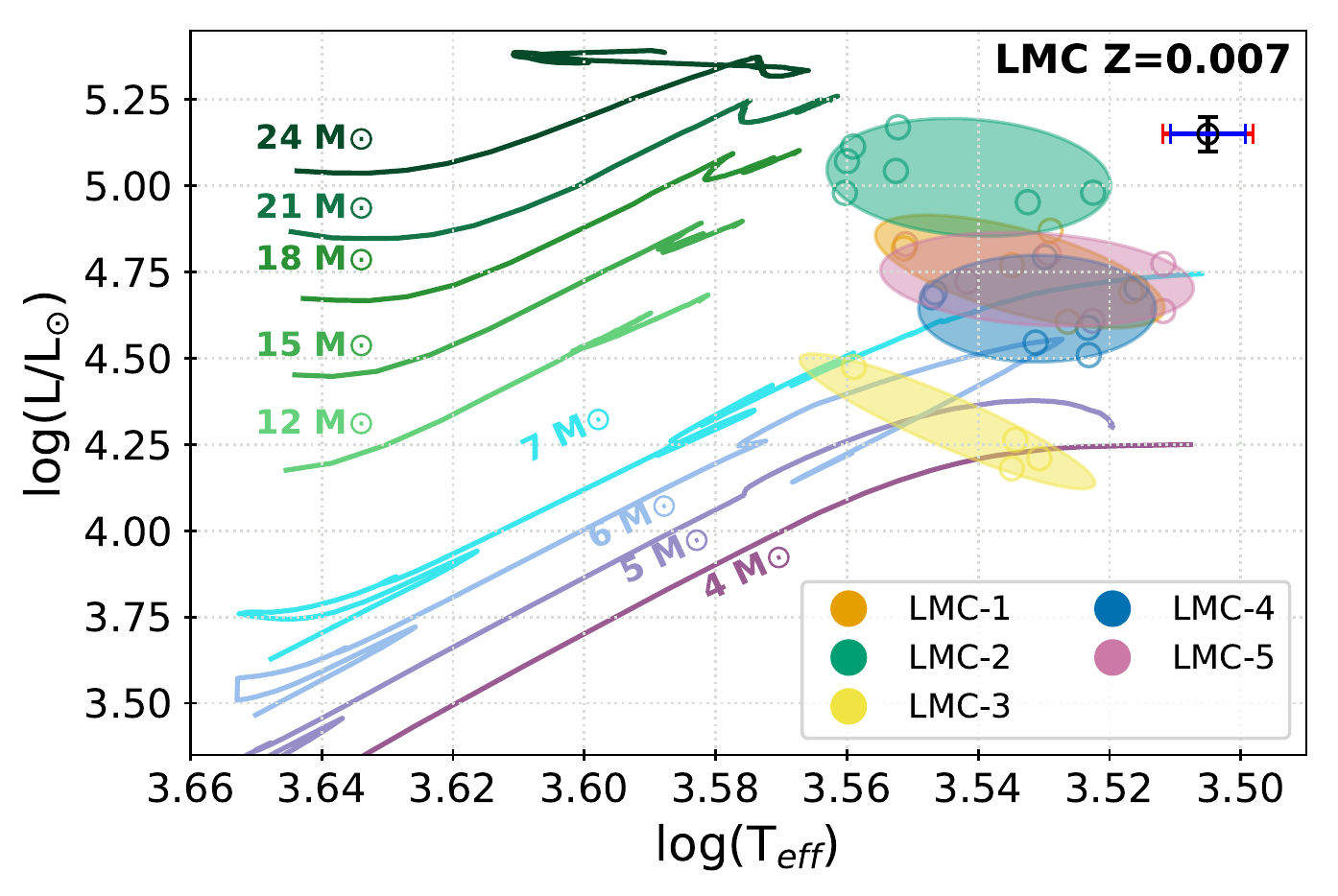}
    \plottwo{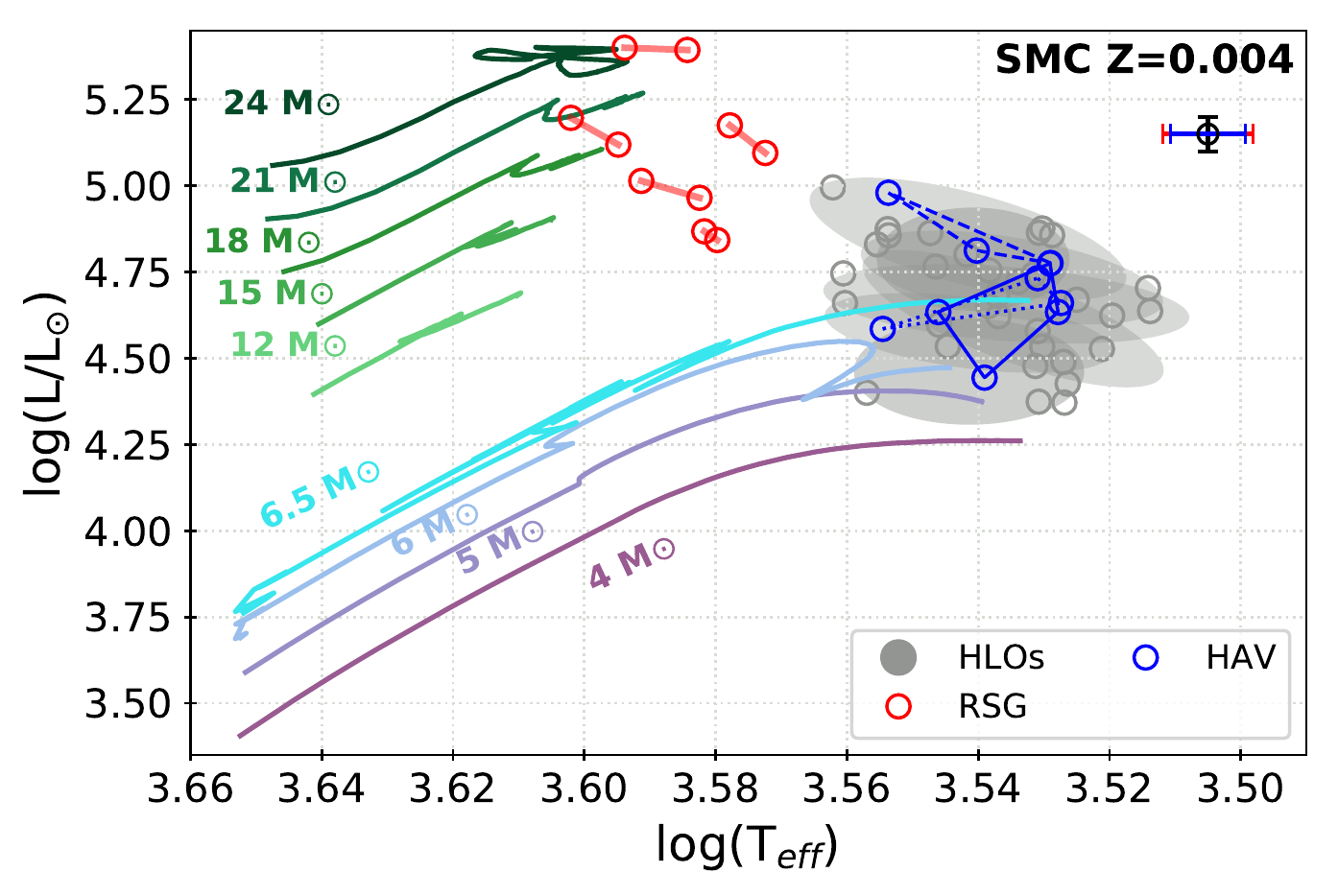}{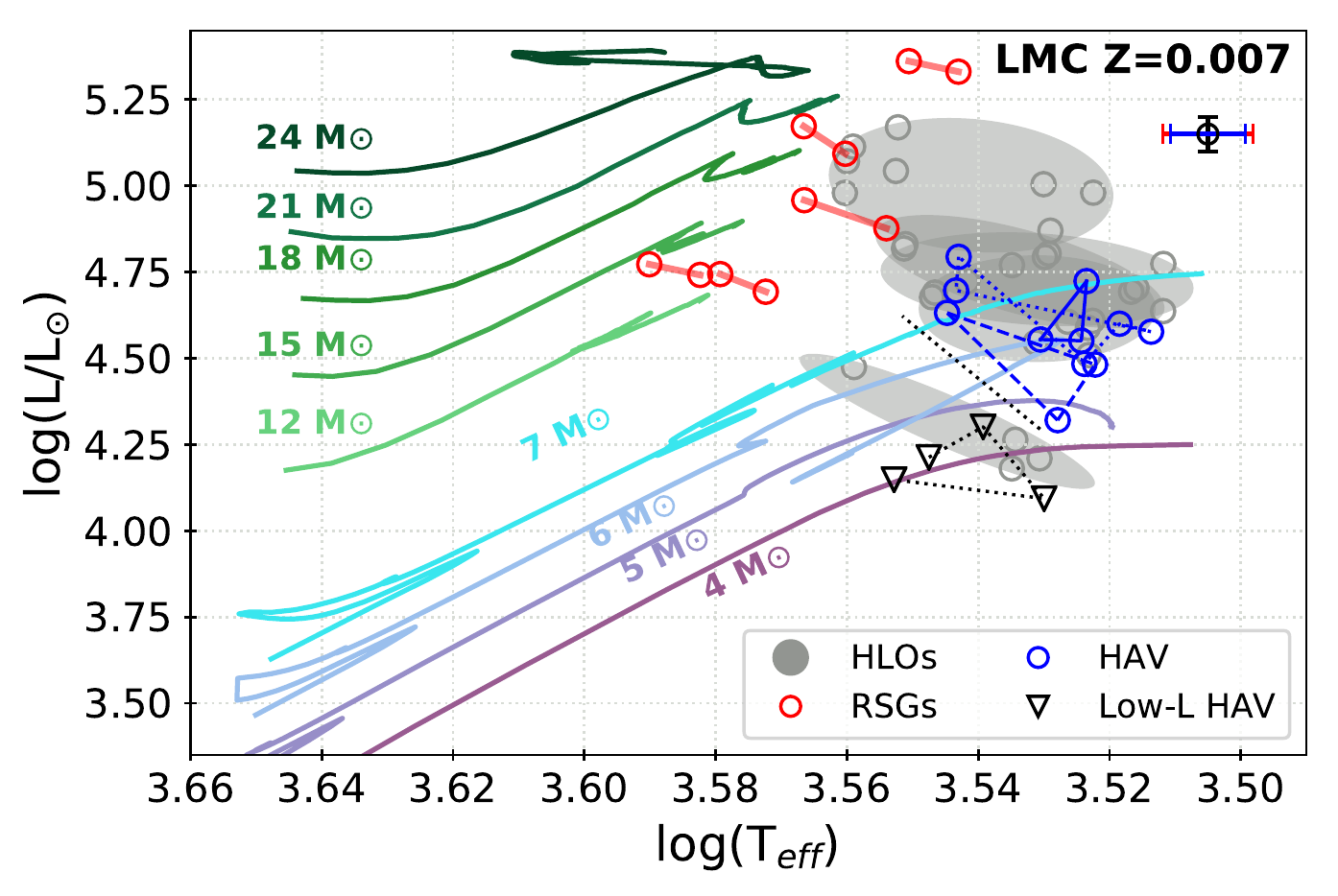}
    \caption{Hertzsprung-Russell diagrams for the SMC (left) and LMC (right), with the same MESA evolutionary models described in Figure~\ref{hv2112_hrd_phase} for the SMC. In the LMC, the AGB models range from 4-7 M$_{\odot}$, and the highest mass track (7 M$_{\odot}$ has been evolved through the carbon burning phase. The small circles correspond to different phases of HLOs, and an oval is overlaid to indicate the overall area occupied by individual stars -- each HLO is represented by a single color. The systematic errors are estimated to be $\pm$ 50K and  log(L/L$_{\odot}$) $\pm$ 0.05, indicated by the example error bar in the top right of each panel. For the log scaling, the red bar corresponds to a $\pm$ 50K error on a temperature of \mytilde 3150K, and the blue bar to \mytilde 3800K. \textit{Bottom:} The same as the top panels, but with a selection of spectroscopically confirmed RSGs (red points) and several HAVs (blue points). A HAV with a K$_{\mathrm{s}}$-band magnitude lower than LMC-3 is also included (black triangles). The HLOs are greyed out for clarity.}
    \label{hrds}
\end{figure*}

Figure~\ref{hrds} (top panels) shows Hertzsprung-Russell diagrams for the SMC and LMC. Each open circle is the temperature and luminosity from one contemporaneous ASAS-SN and NEO-WISE observation of a HLO, with each HLO represented by a single color. Colored ovals are overlaid on the open circles to show the overall range of the temperatures and luminosities of each HLO. Evolutionary tracks for RSGs (12-24 M$_{\odot}$) and AGB stars (4-6.5 or 7 M$_{\odot}$) are also shown. We find that the properties derived above for the HLOs are inconsistent with the RSG evolutionary models -- they are all much cooler than expected for RSGs of similar luminosities -- but are mostly consistent with the more massive AGB tracks. A possible exception is LMC-2, which is the closest HLO to any RSG track.

\subsubsection{Comparison to Observed RSGs}\label{sec_rsg_comp}

While our results in \S\ref{sec_comp_models} show the HLOs do not match with evolutionary models of RSGs, \citet{Levesque.E.2007.MCRSGTooCold} demonstrated that some late-type RSGs in the Magellanic Clouds appear to be colder than evolutionary models suggest is possible. These RSGs are also variable, although with significantly smaller amplitudes than the HLOs. Additionally, it has been shown that the calculated temperatures of RSGs depend on the method used. In particular, while our temperatures are not derived from performing spectral fitting of the TiO absorption bands, these bands do impact the overall V-band flux. It is therefore possible that the effective temperatures found above are somewhat lower than would be found by other methods \citep{Davies.B.2013.RSGTemperaturesLow}. Finally, systematic uncertainties exist within all stellar models, and not every evolutionary code predicts the same temperature for the RSG branch. Thus, to assess whether the HLOs are truly cooler than RSGs, we apply the same SED fitting method to a control sample of RSGs.

From the spectroscopically confirmed RSGs of \citet{Massey.P.2003.RSGinMC} described in \S\ref{sec_phase_space}, we fit the SEDs of 10 stars (5 in each Cloud) using exactly the same procedures. Since the V-band variability of these RSGs is small ($<$ 1 mag), only two points on each light curve were fit, as close to the peak and trough as possible. Plotted on HRDs (Figure~\ref{hrds}, bottom panels) as red open circles, most of these RSGs are slightly colder than the MESA evolutionary tracks. However, the HLOs are even colder than this control sample, showing that they are indeed cooler than observed RSGs, even if the derived temperatures are systematically underestimated.

\subsubsection{Comparison of HLOs and HAVs}\label{sec_mira_comp}

We also want to assess whether the presence of the double peak feature in the light curve of the HLOs is indicative of distinct physical properties from other stars that lack a double peak but otherwise resemble HV2112. Taking our sample of 27 HAVs that are not HLOs, we select 7 stars -- 3 in each of the Clouds with K$_{\mathrm{s}}$-band luminosities similar to that of the HLOs, and 1 LMC source with a K$_{\mathrm{s}}$-band luminosity lower than that of LMC-3 (the lowest luminosity HLO). As with the RSGs, we followed the same procedure as in \S\ref{sec_sed_marcs}. Due to the high variability amplitudes, we sampled as many phases as possible along their light curves. The results are shown in the bottom panels of Figure~\ref{hrds} as blue open circles (black open triangles) for the more (less) luminous HAVs. Unlike the RSG control sample, the more luminous HAVs overlap the same region occupied by most of the HLOs. This suggests that the double peak feature may not indicate a physically distinct class.

\subsection{Luminosity-Period Diagrams}\label{sec_pulsation_comp}\label{sec_lpd}

In both RSGs and variable AGB stars, variability is driven by pulsations. To investigate whether the HLOs are consistent with the pulsational variability expected from these stellar structures, we compare the measured periods and luminosities of the HLOs to evolutionary models. We use GYRE \citep{Townsend.R.2013.GYRE}, a stellar oscillation code designed to couple with MESA to obtain pulsation frequencies at every timestep for the RSG and AGB evolutionary tracks described above. We restrict our analysis to the frequencies of radial fundamental modes, which are both the lowest frequency and the expected modes for large amplitude pulsators.

\begin{figure*}
    \centering
    \plottwo{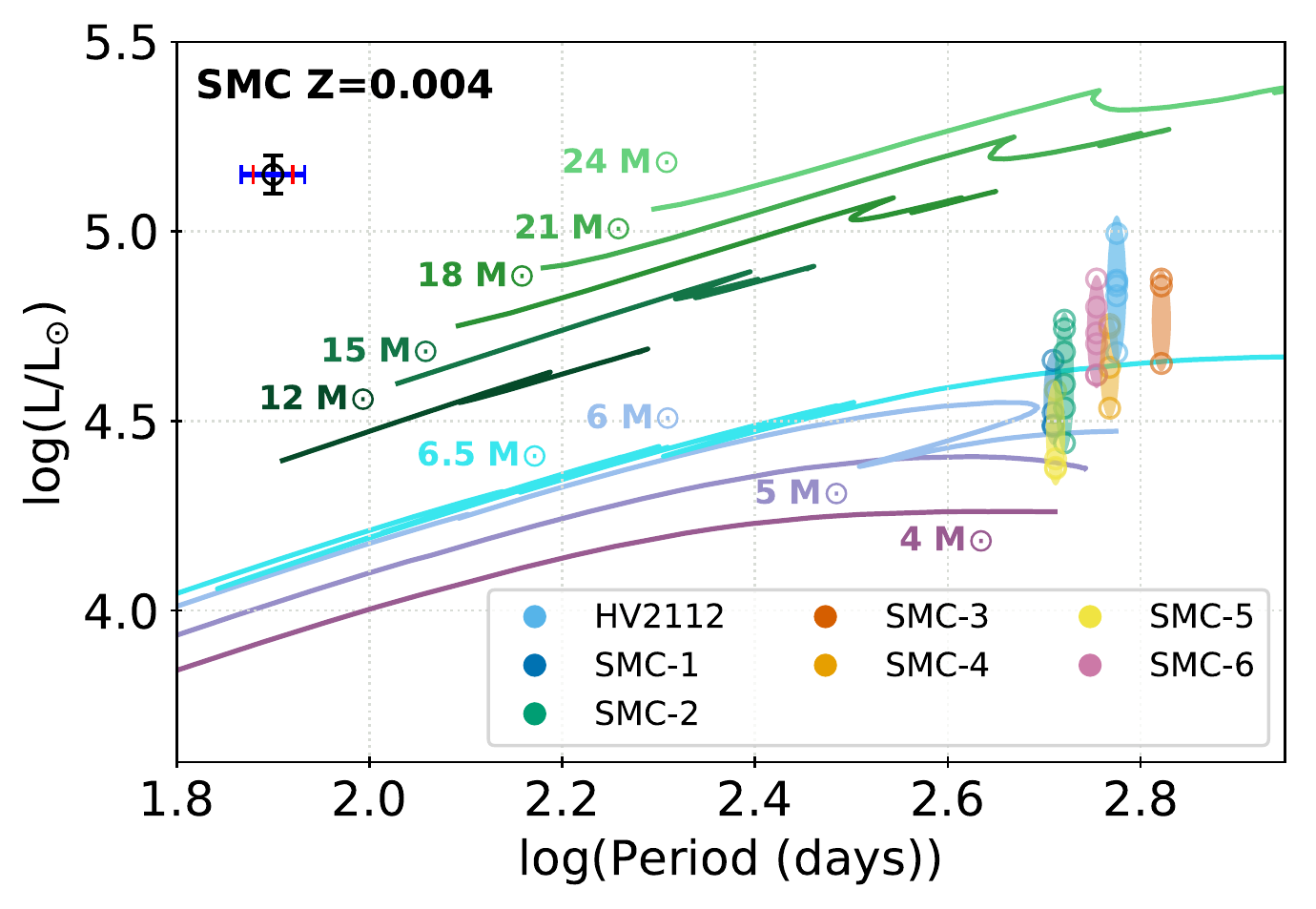}{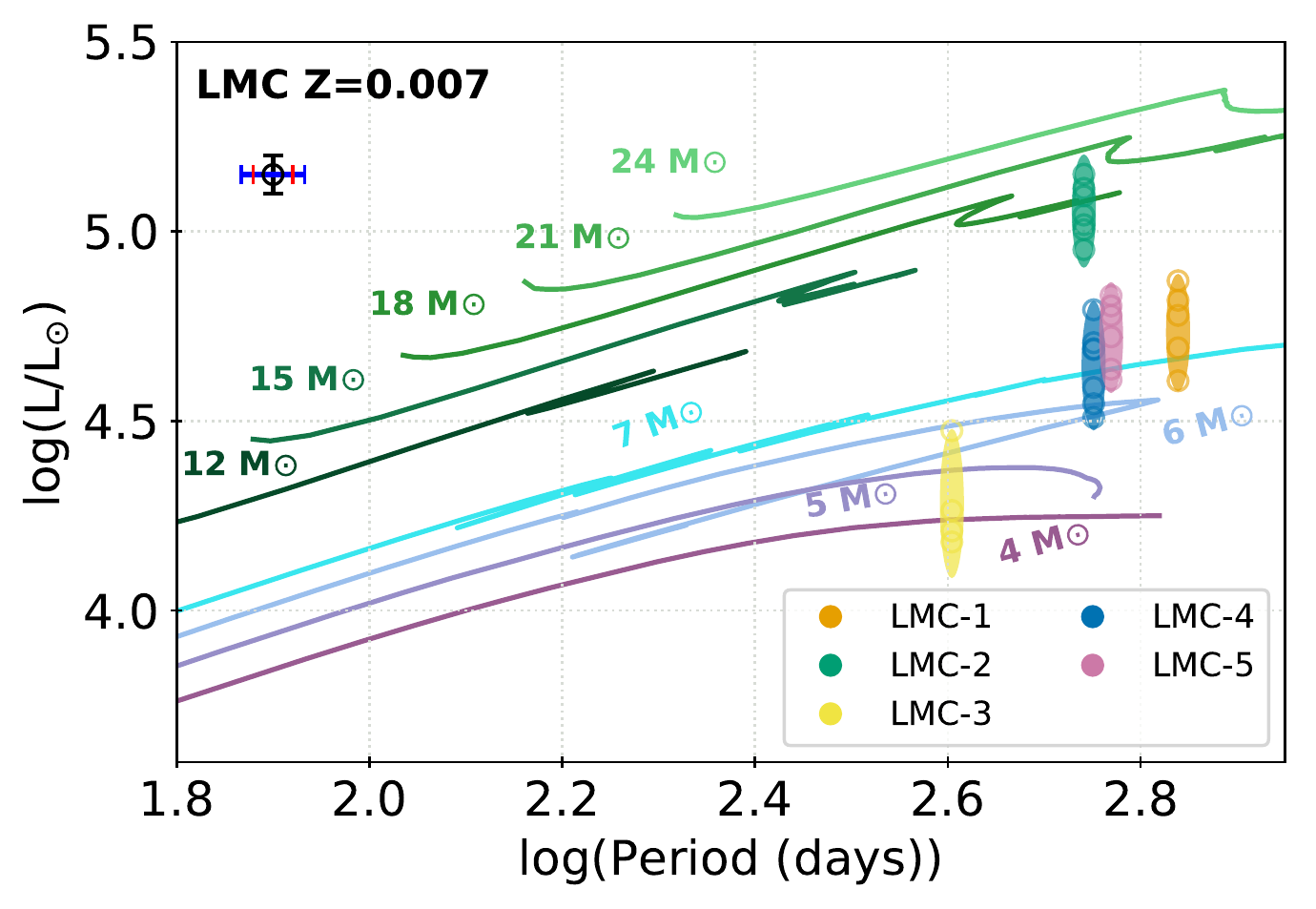}
    \plottwo{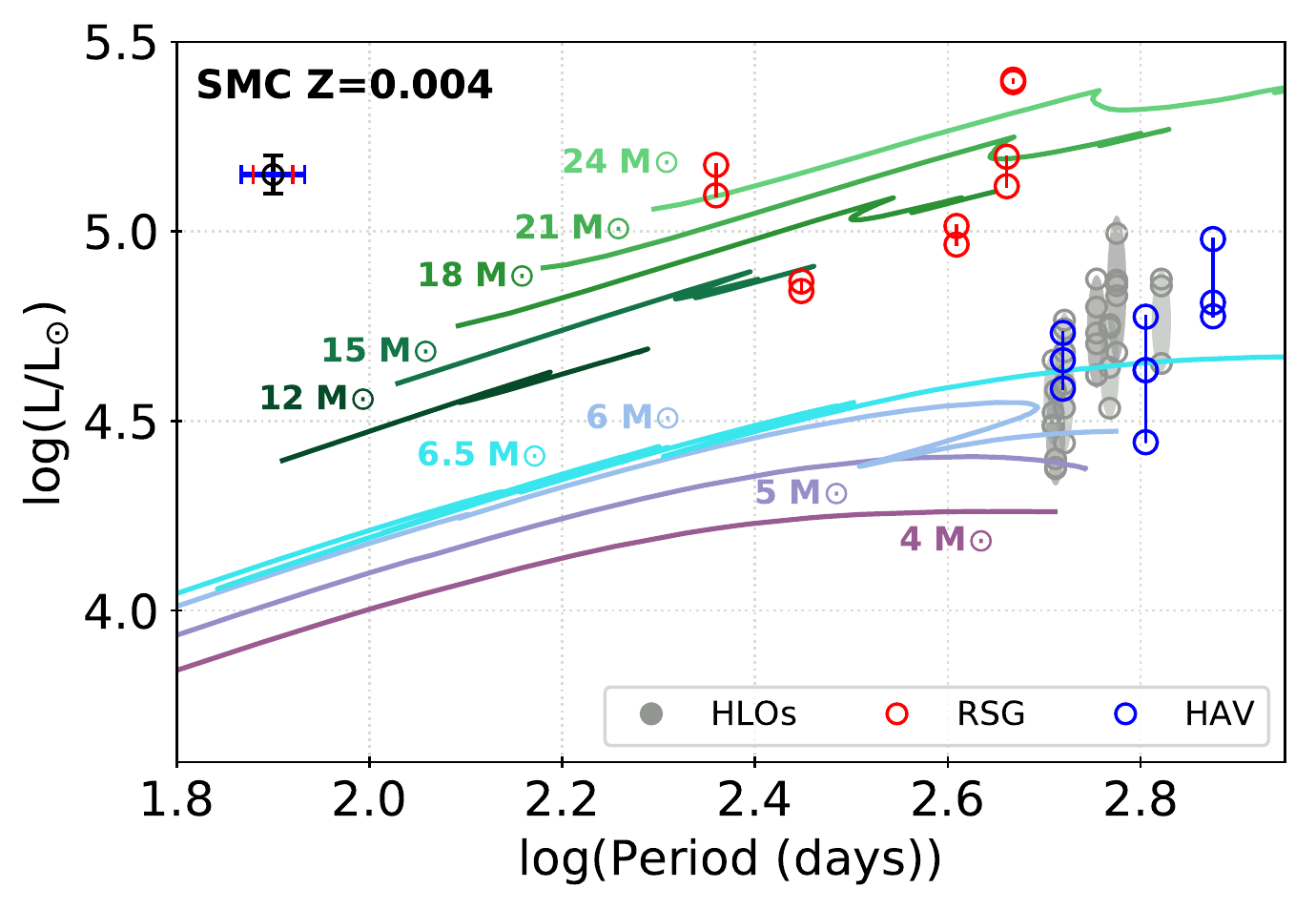}{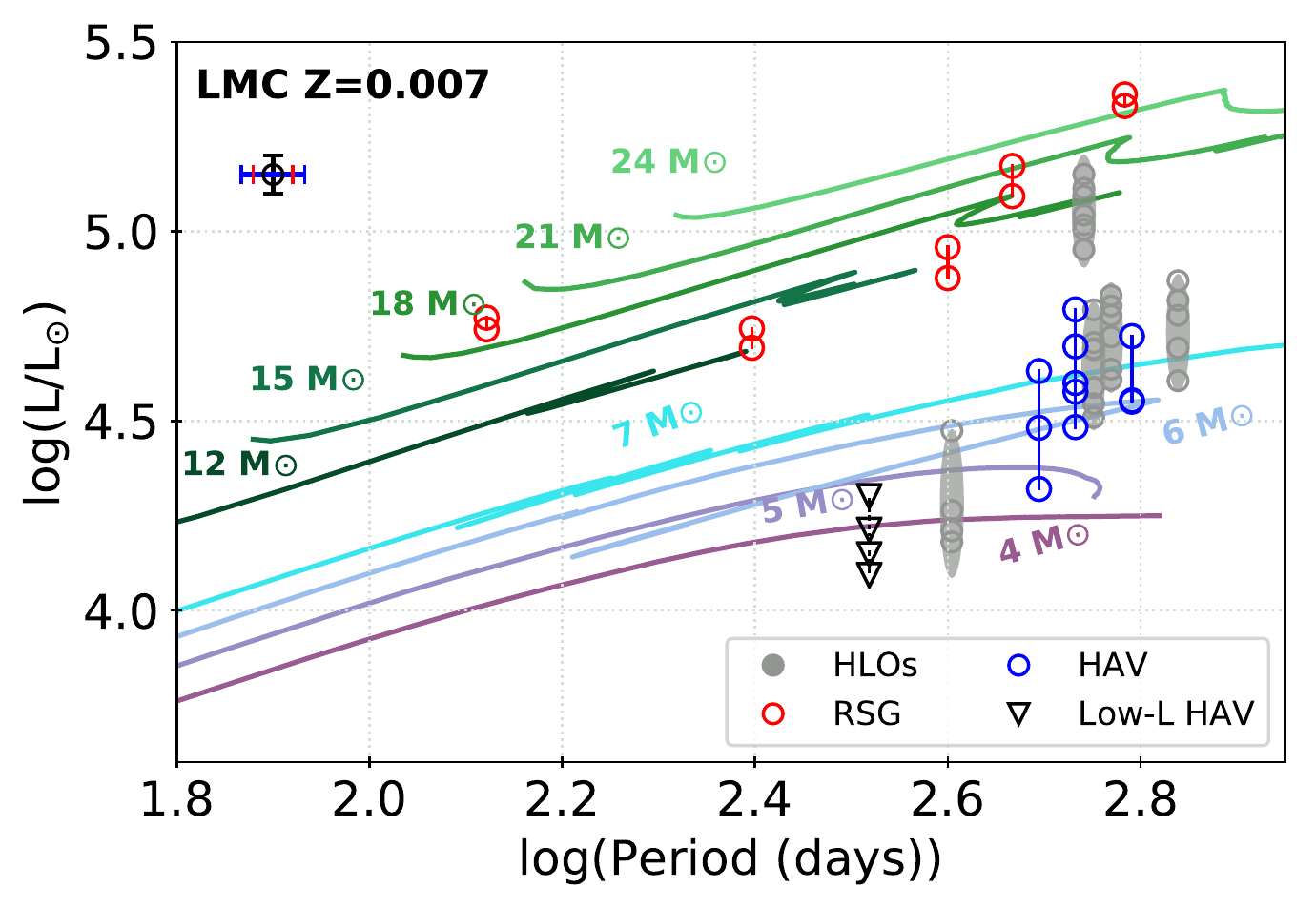}
    \caption{Period-Luminosity diagrams for the SMC (left) and the LMC (right). The evolutionary tracks are the same as in Figure~\ref{hrds}, as are the colors for individual HLOs. The oval along the luminosity axis represents the range of luminosity values of the HLOs as they pulsate. The example error bar has errors of log(L/L$_{\odot}$) = $\pm$ 0.05 and $\pm$ 30 days in period. The red bar corresponds to the scaling at \mytilde 630 days, and the blue at \mytilde 400 days. In the bottom panels, a sample of HAVs and RSGs in the Clouds is also shown.}
    \label{l_vs_p}
\end{figure*}

Figure~\ref{l_vs_p} compares the luminosity and period of the HLOs and selected HAVs to RSG and AGB evolutionary tracks in luminosity and period space. With one exception, all of the HLOs are inconsistent with the MESA+GYRE RSG models. The periods are far too long for their luminosities if these stars had the typical mass and structure of RSGs. Notably, the one star that is consistent with the RSG models in this phase space is LMC-2, which is dissimilar to the other HLOs in numerous characteristics (see \S\ref{sec_j81}). The other HLOs are again consistent with the highest mass AGB tracks.

The control samples described in Sections \ref{sec_rsg_comp}-\ref{sec_mira_comp} are also shown in the bottom panels of Figure~\ref{l_vs_p}. Here we see the RSG control sample agrees with the RSG evolutionary tracks, and that the HAV sample generally overlaps with the location of the HLOs. From these results, we can see that the HLOs are inconsistent in more than just their pulsation amplitude with the properties of typical RSGs. Their variability period in relation to their luminosity, and their placement on the HRD are also inconsistent. In contrast, their pulsations properties appear to be consistent with AGB models for stars with initial masses of $\gtrsim$ 6 M$_{\odot}$.

\subsection{Constraints on Current Mass}\label{sec_pulsation_mass}

\begin{figure*}
    \centering
    \includegraphics[width=0.9\textwidth]{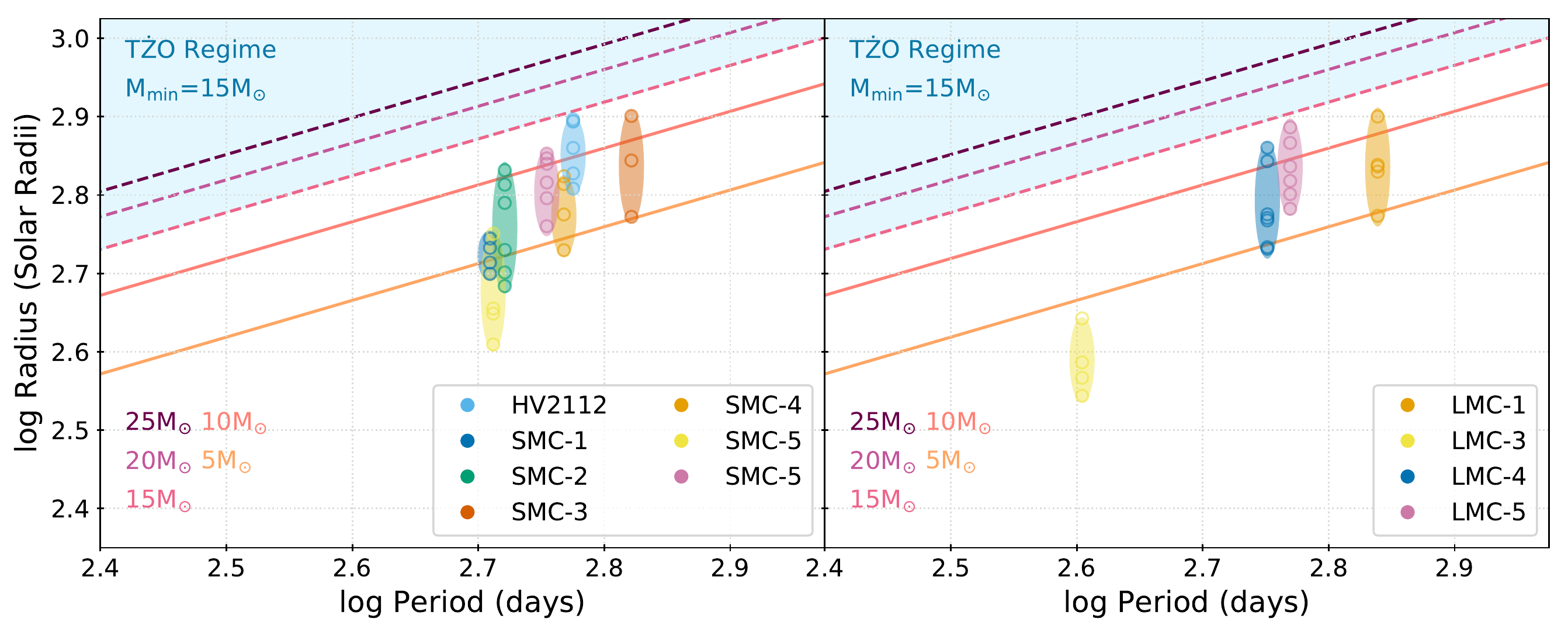}
    \caption{Period - Radius diagrams for the SMC (left) and the LMC (right). Lines of constant mass from 5 to 25 M$_{\odot}$ (in increments of 5 M$_{\odot}$) are shown as lines from orange to dark violet. The shaded blue region represents masses above 15 M$_{\odot}$, the hypothesized minimum mass for T\.ZOs to continue nuclear fusion \citep{Cannon.R.1993.TZOStructure}. Lines of constant mass within this region are dashed, while lines of constant mass lower than 15$M_{\odot}$ are solid. For each HLO, the radii calculated from the SED models as a function of phase are shown at their estimated periods, with the same color scheme as Figures \ref{hrds} and \ref{l_vs_p}.}
    \label{radius_period_plots}
\end{figure*}

The MESA models in Figures \ref{hrds}-\ref{l_vs_p} show the evolution of single stars with a range of initial masses. There are no T\.ZOs models where we can make a direct comparison. However, constraints on the current mass of the HLOs could provide a critical differentiation between their nature as T\.ZOs or s-AGB stars. In particular, s-AGB stars should be exclusively $\lesssim$11 M$_\odot$, while massive T\.ZOs are predicted to be a stable stellar configuration only for masses $\gtrsim$15 M$_\odot$ (a \mytilde14 M$_{\odot}$ envelope minimum plus a \mytilde 1 M$_{\odot}$ neutron star core). Below this mass, nuclear fusion to support the T\.ZO will cease \citep{Cannon.R.1993.TZOStructure}. Fortunately, fundamental radial mode pulsations are sensitive to the mean density of a star, and hence can be used to probe its current stellar mass, $M_{tot}$, if information on the current radius, $R$, and internal structure are known. In particular, the period of the fundamental mode, $P$, can be generally described as

\begin{equation}\label{eq1}
    P = \alpha \sqrt{\frac{R^{3}}{GM_{tot}}}  \bigg(\frac{\langle \rho \rangle}{\rho_{o}}\bigg)^{\beta}, 
\end{equation}

\noindent where $\langle \rho \rangle$ is the average density of star, $\rho_{o}$ is an arbitrary reference density, and $\alpha$ and $\beta$ are normalizing constants that are determined by the structure of the star. This equation is a modification of the standard pulsation constant formalism which postulates that the fundamental-mode period is proportional to the sound crossing time. For a given structure, $\alpha$ and $\beta$ should be constant, and we expect $\beta$ to be a small number.

While the results above indicate the HLOs are inconsistent with RSGs with initial masses $\geq$ 15 M$_{\odot}$, the internal structures of T\.ZOs are very different from RSGs. While we lack T\.ZO models to calculate these directly, the structure of a T\.ZO can roughly be described by a dense, \mytilde1 M$_{\odot}$, degenerate core and a large, fully convective, envelope. \emph{Broadly, this is the same stellar structure that describes AGB stars}. 

If we assume that T\.ZOs and AGB stars can be coarsely described by the same stellar structure, we can use the MESA AGB models introduced in \S\ref{sec_comp_models} to calculate appropriate values for $\alpha$ and $\beta$ in equation \ref{eq1}. These models directly provide values of period, current mass, and radius as a function of time, from which average density can be computed. We then fit equation \ref{eq1} to these tracks in order to determine the constants $\alpha$ and $\beta$. From fitting the 6 M$_{\odot}$ SMC track, we find values of $\alpha = 5.47$  and $\beta = -0.210$, given a reference density of $\rho_{o} = 5.3 \times 10^{-8}$ g cm$^{-3}$. The difference in results from fitting the evolutionary tracks with different masses and/or metallicities were small, with $\alpha$ and $\beta$ varying by less than 10\%. The effect of such changes are negligible: for a fixed stellar mass, the radius associated with a given period varies by only \mytilde 1.7\% (at $P=$400 days) to \mytilde 2.6\% (at $P=$700 days).

With these values of $\alpha$ and $\beta$, we can place lines of constant mass in the plane of period vs.\ radius (Figure~\ref{radius_period_plots}). Using $L = 4 \pi R^{2} \sigma T^{4}$, we then calculate the radii of the HLOs using the results from the SED fitting. These are also plotted in Figure~\ref{radius_period_plots} using the same color scheme as in Figures \ref{hrds}-\ref{l_vs_p}. Vertical ovals correspond to the range of radii observed for the  HLOs throughout their pulsation periods, as we do not directly constrain the unperturbed radius. LMC-2 is not plotted, as we believe it to be more consistent with a RSG (see Section~\ref{sec_discussion}), and therefore the stellar structure assumed to calculate the lines of constant mass would not be appropriate.

Figure~\ref{radius_period_plots} shows that for this stellar structure, the pulsation properties of the HLOs imply current masses between $\sim$5$-$10 M$_\odot$. Exceptions include LMC-3, which is below 5 M$_\odot$, and HV2112, whose mean radius implies a mass slightly above 10 M$_\odot$. We note that these precise masses depend on the absolute value of our derived temperatures, which have some systematic uncertainty, as discussed in \S\ref{sec_rsg_comp}. To estimate the \emph{maximal} impact this could have on our final masses we also carry out this calculation assuming the HLOs have temperatures 400 K warmer than found with the MARCS/DUSTY modelling above. This leads to an $\sim$20\% decrease in the inferred stellar radii, which results in current mass estimates approximately a factor of two smaller than the values shown in Figure~\ref{radius_period_plots}. The implication of these masses for the possible T\.ZO nature of these sources, in light of the proposed $\sim$15 M$_\odot$ minimum mass for massive T\.ZOs will be discussed in Section~\ref{sec:TZOnature}, below.

\subsection{Mass-Loss Rates and Dust}\label{sec_mlr_dust}

The MCMC fits (\S\ref{sec_model_results}) also provided an estimate of the V-band dust opacity, $\tau_{\text{V}}$, for each HLO. Following the method of \citet{Adams.S.2016.MLRRef}, for a constant velocity wind, the mass loss rate (MLR) is

\begin{equation}
\dot{M} = \frac{4 \pi v_{\mathrm{w}} r_{\mathrm{in}} \tau_{\mathrm{V}}}{\kappa_{\mathrm{V}}}\bigg(1-\frac{r_{\mathrm{in}}}{r_{\mathrm{out}}}\bigg)^{-1}, 
\end{equation}

\noindent where the wind velocity $v_{\mathrm{w}}$ is taken to be 15 km s$^{-1}$, a typical value for the circumstellar envelope around AGB stars \citep{Hofner.S.2018.MassLossinAGBReview}, and $r_{\mathrm{in}}$ and $r_{\mathrm{out}}$ are the inner and outer radii of the dust shell. $\kappa_{\text{V}}$, the V-band opacity, is related to the gas-to-dust ratio $r_{\mathrm{gd}}$, the grain size $a$, bulk grain density $\rho$, absorption coefficient $Q$, by:

\begin{equation}
    \kappa_{\mathrm{V}} = \frac{3r_{\mathrm{gd}}Q}{4a\rho}.
\end{equation}

\noindent For $r_{\mathrm{gd}} \approx 0.01$, a $\approx 0.1$ $\mu$m, and $\rho = 3.3$ g cm$^{-3}$, as are appropriate for silicate grains \citep{Draine.B.1984.PropertiesSilicateGrains}, $\kappa_{\mathrm{V}} \approx $ 50 cm$^{2}$ g$^{-1}$. Our adopted value of a $\approx 0.1 \mu$m corresponds to the mass-weighted average of the MRN grain-size distribution, assuming a$_{\mathrm{min}}$ = 0.025 $\mu$m and a$_{\mathrm{max}}$ = 0.25 $\mu$m, as appropriate for silicate grains \citep{Mathis.J.1977.MRNDist}. In addition, we note that mass-loss rates calculated using the full range of possible silicate grain sizes listed above would vary from our quoted values by less than a factor of 3.3.

We calculate the posterior distribution of $\dot{M}$ for each phase by drawing from the full posterior distribution for r$_{\mathrm{in}}$, L, and $\tau_{\text{V}}$, and average the results across all phases for each HLO. In Table \ref{dust_estimates} we show the resulting estimates of $\dot{M}$. The MLRs of the HLOs are all on the order of $10^{-7}$ M$_{\odot}$ yr$^{-1}$, except for LMC-2 which is an order of magnitude higher.

\begin{deluxetable}{cc||cccc}
\tabletypesize{\footnotesize}
\tablecolumns{6} 
\tablecaption{Estimates of dust mass-loss rates using MCMC fitted $\tau_{\text{V}}$ values (column 3) and empirical relations (columns 4-6). $\tau_{\text{V}}$ values are the mean of all fitted phases for each star. See \citet{Hofner.S.2018.MassLossinAGBReview} for empirical relation references.\label{dust_estimates}}

\tablehead{ \colhead{1} & \multicolumn{1}{c||}{2} & \colhead{3} & \colhead{4} & \colhead{5} & \colhead{6} \\\hline
\colhead{Star} & \multicolumn{1}{c||}{Mean } & \multicolumn{4}{c}{Inferred \.M (10$^{-7}$ M$_{\odot}$ yr$^{-1}$)} \vspace{-0.2cm} \\ 
& \multicolumn{1}{c||}{$\tau_{\text{V}}$}  & \multicolumn{4}{c}{Methods Used}
}
\startdata
   & &   $\tau_{\mathrm{V}}$ MLR & $K_{\mathrm{s}}-[8]$ & $K_{\mathrm{s}}-[24]$ & $[3.6]-[8.0]$   \\\hline
   HV2112 & 0.25$^{+0.04}_{-0.03}$ & 3.8$^{+0.9}_{-0.5}$ & 2.7$\pm$0.23 & 0.38$\pm$0.03 & N/A \\ 
   SMC-1 & 0.14$^{+0.04}_{-0.03}$ & 3.1$^{+1.1}_{-0.8}$ & 2.8$\pm$0.23 & 0.64$\pm$0.06 & N/A  \\
   SMC-2 & 0.23$^{+0.05}_{-0.04}$ & 1.9$^{+0.3}_{-0.3}$ & N/A & 0.25$\pm$0.03 & N/A \\ 
   SMC-3 & 0.19$^{+0.03}_{-0.03}$ & 4.2$^{+1.3}_{-0.9}$ & 8.8$\pm$0.44 & 1.2$\pm$0.07 & N/A \\ 
   SMC-4 & 0.16$^{+0.03}_{-0.02}$ & 1.6$^{+0.5}_{-0.3}$ & N/A & 0.25$\pm$0.02 & N/A \\ 
   SMC-5 & 0.28$^{+0.07}_{-0.04}$ & 3.1$^{+1.1}_{-0.6}$ & N/A & N/A & N/A \\ 
   SMC-6 & 0.28$^{+0.07}_{-0.04}$ & 4.1$^{+0.8}_{-0.6}$ & 2.6$\pm$0.21 & 0.22$\pm$0.03 & N/A \\ 
   LMC-1 & 0.30$^{+0.07}_{-0.06}$ & 4.1$^{+1.0}_{-0.6}$ & 2.9$\pm$0.23 & 0.8$\pm$0.05 & N/A \\ 
   LMC-2 & 0.71$^{+0.13}_{-0.09}$ & 37.6$^{+11.7}_{-5.8}$ & 5.4$\pm$0.38 & 17$\pm$0.53 & 10$\pm$1.4 \\ 
   LMC-3 & 0.62$^{+0.07}_{-0.07}$ & 8.7$^{+2.5}_{-1.4}$ & N/A & N/A & N/A \\
   LMC-4 & 0.22$^{+0.04}_{-0.03}$ & 3.5$^{+0.8}_{-0.6}$ & N/A & 1.3$\pm$0.08 & N/A \\
   LMC-5 & 0.21$^{+0.03}_{-0.03}$ & 6.4$^{+2.0}_{-1.1}$ & 4.1$\pm$0.34 & 1.9$\pm$0.11 & N/A \\
\enddata
\end{deluxetable}

In addition to estimating the MLR from the SED fits, we can estimate the MLR of the HLOs using their photometric colors. \citet{Hofner.S.2018.MassLossinAGBReview} aggregated a number of color estimation methods. We use the estimates for O-rich AGB stars and RSGs in the Magellanic Clouds, since all HLOs have the infrared colors of one of these source types (Figure~\ref{boyer_cmd}). In Table \ref{dust_estimates} we give the result for each HLO (columns 4-6). These results are mostly consistent with the estimates from the SED fitting, though the $K_{\mathrm{s}}-[24]$ estimate is around an order of magnitude lower. If an entry is `N/A', the colors for that particular estimate were outside the bounds described in \citet{Hofner.S.2018.MassLossinAGBReview}. We note that all of the photometry used here is single-epoch, so these results do not take the variability of the HLOs into account.

AGB stars in the thermally pulsing phase have typical mass-loss rates ranging from $10^{-7}$ to $10^{-5}$ M$_{\odot}$ yr$^{-1}$ \citep{Hofner.S.2018.MassLossinAGBReview}. \citet{Groenewegen.M.2009.MCRSGAGBStars2112SAGB} found that mass-loss rates for O-rich AGB stars were all $<$ $10^{-6}$ M$_{\odot}$ yr$^{-1}$. RSGs display mass-loss rates of 10$^{-7}$ to 10$^{-4}$ M$_{\odot}$ yr$^{-1}$ \citep{Mauron.N.2011.MLRforRSGdeJ}. Thus, the estimated mass-loss rates for the HLOs are comparable to both those of O-rich TP-AGB stars and RSGs, and none show signs of extremely enhanced or `superwind' (\mytilde 10$^{-4}$ M$_{\odot}$ yr$^{-1}$) mass loss rates. This may be a selection effect, as stars with significantly high mass-loss rates would be so visually obscured as to not appear in the ASAS-SN data.

\section{Stellar Lifetime Implications}\label{sec_lifetimes}

Here, we assess the evolutionary lifetime implied by the size of our HLO sample. We consider both s-AGB stars and T\.ZOs, in turn, and discuss whether the resulting lifetimes can form a self-consistent physical picture with the assumed origin. 

We assume that all HLOs come from the same physical class, with the exceptions of LMC-2 (which we believe is consistent with being a highly variable RSG; see \S\ref{sec_j81}) and LMC-3 (which is more consistent with a lower mass---M $<$ 5 M$_\odot$---AGB star). In addition, as detailed in \S\ref{sec_mira_comp}, it appears that some of the luminous HAVs display the same physical properties as the HLOs, lacking only the double peak feature in their light curves. In particular, while 7 of the HAVs have K$_{\mathrm{s}}$-band magnitudes dimmer than LMC-3 (and are thus likely also lower-mass AGB stars; see Figures~\ref{hrds}-\ref{l_vs_p}), 20 overlap in bulk properties with the HLOs. To account for the possibility that these 20 HAVs are of the same class as the HLOs, we estimate the lifetimes assuming either that the total population consists only of the 10 remaining HLOs (7 in the SMC, 3 in the LMC), or that the total population consists of 30 stars (10 HLOs $+$ 20 HAVs; 10 in the SMC, 20 in the LMC).

\subsection{Super-AGB Stars}\label{sec_life_sagb}

First, we calculate the lifetime of the HLO evolutionary phase assuming that the HLOs come from stars with initial masses between 6.5-10 M$_\odot$ stars (as expected for s-AGB stars). We do this by comparing the number of sources in our population to the total number of known AGB stars in the Clouds, taking into account both the lifetimes of the AGB phase from stellar models and the initial mass function (IMF).

Assuming continuous star formation, we can relate the total number of stars, $N$, in an evolutionary state to the lifetime of that phase and the IMF as

\begin{equation}
    N \propto \int_{M_1}^{M_2} m^\Gamma \tau_m dm,
\end{equation}

\noindent where $\Gamma$ is the slope of the IMF (taken here to be -1.35; \citealt{Salpeter.E.1955.SalpeterIMF}), and $\tau_m$ is the lifetime of a star of mass $m$ in that evolutionary state. For the HLOs, we are interested in assessing the average lifetime of the phase, $\langle \tau \rangle_{HLO}$, given the number of observed objects and can thus remove $\tau_{m}$ from the integral. We can then express this mean lifetime in terms of the number of known AGB stars in the Clouds as,

\begin{equation}\label{eq:HLO2}
   \langle \tau \rangle_{HLO} = \frac{N_{HLO}}{N_{AGB}} \int_{M_1}^{M_2} m^\Gamma \tau_{AGB} dm  \left( \int_{6.5}^{10} m^\Gamma dm \right)^{-1}.
\end{equation}

\noindent In order to complete this calculation, we require a region of the color-magnitude diagram where (a) a complete sample of AGB stars is known and (b) the time that AGB stars of various masses spend in this region is also known.

\begin{figure}
    \centering
    \includegraphics[width=\linewidth]{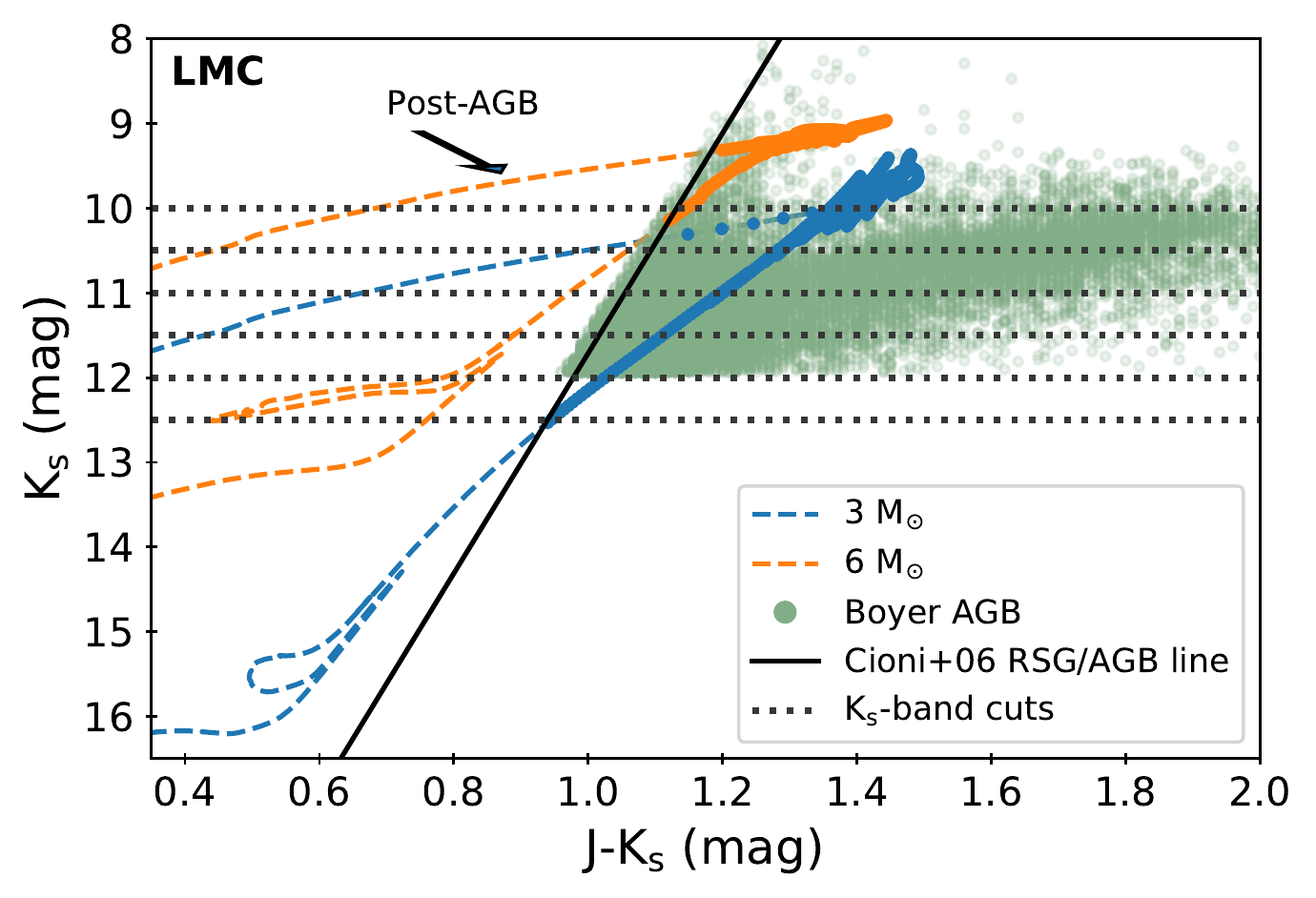}
    \caption{Examples of color-magnitude evolution for 3 and 6 M$_{\odot}$ LMC tracks with the \citet{Cioni.M.2006.ColorcutSAGBs} RSG-AGB color-cut line as a solid black line and the K$_{\mathrm{s}}$-band cuts described in the text as dashed dark grey lines. AGB stars classified in \citet{Boyer.M.2011.SAGE.MC.Photom} are shown as green circles. Also noted is the post-AGB section of the track.}
    \label{lifetime_example}
\end{figure}

For the observed population of AGB stars in the Clouds, we use the sample of \citet{Boyer.M.2011.SAGE.MC.Photom}, which is claimed to be complete. Within the region that we conducted our search for HLOs (\S\ref{sec_phot_select}) there are 4799 AGB stars in the SMC and 23519 in the LMC. This sample uses a color cut from \citet{Cioni.M.2006.ColorcutSAGBs} to differentiate AGB stars from RSGs as a function of K$_{\mathrm{s}}$-band luminosity. This line, as well as the observed population of AGB stars in the LMC, are shown in Figure~\ref{lifetime_example} (see also Figure~\ref{boyer_cmd}). 

For the AGB lifetimes, we use the MIST stellar evolutionary tracks \citep{Dotter.A.2016.MISTPaperI,Choi.J.2016.MISTPaperII}, which include synthetic photometry based on the ATLAS12 model atmospheres \citep{Kurucz1993}. Metallicities of [Fe/H] = $-$0.37 and [Fe/H] = $-$0.95 were used for the LMC and SMC, respectively \citep{Choudhury.S.2016.LMCMetallicity,Choudhury.S.2018.SMCMetallicity}. For each stellar track between 0.7 M$_{\odot}$ to 1 M$_{\odot}$ (step size of 0.1 M$_{\odot}$) and 1 M$_{\odot}$ to 10 M$_{\odot}$ (step size of 1 M$_{\odot}$) we calculate the lifetime spent within the region of color-magnitude space from which the observed sample of AGB stars was selected (Figure~\ref{lifetime_example}).

We calculate the AGB lifetimes based on a series of flat K$_{\mathrm{s}}$-band cuts (dashed lines in Figure~\ref{lifetime_example}) and variations of the \citet{Cioni.M.2006.ColorcutSAGBs} RSG/AGB line, which we shift red-ward by up to 0.2 mag in J-K$_{\mathrm{s}}$. These lifetimes, combined with the number of AGB stars satisfying the same limits, are then input into Equation \ref{eq:HLO2} to give a range of possible mean lifetimes for the HLO evolutionary stage. These ranges account for possible discrepancies between the model and observed colors as well as effects due to variations in the star formation rate of the SMC/LMC of approximately a factor of 5 over the $\sim$5 Gyr lifetime of a 1 M$_\odot$ star \citep{Harris.J.2004.SFRinSMC,Harris.J.2009.SFRinLMC}.

For M$>$6 M$_{\odot}$ in the SMC and M$>$7 M$_{\odot}$ in the LMC, the MIST tracks terminate either before entering or in the region of interest due to issues with model convergence during the thermally pulsing AGB phase \citep{Dotter.A.2016.MISTPaperI}. In order to investigate the impact that this uncertainty has on our final HLO lifetime estimates, we test three different methods for quantifying the AGB lifetime for these high mass tracks: (i) setting them to 0 years, (ii) setting them equal to the lifetime of the last successfully converged mass track, and (iii) determining a linear relationship between mass and lifetime for the converged tracks and extrapolating to higher masses. All three choices have negligible results on our final HLO lifetime estimate due to the steepness of the IMF.

Putting all of these components together we calculate a mean lifetime for the HLO evolutionary phase, if they are produced by 6.5--10 M$_{\odot}$ stars, of \mytilde ($0.5$ -- $7.0$)$ \times10^{4}$ yr in the SMC and \mytilde ($0.3$ -- $0.8$)$ \times10^{4}$ yr in the LMC, if the 10 HLOs represent the full population. If we also include the 20 high luminosity HAVs described above, these numbers increase to \mytilde ($0.8$ -- $9.9$)$ \times10^{4}$ yr for the SMC and \mytilde ($1.7$ -- $5.8$)$ \times10^{4}$ yr for the LMC.

Overall these numbers are consistent with the expected lifetimes of the s-AGB phase. \citet{Doherty.C.2017.SAGBStarsECSNE} estimate the lifetime of the thermally pulsing phase to be \mytilde $10^{4}$ -- $10^{5}$ years. However, if the 10 HLOs are the full population of stars of this class, then the lifetime estimated from the LMC is approximately an order of magnitude lower than this prediction, possibly indicating that formation or lifetime of the HLO class favors lower metallicity. This may also be due to a selection effect, or the star formation history of the LMC. In contrast, if the high luminosity HAVs are also included, we find similar lifetimes between the galaxies.

\subsection{Thorne-\.Zytkow Objects}\label{sec_life_tzo}

Next we investigate implications for the implied rate and lifetime of the T\.ZO phase if all the HLOs are of this origin. The only current estimate for the number of T\.ZOs visible in the Magellanic Clouds was completed by \citet{Tout.C.2014.HV2112SAGB}. Using the binary population synthesis code BSE and a rough estimate for the stellar mass in clusters with ages of $\sim$10$^7$ years, they estimate a 10\% probability of finding 1 T\.ZO in the SMC. At face value, this would make the probability of all 10 HLOs (and 20 high luminosity HAVs) being of a T\.ZO origin very small.

However, both the T\.ZO birthrate and subsequent lifetime of the T\.ZO phase are uncertain. The birthrate critically depends on uncertain outcomes of the common envelope phase, while the lifetime is typically estimated based on expectations for when the T\.ZO would exhaust its rp-process seed elements or undergo enough mass loss to reduce the envelope mass below the minimum required for fusion. These inferred lifetimes range from 10$^{5}$ $-$10$^{6}$ years for standard RSG winds \citep{Cannon.R.1993.TZOStructure,Biehle.G.1994.TZOObservational} down to 10$^{4}$ years if T\.ZOs enter an AGB-like super-wind phase \citep{Tout.C.2014.HV2112SAGB}. In order to investigate whether the uncertainties in these parameters are large enough to accommodate a T\.ZO origin for the HLOs, we perform an independent estimate for the birthrate of T\.ZOs in the Magellanic Clouds via both the common envelope evolution and supernova kick formation channels. We update the methodology of \citet{Podsiadlowski.P.1995.TZOEvolution} for the Magellanic Clouds. The inferred number of T\.ZOs is then simply the estimated birthrate multiplied by an average T\.ZO lifetime, $\langle \tau \rangle$.

For the common envelope channel, we follow \citet{Podsiadlowski.P.1995.TZOEvolution} and \citet{Taam.R.1978.TZOXRBperiods} who estimate that all HMXBs with periods $<$ 100 days will eventually enter a common envelope phase but fail to eject the envelope, thus producing a T\.ZO. We use recent X-ray binary catalogs to find the approximate number HMXBs in the Magellanic Clouds. \citet{Haberl.F.2016.HMXBinSMC} find 148 HMXBs in the SMC, of which 53 have reported orbital periods and 38 (72\%) of those have periods $\leq$ 100 days. In the LMC, \citet{Antoniou.V.2016.HMXBinLMC} find 42 HMXBs, 13 with a reported period, and 10 (78\%) with periods $\leq$ 100 days. We make no assumptions about the ratio of long- to short-period HMXBs over the entire population, as period measurements may be biased towards short-period systems. Instead, we calculate a minimum formation rate (assuming all P $\leq$ 100 day HMXBs form T\.ZOs) based on 38 HMXBs in the SMC and 10 HMXBs in the LMC. These numbers are combined with an estimate for the lifetime of the HMXB phase of $\sim$10$^5$ years \citep{Podsiadlowski.P.1995.TZOEvolution} to give a T\.ZO birthrate for the common envelope channel of $\geq 4\times10^{-4}$ yr$^{-1}$ in the SMC and $\geq 1\times10^{-4}$ yr$^{-1}$ in the LMC.

For SN kick formation, we adopt the binary system and geometric assumptions of \citet{Podsiadlowski.P.1995.TZOEvolution}: that $\sim$25\% of NSs are born in close binaries, that 25\% of those systems survive their SN as a bound system, that $\sim$25\% of those will receive a kick such that they spiral into the companion, and that $\sim$50\% of those will have a massive enough companion envelope to form a stable T\.ZO configuration. These are then combined with a SN rate for the Magellanic Clouds to yield a T\.ZO birthrate. Assuming that all stars with M $>$ 8 M$_\odot$ will explode as a SN, we adopt the number of core-collapse SN per unit stellar mass to be $0.01 M_{\odot}^{-1}$ \citep{Maoz.D.2010.SNeRateinClouds}. Note that this is conservative, as 10--30\% of core collapses result in failed SN \citep{Adams.S.2017.SearchforFailedSNE7years}. This is combined with average current star formation rates of 0.06 M$_{\odot}$ yr$^{-1}$ in the SMC and 0.25 M$_{\odot}$ yr$^{-1}$ in the LMC \citep{Chandar.R.2015.LMCSMCSFR}. Together, these yield a T\.ZO birthrate from SN kicks of \mytilde $5\times10^{-6}$ yr$^{-1}$ in the SMC and \mytilde $2\times10^{-5}$ yr$^{-1}$ in the LMC.

Based on these assumptions, we find that the common envelope formation channel dominates over the SN kick route in the Magellanic Clouds. This is in contrast to \citet{Podsiadlowski.P.1995.TZOEvolution} who estimated that the two channels were comparable in the Milky Way. This discrepancy is due to the lower SN rates in the Clouds, coupled with the large number of observed HMXBs.
Taking the combined birthrates of T\.ZOs in the Magellanic Clouds ($\geq$ $4.1\times10^{-4}$ yr$^{-1}$ in the SMC; $\geq$ $1.2\times10^{-4}$ yr$^{-1}$ in the LMC) we estimate maximum lifetimes for the T\.ZO phase of $\leq$ $1.7\times10^{4}$ yrs and $\leq$ $2.5\times10^{4}$ yrs for the SMC and LMC, respectively, if the 10 HLOs represents the full population. If we also include the 20 high luminosity HAVs, these maximum lifetimes increase to $\leq$ $2.4\times10^{4}$ yrs (SMC) and $\leq$ $1.7\times10^{5}$ yrs (LMC). Conversely, if we also include all HMXBs without reported periods when calculating the T\.ZO birthrate, we find \emph{minimum} lifetimes for the T\.ZO phase of $5.3\times10^{3}$ yrs (SMC) and $7.2\times10^{3}$ yrs (LMC).

These results are on the extreme low end of previous theoretical T\.ZO lifetime calculations. Adopting our updated Magellanic Cloud T\.ZO birthrates and the ``cannonical'' lifetimes of $\sim$10$^5$ to 10$^{6}$ years would yield a prediction of 40-400 T\.ZOs in the SMC and 12-120 T\.ZOs in the LMC, even more numerous than our current population of HLOs and HAVs. Thus, although our sample of HLOs and HAVs is much larger than some previous estimates for the number of T\.ZOs that should be present in the Clouds \citep{Tout.C.2014.HV2112SAGB}, we find that the number of HLOs cannot be used to exclude a T\.ZO origin as a possibility. This is a direct result of the large numbers of HMXBs in the Clouds, coupled with uncertainties associated with common envelop evolution and mass loss in cool and luminous stars.

\section{Discussion}\label{sec_discussion}

The goal of this work was to determine whether there is a population of objects with properties like that of HV2112 in the Magellanic Clouds, and to assess the nature of such a population. Our criteria of red, luminous stars with high amplitudes and double peak features in their light curves identified 11 candidates stars from the original ASAS-SN sample of over a million stars in the Magellanic Clouds. 

We will now discuss the HLOs as a class, highlight individual sources of note, and assess the implications of possible stellar identities.

\subsection{HLOs as a Class}\label{sec_hlo_class}

We defined as criteria for the HLOs that they must have similar variability properties to HV2112 -- specifically a high amplitude and a double peaked light curve shape. Through our photometric analysis in Sections 4 and 5, we have shown that the physical properties of the HLOs are broadly consistent with each other, with two exceptions (LMC-2, see \S\ref{sec_j81}, and LMC-3, see \S\ref{sec_havs_class}). All of the remaining HLOs have periods longer than 500 days. They are clustered in all of their photometric properties (Figure~\ref{phase_space_plots}).  Their mid-IR properties place most at the tip of the O-rich AGB branch (Figure~\ref{boyer_cmd}), and the HLOs all exhibit similar color evolution (Figure~\ref{phase_motion_diagrams}). There is no evidence for superwinds in any of the HLOs (\S\ref{sec_mlr_dust}). Finally, these stars are clustered together on the Hertzsprung-Russell diagram in Figure~\ref{hrds} and the period-luminosity diagram in Figure~\ref{l_vs_p}. Therefore, these stars all appear belong to the same broad class.

\subsubsection{Inclusion of HAVs in this Class}\label{sec_havs_class}

As seen in the phase space diagram (Figure~\ref{phase_space_plots}), the Hertzsprung-Russell diagrams (Figure~\ref{hrds}), and the luminosity-period diagrams (Figure~\ref{l_vs_p}), some of the high amplitude variables display similar physical properties to the HLOs. The 20 HAVs with higher luminosities and longer periods likely belong to the same class as the HLOs. This would imply that the double peak feature observed in the HLOs, sometimes attributed to shocks propagating through the stellar atmosphere \citep{Kudashkina.L.1994.MiraDPShocks}, may not be an indication of a unique evolutionary state. Additionally we note that some of the HAVs may have double peak features below the ASAS-SN detection limit. The 7 remaining, less luminous and shorter period HAVs are likely standard Mira variables. LMC-3 likely also falls into this category, as it overlaps with the evolutionary tracks of AGB stars with initial masses of 5 M$_{\odot}$ or less.

\subsubsection{HV2112}\label{sec_hv2112_specific}

Compared to the other HLOs (excluding LMC-2), HV2112 displays extreme properties. It has the highest V-band variability amplitude, the brightest mean V-band and 2MASS K$_{\mathrm{s}}$-band magnitudes, the highest mean luminosity, and the highest estimated current mass. It is the only HLO with an estimated mass $\gtrsim$10 M$_\odot$. Through our systematic analysis, we have demonstrated that there is no other star with as extreme properties in these regards in the SMC/LMC. Thus, while we believe that the overall conclusions on the nature of the HLO class (see \S\ref{sec_nature}) also apply to HV2112, it is clearly an extreme member.

Previous debate over the identity of HV2112 has included whether its luminosity was too high to be consistent with an AGB origin. \citet{Levesque.E.2014.HV2112disc} found a luminosity of log(L/L$_{\odot}$) = 5.02 through spectroscopic fitting, while \citet{Beasor.E.2018.HV2112AGB} integrated under an optical to IR SED and estimated a luminosity of 4.70 $<$ log(L/L$_{\odot}$) $<$ 4.91. Through our fitting of contemporaneous SEDs, we have shown that the luminosity of HV2112 oscillates over 4.68 $<$ log(L/L$_{\odot}$) $<$ 5.00 over its pulsation cycle, broadly consistent with both previous estimates.

The maximum luminosity for AGB stars is very model dependent. From \citet{Paczynski.B.1970.AGBLumLimit}, the maximum bolometric luminosity for an AGB star is $M_{bol}$ $\simeq -7.1$ mag, which corresponds to $\log(L/L_{\odot}) = 4.74$. The maximum luminosity of HV2112 is above this limit. More recently however, \citet{Eldridge.J.2009.PopSynthMassive} predicted that AGB stars can be as luminous as $\log(L/L_{\odot}) \simeq 5$ ($M_{bol}$ \mytilde $-8.2$ mag). Additionally, s-AGB stars are predicted to be more luminous than classical AGB limits. Lower metallicity models, in particular, can reach $\gtrsim 10^{5}$ L$_{\odot}$ \citep{Doherty.C.2017.SAGBStarsECSNE}. Therefore, the luminosity of HV2112 does not rule out a s-AGB origin. Further discussion of s-AGB stars as a plausible identity for the HLOs will follow in \S\ref{sec_disc_agb}.

\vspace{2em}

\subsubsection{LMC-2}\label{sec_j81}


LMC-2 has been the odd star out in several of our analyses. It is the most luminous star in our sample, with log(L/L$_{\odot}$) = 4.95-5.15, which would rule out an AGB identity. It is the only star to agree with RSG evolutionary tracks in the luminosity-period diagram (Figure~\ref{l_vs_p}), and at its highest luminosity and temperature, LMC-2 lies close to the RSG tracks in the HRD (Figure~\ref{hrds}). It is also the HLO with the highest mass loss rate and far-IR flux excess. It was classified as a `far-IR' object by \citet{Boyer.M.2011.SAGE.MC.Photom}. LMC-2 sits in the area of the color-magnitude diagram (Figure~\ref{boyer_cmd}) occupied by RSGs, so we posit that it would have been classified as an RSG if not for its dust. 

Finally, the light curve of this object has been slowly becoming less similar to our HLO criteria over time. While it initially appeared to have an exaggerated double peak light curve structure, the variability pattern of LMC-2 has changed in the more recent g-band photometry (see LMC-2 in Figure~\ref{all_hlos}), something not seen in the other HLOs. We conclude that LMC-2 likely does not belong in the HLO class, and it will not be included in the following discussion. LMC-2 could be an example of a late-stage RSG undergoing large amplitude pulsations, which has been theorized to occur in RSGs with large luminosity-to-mass ratios \citep{Heger.A.1997.RSGPulsationsOrig,Yoon.C.2010.RSGPulsationMassLoss}.

\subsection{Nature of Sources}\label{sec_nature}

\begin{deluxetable*}{cc||c|c|c|c||c|c|c|c||c|c|c|c}
\tabletypesize{\footnotesize}
\tablecolumns{14}
\tablecaption{A summary of whether various properties are consistent with AGB stars, super-AGB stars, RSGs, or T\.ZOs. LMC-2 is separated from all other HLOs as it consistently displays different properties, and LMC-3 is included with the low luminosity HAVs. Any divergent behavior amongst the rest of the HLOs is footnoted. A horizontal line separates observed properties from inferred properties. \label{hlo_classes_small}} 
\tablehead{\colhead{Property} & \multicolumn{1}{c||}{Section} & \multicolumn{4}{c||}{HLOs + High L$_{\odot}$ HAVs} & \multicolumn{4}{c||}{LMC-2} & \multicolumn{4}{c}{LMC-3 + Low L$_{\odot}$ HAVs}}
\startdata
    & & AGB & s-AGB & RSG & T\.ZO & AGB & s-AGB & RSG & T\.ZO & AGB & s-AGB & RSG & T\.ZO  \\\hline\hline
   Variability & \ref{sec_phase_space} & \textcolor{blue}{\checkmark} & \textcolor{blue}{\checkmark} & \textbf{\textcolor{red}{X}} & \textbf{?} & \textcolor{blue}{\checkmark} & \textcolor{blue}{\checkmark} & \textcolor{blue}{\checkmark}\textcolor{black}{\tablenotemark{a}} & \textbf{?} & \textcolor{blue}{\checkmark} & \textcolor{blue}{\checkmark} & \textbf{\textcolor{red}{X}} & \textbf{?} \\
   Mean V- Magnitude & \ref{sec_phase_space} & \textbf{\textcolor{red}{X}} & \textbf{?} & \textbf{\textcolor{red}{X}} & \textcolor{black}{\textbf{?}} & \textbf{\textcolor{red}{X}} & \textbf{?} & \textcolor{blue}{\checkmark} & \textcolor{black}{\textbf{?}} & \textbf{\textcolor{red}{X}} & \textbf{?} & \textbf{\textcolor{red}{X}} & \textcolor{black}{\textbf{?}} \\ 
   K$_{\mathrm{s}}$- Magnitude & \ref{sec_phase_space} & \textcolor{blue}{\checkmark} & \textbf{\textcolor{blue}{\checkmark}} & \textcolor{blue}{\textbf{\checkmark}} & \textbf{\textcolor{blue}{\checkmark}} & \textbf{\textcolor{blue}{\checkmark}} & \textbf{\textcolor{blue}{\checkmark}} & \textcolor{blue}{\checkmark} & \textcolor{blue}{\checkmark} & \textcolor{blue}{\checkmark} & \textbf{\textcolor{blue}{\checkmark}} & \textcolor{red}{\textbf{X}} & \textbf{\textcolor{red}{X}} \\ 
   J$-$K$_{\mathrm{s}}$ Colors & \ref{sec-cmag-jk} & \textcolor{blue}{\checkmark} & \textcolor{blue}{\checkmark} & \textbf{\textcolor{red}{X}\textcolor{black}{\tablenotemark{b}}} & \textcolor{blue}{\textbf{\checkmark}} & \textcolor{red}{\textbf{X}} & \textcolor{red}{\textbf{X}} & \textbf{\textcolor{blue}{\checkmark}} & \textcolor{blue}{\textbf{\checkmark}} & \textcolor{blue}{\checkmark} & \textcolor{blue}{\checkmark} & \textbf{\textcolor{red}{X}\textcolor{black}{\tablenotemark{c}}} & \textcolor{blue}{\textbf{\checkmark}} \\
   \hline
   Position on HRD & \ref{sec_hrd} & \textbf{\textcolor{red}{X}} & \textbf{\textcolor{blue}{\checkmark}} & \textbf{\textcolor{red}{X}} & \textcolor{blue}{\checkmark} & \textbf{\textcolor{red}{X}} & \textbf{\textcolor{red}{X}} & \textbf{\textcolor{blue}{\checkmark}} & \textcolor{blue}{\textbf{\checkmark}} & \textbf{\textcolor{blue}{\checkmark}} & \textbf{\textcolor{red}{X}} & \textbf{\textcolor{red}{X}} & \textcolor{red}{X}  \\ 
   Position on LPD & \ref{sec_lpd} & \textbf{\textcolor{red}{X}} & \textbf{\textcolor{blue}{\checkmark}} & \textbf{\textcolor{red}{X}} & \textcolor{black}{\textbf{?}} & \textbf{\textcolor{red}{X}} & \textbf{\textcolor{red}{X}} & \textbf{\textcolor{blue}{\checkmark}} & \textcolor{black}{\textbf{?}} & \textbf{\textcolor{blue}{\checkmark}} & \textbf{\textcolor{red}{X}} & \textbf{\textcolor{red}{X}} & \textcolor{black}{\textbf{?}} \\ 
   Pulsation Mass & \ref{sec_pulsation_mass} & \textbf{\textcolor{red}{X}} & \textbf{\textcolor{blue}{\checkmark}} & \textbf{\textcolor{red}{X}} & \textbf{\textcolor{red}{X}} & N/A & N/A & N/A & N/A & \textbf{\textcolor{blue}{\checkmark}} & \textbf{\textcolor{red}{X}} & \textbf{\textcolor{red}{X}} & \textbf{\textcolor{red}{X}} \\ 
   Dust MLR\tablenotemark{d} & \ref{sec_mlr_dust} & \textbf{\textcolor{blue}{\checkmark}}\textcolor{black}{\tablenotemark{e}} & \textbf{\textcolor{blue}{\checkmark}}\textcolor{black}{\tablenotemark{e}} & \textbf{\textcolor{blue}{\checkmark}}\textcolor{black}{\tablenotemark{e}} & \textbf{\textcolor{blue}{\checkmark}}\tablenotemark{f} & \textbf{\textcolor{blue}{\checkmark}}\textcolor{black}{\tablenotemark{e}} & \textbf{\textcolor{blue}{\checkmark}}\textcolor{black}{\tablenotemark{e}} & \textbf{\textcolor{blue}{\checkmark}}\textcolor{black}{\tablenotemark{e}} & \textbf{\textcolor{blue}{\checkmark}}\tablenotemark{f} & \textbf{\textcolor{blue}{\checkmark}}\textcolor{black}{\tablenotemark{e}} & \textbf{\textcolor{blue}{\checkmark}}\textcolor{black}{\tablenotemark{e}} & \textbf{\textcolor{blue}{\checkmark}}\textcolor{black}{\tablenotemark{e}} & \textbf{\textcolor{blue}{\checkmark}}\tablenotemark{f} \\ 
   Lifetime/Rates & \ref{sec_lifetimes} & N/A & \textcolor{blue}{\checkmark}\tablenotemark{g} & N/A & \textbf{\textcolor{blue}{\checkmark}}\tablenotemark{h} & N/A & N/A & N/A & N/A & N/A & N/A & N/A & N/A  \\
\enddata
\tablenotetext{a}{Large amplitude growth possible in RSGs, see Sec \ref{sec_j81} and \citet{Yoon.C.2010.RSGPulsationMassLoss}}
\vspace{-0.05in}
\tablenotetext{b}{LMC-4 has J$-$K$_{\mathrm{s}}$ colors more consistent with RSGs}
\vspace{-0.05in}
\tablenotetext{c}{LMC-3 has J$-$K$_{\mathrm{s}}$ colors more consistent with RSGs}
\vspace{-0.05in}
\tablenotetext{d}{Dust MLRs are not calculated for HAVs}
\vspace{-0.05in}
\tablenotetext{e}{Consistent with lower end of MLR range for RSGs and pre-superwind AGBs}
\vspace{-0.05in}
\tablenotetext{f}{If we assume T\.ZO MLRs are similar to RSG rates}
\vspace{-0.05in}
\tablenotetext{g}{Lifetimes are more consistent with theory and SMC/LMC population if high luminosity HAVs are also included}
\vspace{-0.05in}
\tablenotetext{h}{Despite the large number of HLOs+HAVs compared to previous estimates, given the uncertainty in the CE phase and the number of HMXBs, this number cannot exclude a T\.ZO origin (see \S\ref{sec_life_tzo})}
\end{deluxetable*}

We now discuss the possible nature of the HLOs. In Table \ref{hlo_classes_small}, we tabulate various properties that we have analyzed throughout this paper, both observed and inferred. LMC-2 is separated due to its consistently different properties. LMC-3 and the low luminosity HAVs are also separate. Any other differences between individual HLOs are noted in the table. Columns correspond to possible origins for the HLOs: AGB stars, s-AGB stars, RSGs, and T\.ZOs. In each cell, we indicate whether the HLOs (or LMC-2 or LMC-3) are consistent with each stellar class, with a blue check mark (\textcolor{blue}{\checkmark}) for a consistent result and a red X (\textcolor{red}{X}) for an inconsistent result. A question mark (\textbf{?}) indicates where there is uncertainty. These are mostly due to the lack of theoretical T\.ZO model predictions. Below, we discuss the HLOs' relation to these classes in more detail.

\subsubsection{Red Supergiants}\label{sec_disc_rsg}

The properties of the HLOs are largely inconsistent with both the observed properties (e.g. variability and colors; see Figures \ref{phase_space_plots} and \ref{boyer_cmd}, respectively) and theoretical predictions (Figures \ref{hrds} and \ref{l_vs_p}) for RSGs. While the overall luminosities and mass loss rates (Table \ref{dust_estimates}) are consistent with observations of RSGs, we find that the HLOs diverge significantly from both theoretical predictions of RSG evolution and control samples of RSGs in the Magellanic Clouds. Therefore, despite being very luminous and red stars, it is clear that the HLOs likely have a different physical origin.

\subsubsection{Asymptotic Giant Branch Stars}\label{sec_disc_agb1}

In terms of observed properties, such as variability and infrared colors (Figures~\ref{phase_space_plots} and \ref{boyer_cmd}), the HLOs are broadly similar to pulsating red giant stars (Mira variables). In particular, LMC-3 appears to be consistent with thermally pulsing AGB properties. Its position on the HRD (Figure~\ref{hrds}), luminosity-period diagram (Figure~\ref{l_vs_p}) and period-radius diagram (Figure~\ref{radius_period_plots}) are all consistent with an AGB star with initial mass of \mytilde 4-5 M$_{\odot}$. In addition, 7 of the HAVs have both K$_{\mathrm{s}}$-band magnitudes and periods similar to LMC-3. The characteristics of these stars conform with expectations for normal AGB stars (e.g. see the ``Low-L HAV'' in Figures~\ref{hrds}-\ref{l_vs_p}).

However, LMC-3 exhibited the lowest luminosity and shortest period of the HLOs.  For the other 10 HLOs and 20 more luminous HAVs, their properties differ from many of the properties of normal AGB stars. With maximum luminosities ranging over 4.58 $<$ log(L/L$_{\odot}$) $<$ 5.00 and estimated current masses $\sim6-11$ M$_\odot$, almost all HLOs have luminosities and masses greater than the classical AGB limit. 

\subsubsection{Super-Asymptotic Giant Branch Stars}\label{sec_disc_agb}

The high luminosities and inferred masses of the HLOs are, however, consistent with predictions for s-AGB stars \citep[e.g.][]{Siess.L.2010.SuperAGBEarly,Doherty.C.2015.SAGBIV}. This interpretation is further bolstered by the location of the HLOs on the HRD and period-luminosity diagram. We have shown that their placement can be reproduced by MESA models of intermediate mass ($\geq$ \mytilde 6M$_{\odot}$) stars \emph{that have entered the carbon-burning phase}. This off-center carbon burning is the evolutionary characteristic that distinguishes s-AGB from normal AGB stars (M $\lesssim$ 6M$_{\odot}$). The latter only progress to helium burning and cannot produce the combination of high luminosities and long periods observed in the HLOs (Figure~\ref{l_vs_p}).

We note that the late stages of s-AGB evolution are typically associated with very high mass-loss rates or ``superwinds'' with \.M $>$ 10$^{-5}$ M$_\odot$ yr$^{-1}$ \citep{Doherty.C.2015.SAGBIV}. This is not observed in the HLOs, which instead exhibit mass-loss rates on the order of 10$^{-8}$ $-$ 10$^{-7}$ M$_{\odot}$ yr$^{-1}$. However, multiple observations, including the identification of O-rich AGB stars with long periods ($\gtrsim$750 days) that are not self-extincted due to strong mass loss, have been used to argue for a delayed onset of the superwind phase \citep{Vassiliadis.E.1993.AGBEvowithMassLoss}.

Indeed, \citet{Doherty.C.2014.SAGBMassLossCrit} implement a pair of criterion for the onset of a superwind phase: an s-AGB star must either have a period longer than 850 days or a period longer than 500 days and a C/O ratio greater than 1. The latter accounts for the fact that higher mass-loss rates are expected when the atmosphere of a s-AGB star becomes carbon-rich, due to a change in opacity \citep{Marigo.P.2002.AGBCORatio,Cristallo.S.2007.AGB3DUCO}. Given that the HLOs all have periods $<$ 850 days \emph{and} have IR colors consistent with an oxygen-rich atmosphere, we suggest that they may be the population of s-AGB stars that have begun carbon burning but have not yet reached the superwind phase. Prior to this phase, mass loss should be on the order of $10^{-7}$ M$_{\odot}$ yr$^{-1}$ \citep{Doherty.C.2015.SAGBIV}, consistent with the HLOs (Table \ref{dust_estimates}).

Both the current mass and lifetime estimates for the HLOs are consistent with this scenario. \citet{Doherty.C.2015.SAGBIV} find that s-AGB stars of masses 6$-$10 M$_\odot$ should lose only $\sim$0.1$-$0.5 M$_\odot$ prior to onset of the thermally-pulsing or superwind phase. The initial masses of the stars would be therefore be close to the current masses we derive in \S\ref{sec_pulsation_mass}, and within the broad range expected for s-AGB stars. Similarly, the lifetime calculated in \S\ref{sec_life_sagb} overlaps with the lower end of the predicted s-AGB lifetime \citep[2$-$20$\times$10$^4$ years;][]{Doherty.C.2015.SAGBIV}. This is consistent with expectations if the HLOs (and possibly luminous HAVs) represent the population of s-AGB stars only in the pre-superwind evolutionary phase.

In this picture, the HLOs would subsequently evolve \emph{into} heavily dust obscured stars, before ending their lives as either O-Ne white dwarfs or exploding as electron capture (EC) SNe \citep{Miyaji.S.1980.OriginalecSN,Poelarends.A.2008.ECSneSAGBChannel}. Critically, while we do not directly constrain unperturbed radii and significant spread exists, the mean radii of 4 HLOs imply current masses $\gtrsim$8 M$_\odot$, at which point various models predict an EC SNe or even subsequent evolution to a core-collapse SN may be possible. In particular, with an estimated current mass of $\sim$10-11 M$_\odot$, HV2112 falls within the regime where neon may ignite off-center under degenerate conditions \citep[a ``hyper-AGB'' star]{Doherty.C.2015.SAGBIV}, and its existence would put strong constraints on s-AGB models. In the case of an explosive fate, the HLOs may represent direct evolutionary precursors to the dust-enshrouded progenitors of two peculiar low-luminosity transients discussed by \citet{Thompson.T.2009.LowMassLowLumTransients}. 

Thus, we find that the HLOs are consistent with predictions for s-AGB stars, so long as they are in the pre-superwind carbon burning phase. If these HLOs are s-AGB stars, it would significantly increase the population of known objects, as only one strong candidate has been identified to date \citep{Groenewegen.M.2009.MCRSGAGBStars2112SAGB}.

\subsubsection{Thorne-\.Zytkow Objects}\label{sec:TZOnature}

Our initial criteria used to define the HLOs were based on the variability of HV2112, which is considered the strongest Thorne-\.Zytkow Object candidate to date \citep{Levesque.E.2014.HV2112disc}. While spectroscopy/abundances has often been a key discriminant, here we discuss constraints that the photometric properties and variability of the HLOs can put on a possible T\.ZO origin. 

The temperatures, luminosities, and mass loss rates of the HLOs are broadly consistent with expectations for T\.ZOs, which are predicted to be among the coolest RSGs \citep{Thorne.K.1977.TZOstructure} and to have luminosities of log(L/L$_{\odot}$) \mytilde 4.8 $-$ 5.5 \citep{Cannon.R.1992.TZOStrucEvo}. The location of the HLOs on the HRD (Figure~\ref{hrds}) confirm that they are cooler than most RSGs, and while they are distinctly on the faint end of the distribution, most have peak observed luminosities that overlap with the range predicted by \citet{Cannon.R.1992.TZOStrucEvo}. In addition, although \citet{vanParadijs.J.1995.TZODetectionStrats} suggest that T\.ZOs may be dust enshrouded due to strong winds---which is not observed in this sample---the HLOs do have mass-loss rates consistent with RSGs of similar luminosities. 

However, one major prediction from T\.ZO theory is seemingly at odds with observations of the HLOs: that a minimum mass of $\sim$15M$_\odot$ is required for a T\.ZO to be a stable stellar structure \citep[e.g.][]{Cannon.R.1993.TZOStructure}. By combining temperatures, luminosities and periods of the HLOs with pulsation models we estimate current masses of \mytilde 6-11 M$_\odot$, distinctly below this limit. While there are stable ``giant'' T\.ZO models with total masses of $\sim$3.5-8.5 M$_\odot$ \citep{Cannon.R.1993.TZOStructure}, there is a predicted mass gap between these solutions (which are supported by accretion onto the NS) and the massive T\.ZO solutions (which are supported primarily by the irp-process at the base of the convective envelope). When a massive T\.ZO---as required to produce the abundance anomalies in HV2112---falls below the minimum mass to sustain nuclear fusion, it will undergo neutrino runaway and destabilize \citep[e.g.][]{Podsiadlowski.P.1995.TZOEvolution}. Thus, three options exist: (i) the lower mass limit for T\.ZOs needs to be modified, (ii) our current masses are underestimated, or (iii) the HLOs are not T\.ZOs. 

The mass lower limit for T\.ZOs is derived from the requirement that the temperature at the base of the convective envelope is 2$-$3$\times$10$^9$ K \emph{and} that the envelope is able to remain completely convective \citep{Cannon.R.1993.TZOStructure}. As a result, this limit depends sensitively on the convective efficiency, mixing length (MLT) parameters, and the mass of the NS (which influences the radiative temperature gradient). In the context of this model, either \emph{increasing} the MLT parameter, $\alpha$, or \emph{decreasing} the NS mass will decrease the T\.ZO mass limit.  \citet{Cannon.R.1993.TZOStructure} finds that assuming M$_{\rm{NS}}$ $=$ 1.4 M$_\odot$, $\alpha$ $=$ 1.5, and standard MLT convective velocities, the T\.ZO mass limit can be lowered to $\sim$10$-$11 M$_\odot$. In contrast, for $\alpha = 1$, the minimum envelope mass is $\gtrsim$20 M$_\odot$. Thus while reconciliation with the mass of HV2112 may be possible, explaining the population of HLOs as a whole would require significant modifications to the model for convection. We also note that if the minimum T\.ZO mass is lowered, then the predicted lifetime for this phase would increase, exacerbating the tension with the lifetime of the HLO phase estimated in \S\ref{sec_life_tzo}.

We emphasize that no direct model predictions for the variability or pulsation periods of T\.ZOs have yet been produced. Our mass estimates relied on s-AGB models to determine the normalizing constants in Equation~\ref{eq1}. However, the pulsation period should be primarily dependent on the mean density in the envelope and the HLOs also appear consistent with the location of the Hayashi track for low-metallicity, intermediate-mass, stars (Figure~\ref{hrds}). Similarly, while there are uncertainties in our derived physical properties, systematically shifting all HLOs to be $>$15 M$_\odot$ would require our measured luminosities to be systematically underestimated by a factor of 2, temperatures to be overestimated by $\lesssim$ 500K, or periods to be overestimated by more than a factor of 2, all inconsistent with the observational constraints. Thus, we conclude that for T\.ZOs to remain a viable origin for the HLOs, new T\.ZO models are required that either yield a lower minimum mass, or have detailed stellar structures that are capable of producing long pulsation periods ($>$500 days) at intermediate luminosities (4.5 $\lesssim$ $\log(L/L_\odot)$ $\lesssim$ 5.1).

\section{Summary and Conclusions}\label{sec_summary}

\emph{The Search:} We have performed a systematic search for cool, luminous, and highly variable stars in the Magellanic Clouds. From amongst $\sim$1.5 million objects, we identify 11 additional stars with photometric and variability characteristics extremely similar to that of the Thorne-\.Zytkow Object candidate HV2112, which we designate HV2112-like objects (HLOs), and 27 additional high-amplitude variables (HAVs) which are similar, but lack a distinctive double-peak feature in the rising phase of their light curves.  

\emph{Basic Properties:} The HLOs have V-band amplitudes $>$ 2.5 mag, periods $>$ 400 days, mean absolute V-band magnitudes between $-$2.5 and $-$5 mag, and absolute K$_{\mathrm{s}}$-band magnitudes between $-8$ and $-10$ mag, properties which make them outliers in comparison to both RSGs and Mira variables (Figure~\ref{phase_space_plots}). Their NIR colors are mostly consistent with O-rich AGB stars (Figure~\ref{boyer_cmd}), although they oscillate between the RSG and AGB branch of the optical-IR CMD throughout their pulsation cycle (Figure~\ref{phase_motion_diagrams}).

\emph{Physical Properties:} Through fitting contemporaneous optical-to-IR SEDs, we find the HLOs have temperatures, luminosities, and mass-loss rates of 3250 K $<$ T$_{\rm{eff}}$ $<$ 3600 K, 4.15 $<$ $\log(L/L_\odot)$ $<$ 5.15, and 1$\times$10$^{-7}$ M$_\odot$ yr$^{-1}$ $\lesssim$ \.M $\lesssim$ 4$\times$10$^{-6}$ M$_\odot$ yr$^{-1}$. Throughout their pulsation periods the temperatures of the HLOs vary by $\sim$200$-$400K, luminosities by $\sim$60-95\% and radii by $\sim$25$-$50\%. Combing these physical properties with theoretical pulsation models we estimate current masses for the HLOs of $\sim$5$-$11 M$_\odot$.

\emph{Lifetimes:} By considering the HLOs as descendants of either $\sim$ 6.5-10 M$_\odot$ stars or high-mass X-ray binaries (as expected for s-AGB stars and T\.ZOs, respectively) we derive lifetimes for the HLO phase of a few $\times$10$^4$ years. This is consistent with expectations for the s-AGB phase and approximately an order of magnitude lower than previous estimates for T\.ZOs.

\emph{Nature of the Sources:} We consider four possible origins for the HLOs and HAVs through comparison with theoretical models: RSGs, AGBs, s-AGBs, and T\.ZOs. A majority of HLOs are inconsistent with a RSG origin: appearing at cooler temperatures and, critically, displaying significantly longer periods than predicted for RSG-like stellar structures at these luminosities. However, one star, LMC-2 does have physical and pulsation properties consistent with a RSG origin. In this case, LMC-2 is one of  the highest amplitude pulsing RSGs discovered to date, possibly consistent with late-stage amplitude growth \citep{Yoon.C.2010.RSGPulsationMassLoss}.

In addition, while one HLO (LMC-3) and 7 lower-luminosity HAVs are consistent with expectations for normal AGB stars (M$\lesssim$5 M$_\odot$), the remaining 10 HLOs and 20 high luminosity HAVs are also inconsistent with this origin. Instead, we find that the luminosities, temperatures, pulsation periods, mass loss rates, current mass estimates, and inferred lifetimes of these are stars are all consistent with expectations for s-AGB stars, provided that they have begun carbon-burning, but have not yet entered a superwind phase. This would be the first confirmed population of s-AGB stars.

A detailed comparison to predictions for T\.ZOs is somewhat hampered by a lack of models describing their pulsation properties. However, one major theoretical prediction for T\.ZOs is in apparent conflict with our observations: all of the HLOs have current mass estimates that are below the estimated minimum mass of $\sim$15M$_\odot$ for a T\.ZO to be a stable stellar structure. While this mass limit is strongly dependent on the treatment of convection, in order for T\.ZOs to remain a viable origin for the HLOs, new T\.ZO models are required that either have a lower minimum mass or otherwise have a stellar structure capable of producing long ($>$ 500 day) pulsation periods at intermediate luminosities.

HV2112 remains the most extreme member of the HLO class. It displays the highest variability amplitude, highest luminosity, and largest current mass estimate at $\sim$10$-$11 M$_\odot$. There are no other stars as extreme in these regards in the SMC/LMC. Future spectroscopic observations of the HLOs will further elucidate their nature and connection to HV2112.

\acknowledgements

\noindent \section*{Acknowledgements}

The authors thank John Percy, Emily Levesque, Carolyn Doherty, Martha Boyer, Marten van Kerkwijk, Dae-Sik Moon, Katie Breivik, and Dan Huber for helpful conversations, and Tyler Downey and Miranda Herman for helpful edits. The authors thank the anonymous reviewer for a helpful and constructive referee report.

The authors at the University of Toronto acknowledge that the land on which the University of Toronto is built is the traditional territory of the Huron-Wendat, the Seneca, and most recently, the Mississaugas of the Credit River. They are grateful to have the opportunity to work in the community, on this territory. The Dunlap Institute is funded through an endowment established by the David Dunlap family and the University of Toronto. 

A.O. acknowledges support from the Queen Elizabeth II Graduate Scholarship in Science and Technology, Lachlan Gilchrist Fellowship Fund, and the Walter C. Sumner Memorial Fellowship. M.R.D acknowledges support from the Dunlap Institute at the University of Toronto and the Canadian Institute for Advanced Research (CIFAR). Part of this work was supported through the Hubble Fellowship Grant NSG-HF2-51373 to M.R.D, awarded through the Space Telescope Science Institute, which is operated by the Association of the Universities for Research in Astronomy Inc., for NASA, under contract NAS5-26555. BJS, KZS, and CSK are supported by NSF grants AST-1515927, AST-1814440, and AST-1908570. BJS is also supported by NSF grants AST-1920392 and AST-1911074. B.M.G. acknowledges the support of the Natural Sciences and Engineering Research Council of Canada (NSERC) through grant RGPIN-2015-05948, and of the Canada Research Chairs program. Support for JLP is provided in part by FONDECYT through the grant 1191038 and by the Ministry of Economy, Development, and Tourism's Millennium Science Initiative through grant IC120009, awarded to The Millennium Institute of Astrophysics, MAS. TAT is supported in part by NASA grant 80NSSC20K0531. This research was supported in part by the National Science Foundation under Grant No. NSF PHY-1748958. This research benefited from interactions made possible by the Gordon and Betty Moore Foundation through grant GBMF5076.

We thank the Las Cumbres Observatory and its staff for its continuing support of the ASAS-SN project.  ASAS-SN is supported by the Gordon and Betty Moore Foundation through grant GBMF5490 to the Ohio State University and NSF grant AST-1515927. Development of ASAS-SN has been supported by NSF grant AST-0908816, the Mt. Cuba Astronomical Foundation, the Center for Cosmology and AstroParticle Physics at the Ohio State University, the Chinese Academy of Sciences South America Center for Astronomy (CASSACA), the Villum Foundation, and George Skestos.

This research has made use of: the SVO Filter Profile Service (http://svo2.cab.inta-csic.es/theory/fps/) supported from the Spanish MINECO through grant AYA2017-84089 \citep{2012svo,2013svo}; the SIMBAD database, operated at CDS, Strasbourg, France \citep{SIMBAD.Wenger.M.2000}; and the NASA/IPAC Infrared Science Archive, which is funded by the National Aeronautics and Space Administration and operated by the California Institute of Technology. This research was enabled in part by support provided by Compute Canada (www.computecanada.ca). 

\software{astropy \citep{Astropy.Collab.2018.Astropy}, IRAF \citep{Tody.D.1986.IRAFI,Tody.D.1993.IRAFII}, ISIS \citep{1998ApJ...503..325A,2000A&AS..144..363A}, TOPCAT \citep{Taylor.M.2005.TOPCAT}, MARCS \citep{Gustafsson.B.2008.MARCSModels}, DUSTY \citep{Nenkova.M.2000.DUSTY}}

\appendix

\section{Additional Light Curves of HV2112}\label{hv2112_appendix}

\begin{figure*}[htb!]
    \centering
    \includegraphics[width=0.75\linewidth]{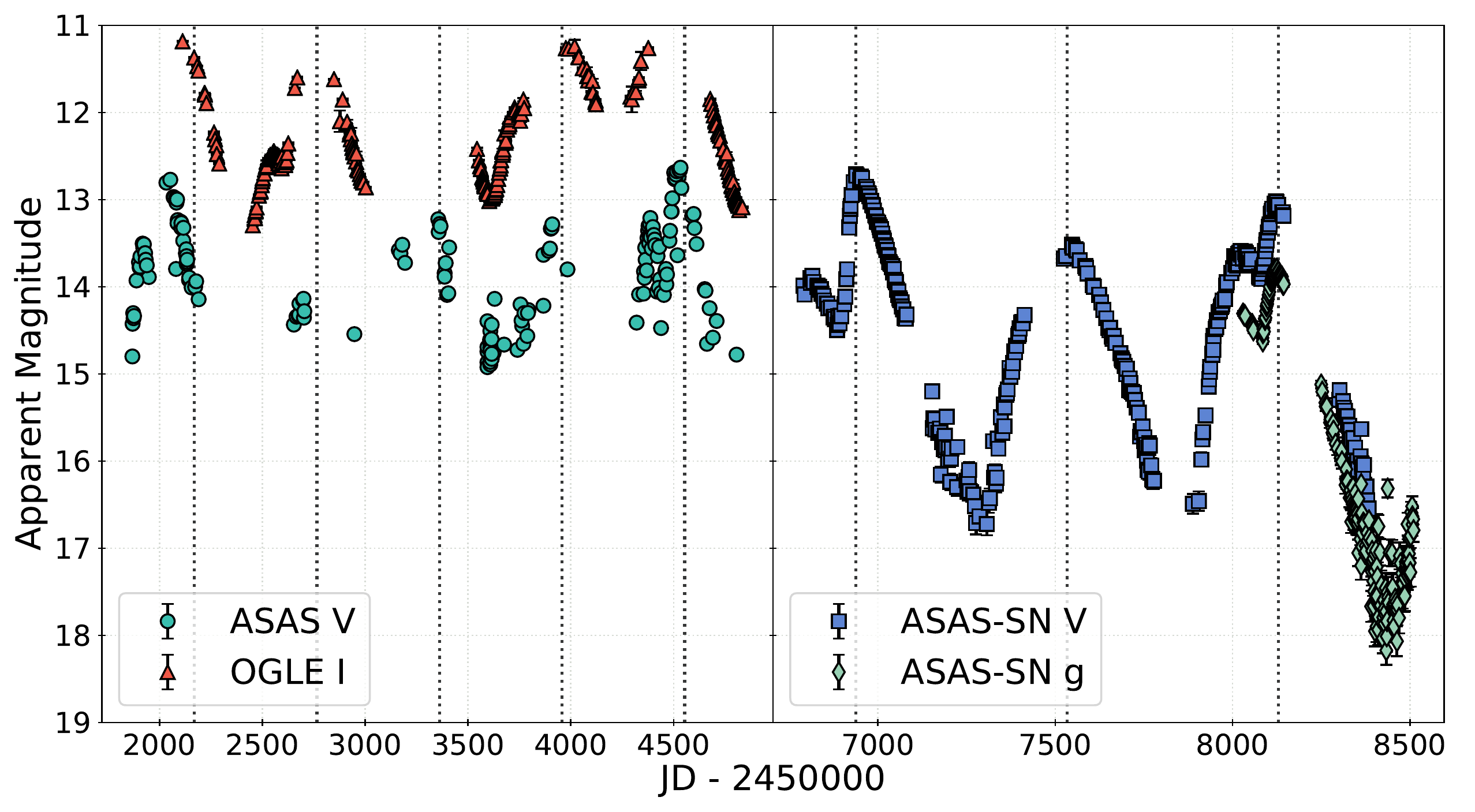}
    \caption{Light curves of HV2112, including ASAS V-band (left panel, teal circles) OGLE i-band (left panel, red triangles), ASAS-SN V-band (right panel, blue squares) and ASAS-SN g-band (right panel, green diamonds). Grey dashed lines indicate the placement of the light curve peaks with a period of 596 days, as derived in Section \ref{sec_lpvs}. While some variation is evident, the period is relatively stable over this 6000 day time period.}
    \label{hv2112_extra_lightcurves}
\end{figure*}

\clearpage

\vspace{-1em}

\section{Example of Anomalous Light Curves}\label{anom_app}

\begin{figure*}[htb!]
    \centering
    \includegraphics[width=0.45\textwidth]{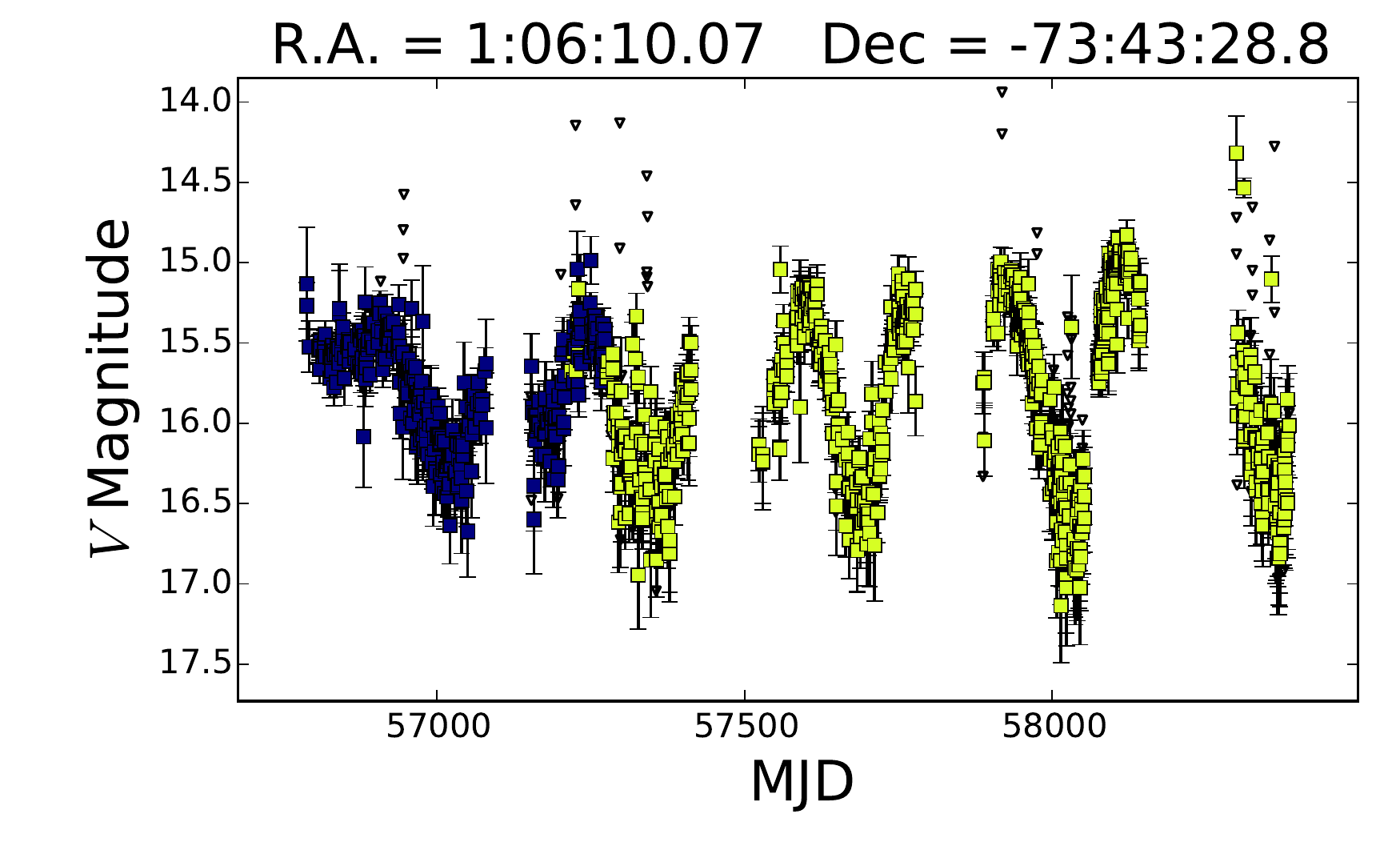}
    \includegraphics[width=0.45\textwidth]{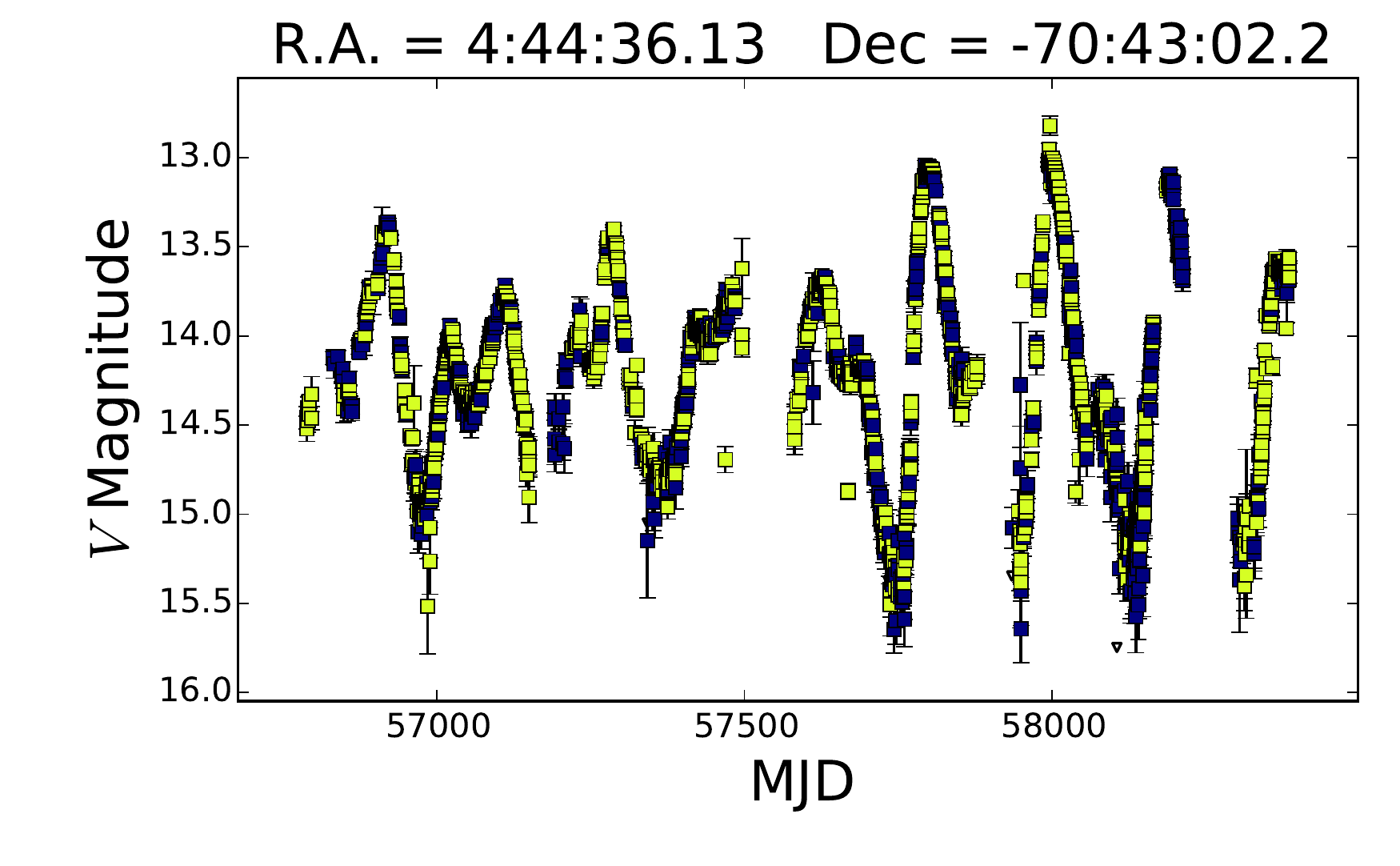}
    \includegraphics[width=0.45\textwidth]{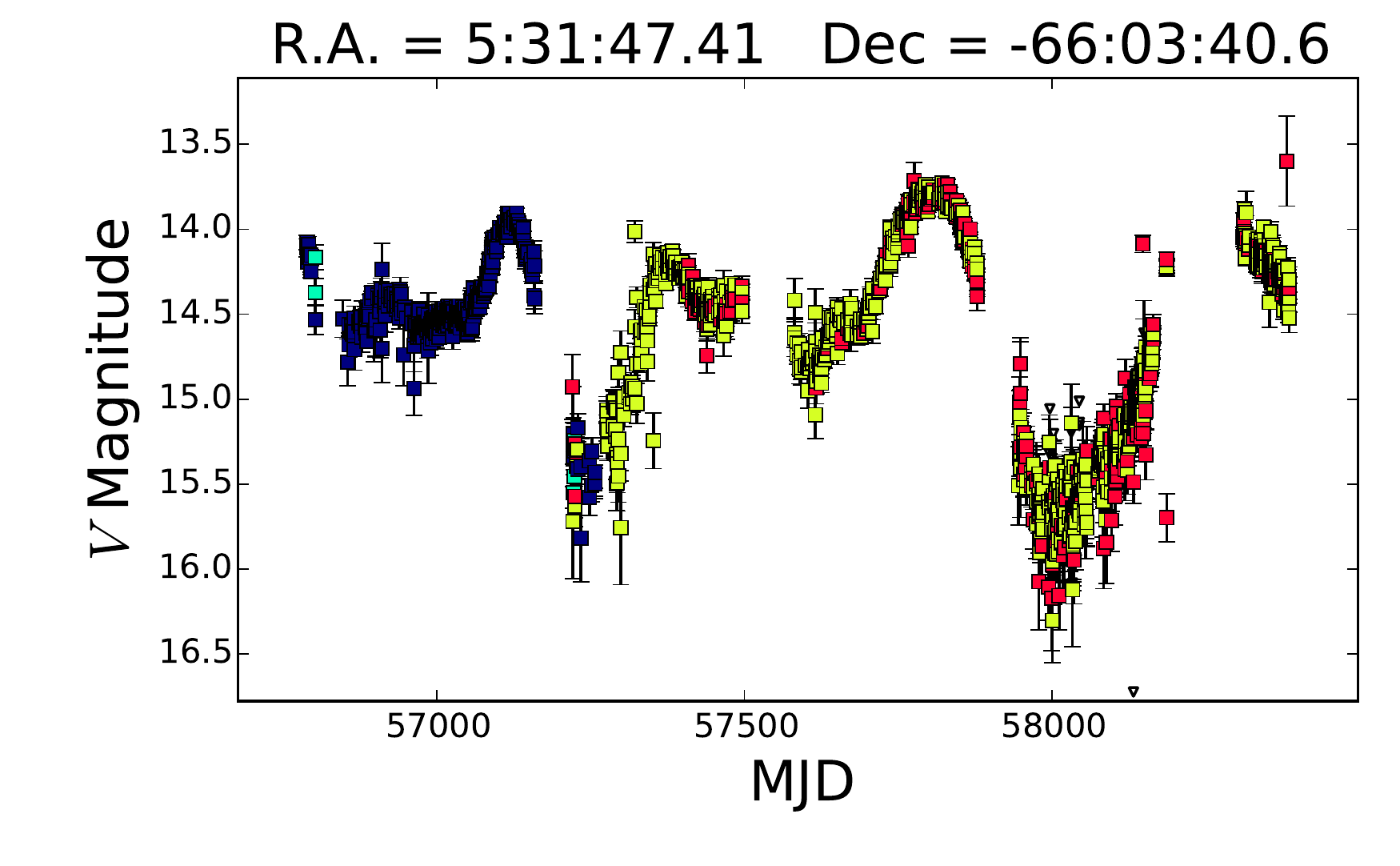}
    \includegraphics[width=0.45\textwidth]{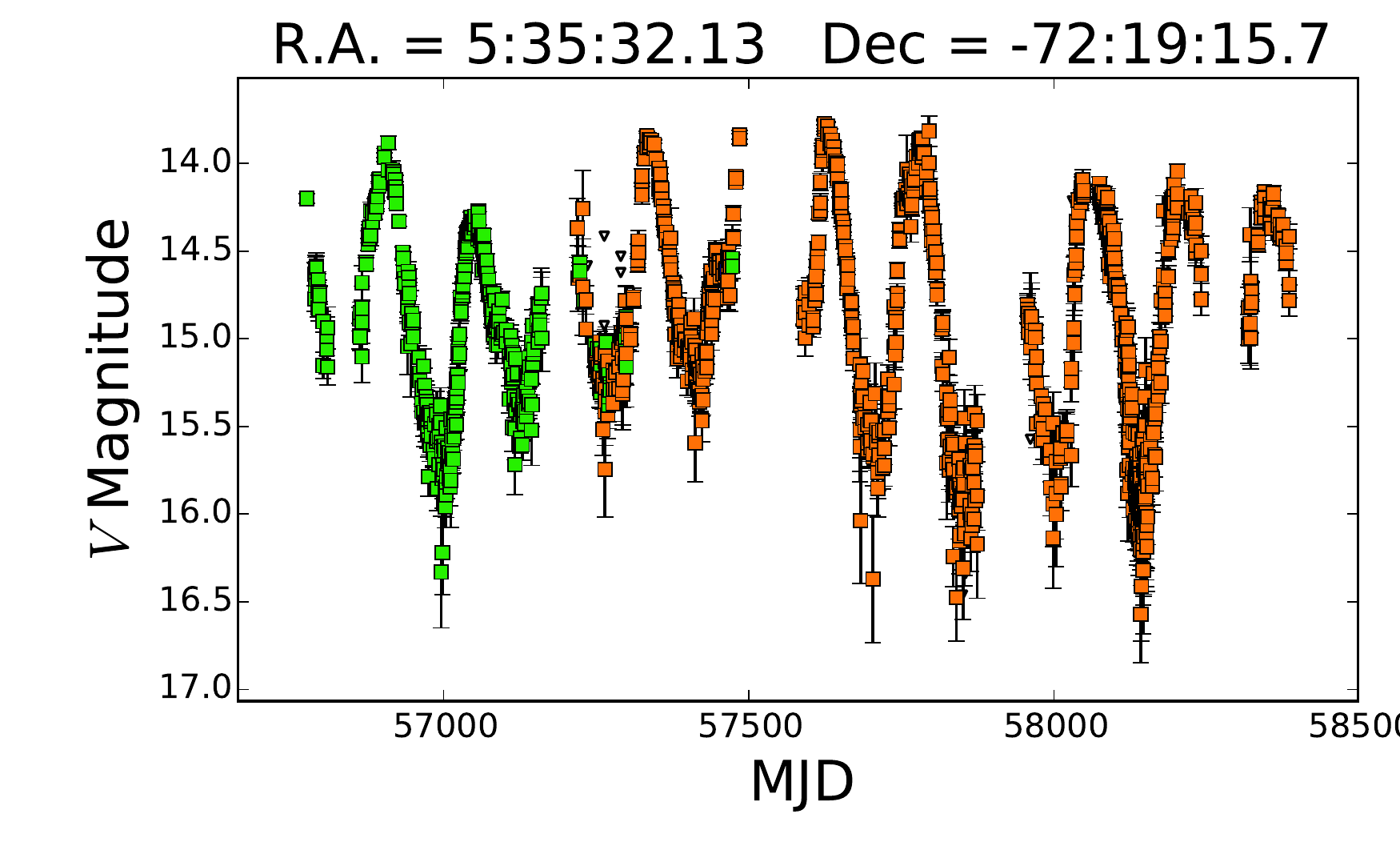}
    \caption{Examples of anomalous light curves.}
    \label{anom_lcs}
\end{figure*}

\clearpage

\section{HAV Data and Light Curves}\label{hav_appendix}

\begin{deluxetable*}{cccrrrr}[htb!]
\tabletypesize{\footnotesize}
\renewcommand\thetable{A}

\tablecaption{Basic Properties of HAVs. Stars marked with a `*' in their DEC column are included in Figures \ref{hrds}-\ref{l_vs_p}. \label{hav_info}}
\tablehead{\colhead{RA} & \colhead{DEC} & \colhead{2MASS} & \colhead{Period\tablenotemark{a}} & \colhead{$\Delta$V\tablenotemark{b}} & \colhead{Mean V\tablenotemark{c}} & \colhead{2MASS} \vspace{-0.275cm}\\
\colhead{J2000} & \colhead{J2000} & \colhead{Name} & \colhead{} & \colhead{} & \colhead{} & \colhead{K$_{\mathrm{s}}$-band\tablenotemark{c}} \vspace{-0.2cm}\\
 & & & \colhead{(days)} & \colhead{(mag)} & \colhead{(mag)} & \colhead{(mag)}
}
\startdata
00:41:21.43 &   -72:50:15.6* & J00412143-7250156    &	640.0 & 	2.6$\pm$0.27 &	15.62$\pm$0.76 & 	9.95$\pm$0.02 \\
00:57:14.48 &	-73:01:21.3* & J00571448-7301213    &	520.0 &  $>$3.5 &	14.59$\pm$0.99 & 	9.90$\pm$0.02 \\	 	
00:58:50.17 &	-72:18:35.6* & J00585016-7218355	&	750.0 &  $>$2.8 &	14.34$\pm$0.71 & 	8.61$\pm$0.02 \\		
04:35:57.34 &   -70:59:50.6 & J04355734-7059505     &	450.0 &  $>$2.7 &	15.71$\pm$0.56 & 	9.77$\pm$0.02 \\		
04:40:26.01 &	-71:39:42.4 & J04402601-7139423	    &	500.0 &  $>$3.6 &	14.98$\pm$0.83 & 	10.07$\pm$0.02 \\	 	
04:53:44.86 &	-68:57:59.3 & J04534486-6857593    &	920.0 & 	3.1$\pm$0.29 &	15.92$\pm$0.68 & 	8.86$\pm$0.02 \\		
04:57:28.85 &	-70:27:29.5 & J04572884-7027294	    &	610.0 & 	2.9$\pm$0.29 &	16.02$\pm$0.71 &  	8.75$\pm$0.02 \\		
04:58:55.65 &	-66:45:41.5* & J04585565-6645414	& 	620.0 &  $>$3.6 &	15.15$\pm$1.02 & 	8.57$\pm$0.02 \\		
05:01:23.99 &	-70:05:53.7 & J05012399-7005536	    &	620.0 &  $>$2.7 &	15.31$\pm$0.75 &   	9.30$\pm$0.02 \\ 		
05:06:04.24 &	-70:16:51.3 & J05060423-7016513	    & 	840.0 & 	2.6$\pm$0.30 &	15.51$\pm$0.64 & 	8.50$\pm$0.02 \\		
05:06:27.68 &	-68:12:03.7 & J05062768-6812036	    &	590.0 &  $>$2.7 &	15.19$\pm$0.57 &   	9.11$\pm$0.02 \\	             	
05:06:39.48 &	-71:35:56.5 & J05063948-7135564	    &	400.0 &  $>$3.1 &	15.36$\pm$0.77 &     10.10$\pm$0.02 \\	 	
05:07:38.30 &	-69:44:09.0 & J05073830-6944089	    &	630.0 & 	2.6$\pm$0.35 &	15.34$\pm$0.56 & 	8.98$\pm$0.02 \\ 		
05:11:59.87 &	-71:36:24.8 & J05115987-7136248	    &	750.0 &  $>$3.4 &	15.18$\pm$0.80 & 	8.98$\pm$0.02 \\		
05:15:40.84 &   -66:04:57.8* & J05154084-6604577	& 	540.0 & 	3.6$\pm$0.30 &	15.62$\pm$1.02 & 	9.33$\pm$0.02 \\		
05:19:10.46 &	-70:58:21.2 & J05191045-7058211	    &	420.0 & 	3.0$\pm$0.30 &	15.55$\pm$0.74 & 	9.95$\pm$0.02 \\		
05:20:01.57 &	-67:34:42.2 & J05200157-6734421	    &	590.0 & 	3.7$\pm$0.26 &	15.82$\pm$0.84 & 	9.36$\pm$0.02 \\		
05:23:10.15 &   -67:50:06.2 & J05231014-6750062	    &	730.0 &  $>$2.7 &	15.54$\pm$0.65 & 	8.44$\pm$0.02 \\		
05:24:22.20 &	-66:06:37.3* & J05242219-6606372	&	500.0 &  $>$3.2 &	15.49$\pm$0.79 &  	9.12$\pm$0.02 \\		
05:24:33.13 &	-70:42:36.1 & J05243313-7042361	    & 	600.0 &  $>$2.5 &	14.88$\pm$0.69 & 	9.43$\pm$0.02 \\ 	             	
05:29:17.70 &	-67:02:34.6 & J05291769-6702345	    &	680.0 &  $>$3.2 &	15.28$\pm$0.80 & 	8.79$\pm$0.02 \\		
05:32:59.92 &	-70:41:23.6 & J05325992-7041235	    & 	330.0 &  $>$2.7 &	15.14$\pm$0.70 & 	10.05$\pm$0.02 \\		
05:40:41.71 &	-66:14:46.8* & J05404170-6614467	&	330.0 &  $>$2.9 &	15.94$\pm$0.69 & 	10.03$\pm$0.02 \\	 	
05:49:13.36 &	-70:42:40.7 & J05491335-7042406	    &	680.0 &  $>$3.1 &	15.07$\pm$0.91 & 	9.45$\pm$0.02 \\		
05:51:55.25 &	-71:04:43.1 & J05515524-7104431	    &	640.0 & 	2.7$\pm$0.26 &	15.15$\pm$0.67 & 	8.79$\pm$0.02 \\		
05:58:44.33 &	-68:26:49.8 & J05584433-6826497    &	480.0 &  $>$3.8 &	14.97$\pm$1.02 & 	9.83$\pm$0.02 \\		
06:05:09.76 &	-72:40:35.2 & J06050976-7240352	    &	540.0 &  $>$3.7 &	15.55$\pm$0.98 &  	9.43$\pm$0.02 \\
\enddata
\tablenotetext{a}{Determination of periods is detailed in \S\ref{sec_lpvs}. Values in this table have been rounded to the nearest tenth; the periods are approximate, as we see cycle-to-cycle variations of the period on the order of \mytilde $\pm$ 15 days.}
\tablenotetext{b}{Determination of amplitudes is detailed in \S\ref{sec_high_amp_var}. ``$>$'' designates a lower limit. Errors are statistical from the data points, not systematic.}
\tablenotetext{c}{Apparent magnitudes, not corrected for extinction}

\end{deluxetable*}

\clearpage

\begin{figure*}[htb!]
    \centering
    \includegraphics[width=0.32\textwidth]{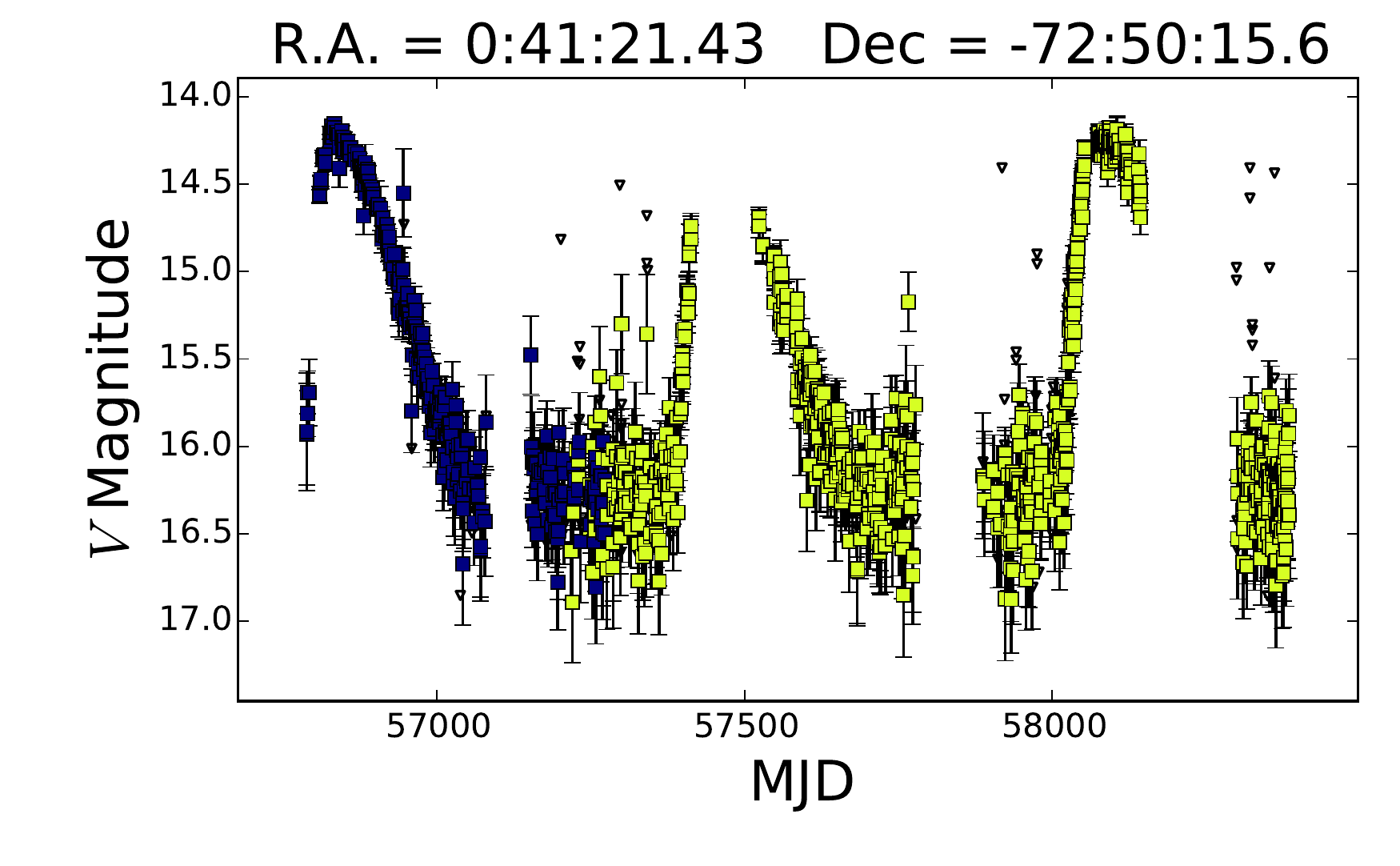}
    \includegraphics[width=0.32\textwidth]{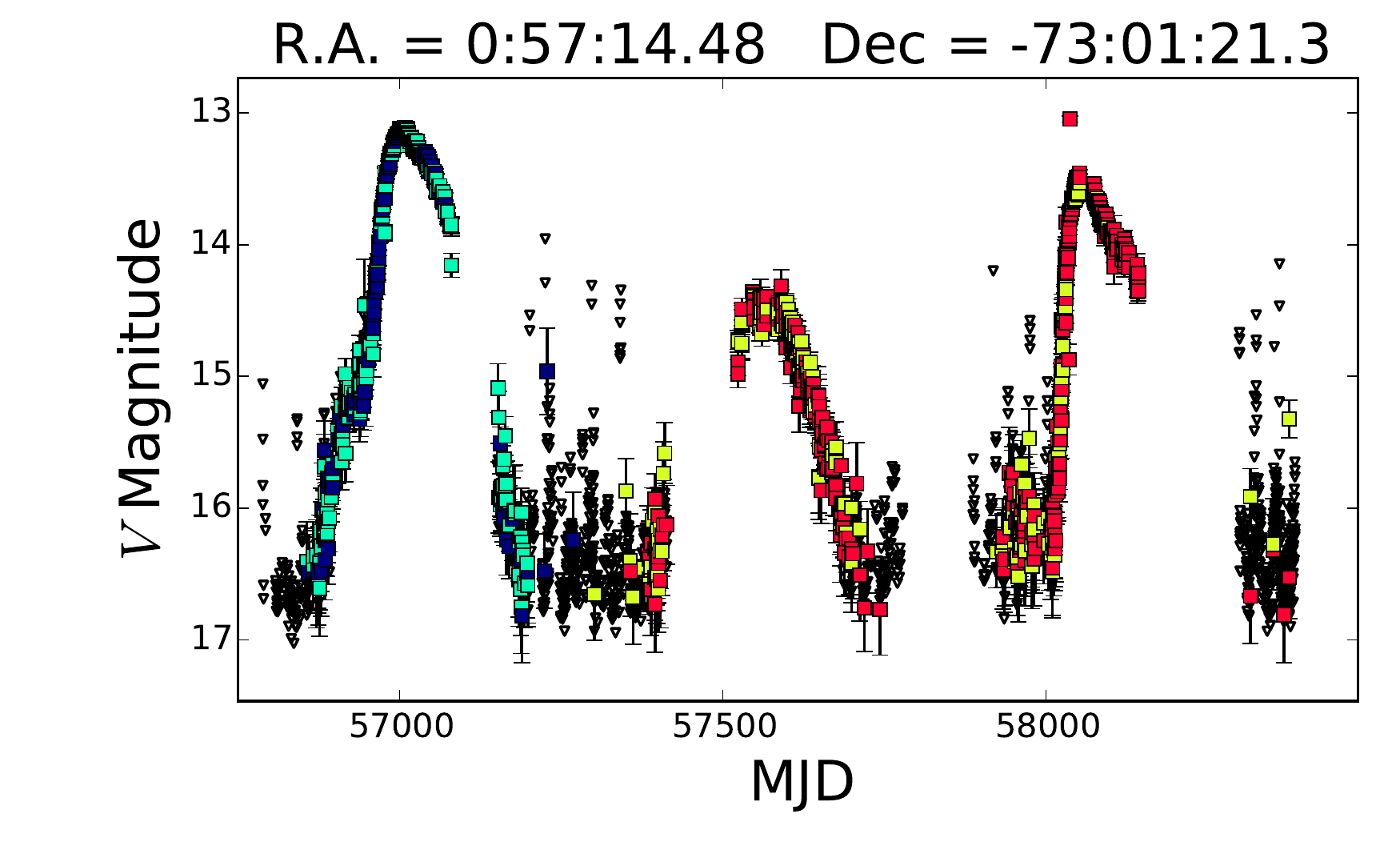}
    \includegraphics[width=0.32\textwidth]{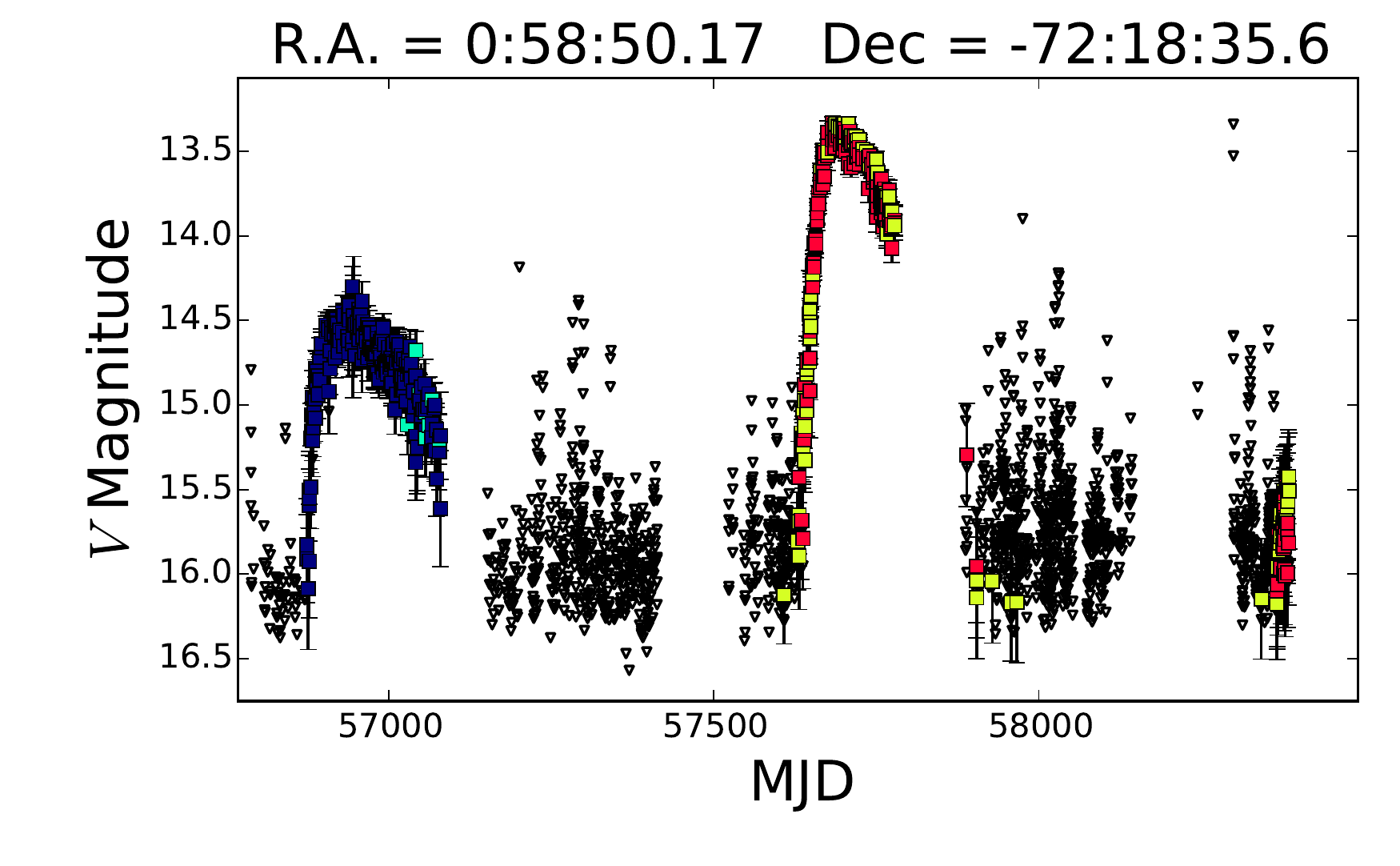}
    \includegraphics[width=0.32\textwidth]{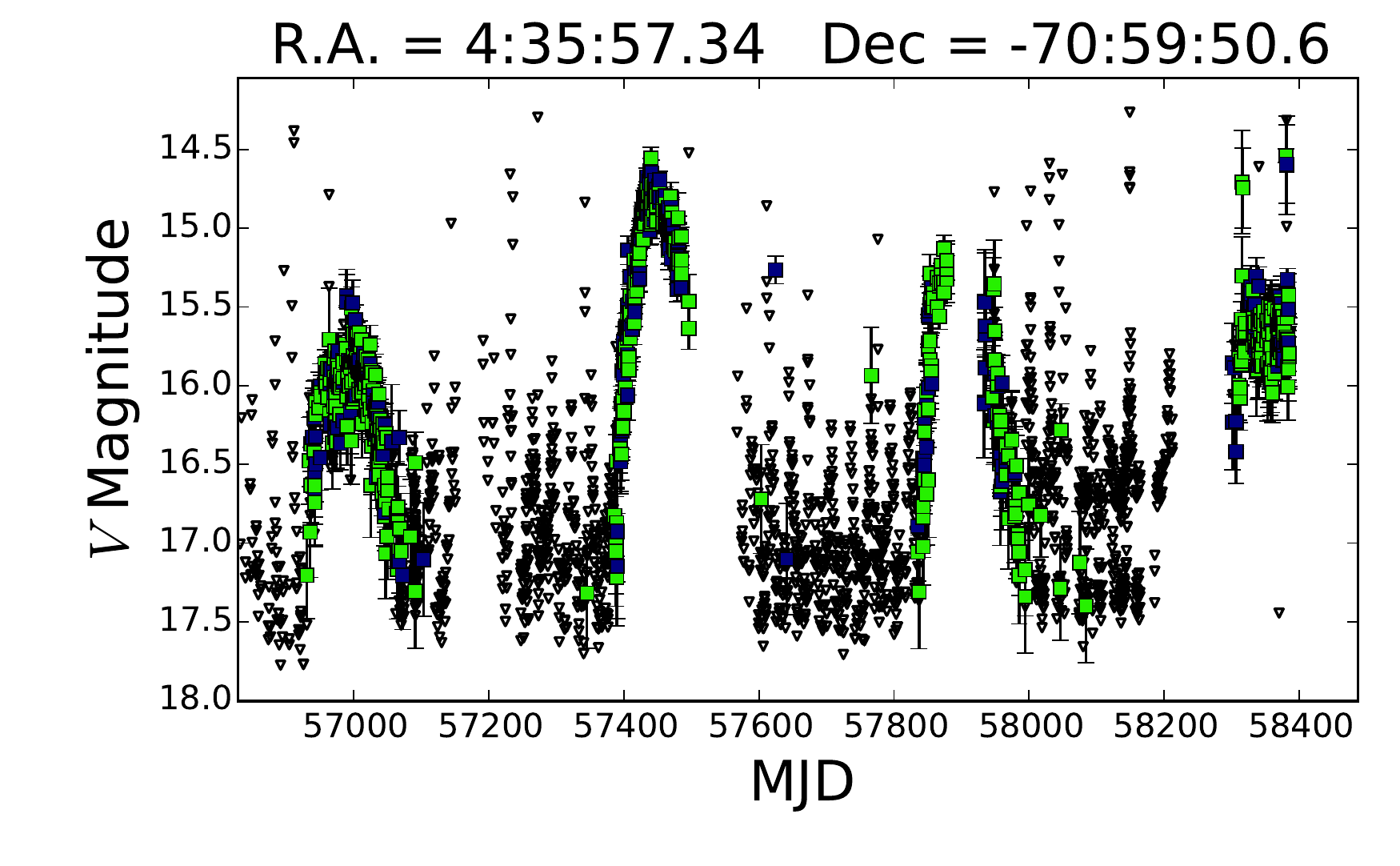}
    \includegraphics[width=0.32\textwidth]{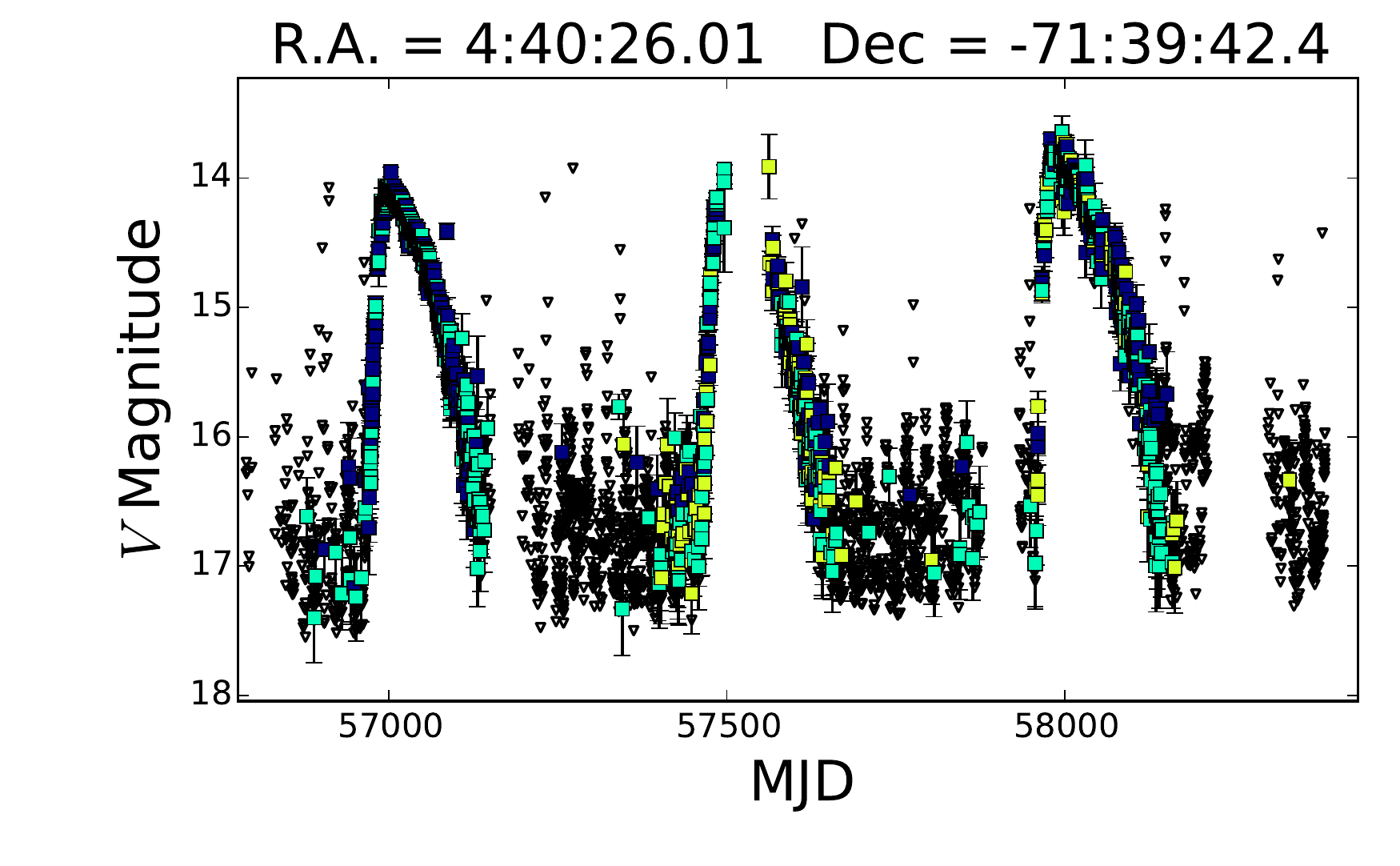}
    \includegraphics[width=0.32\textwidth]{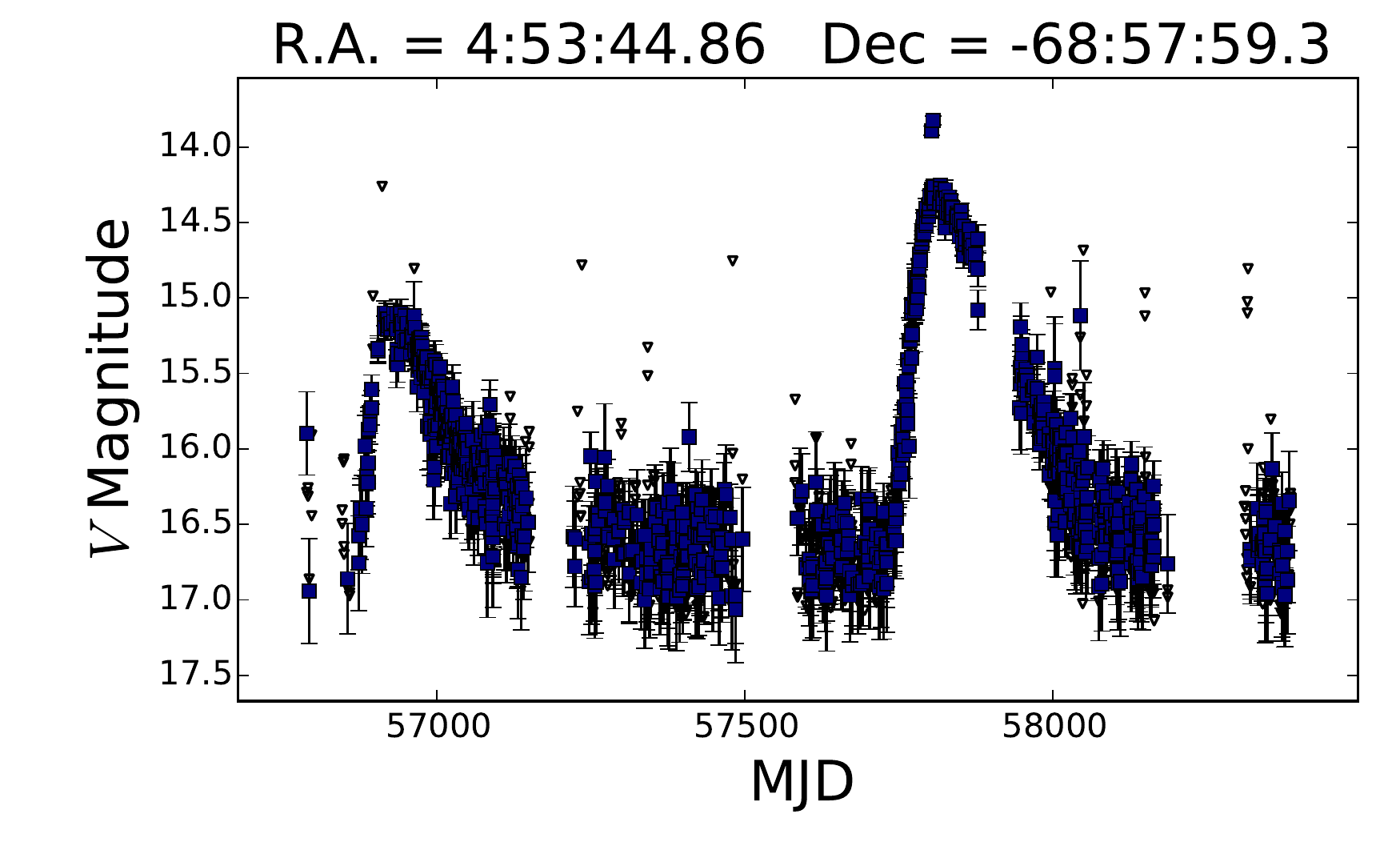}
    \includegraphics[width=0.32\textwidth]{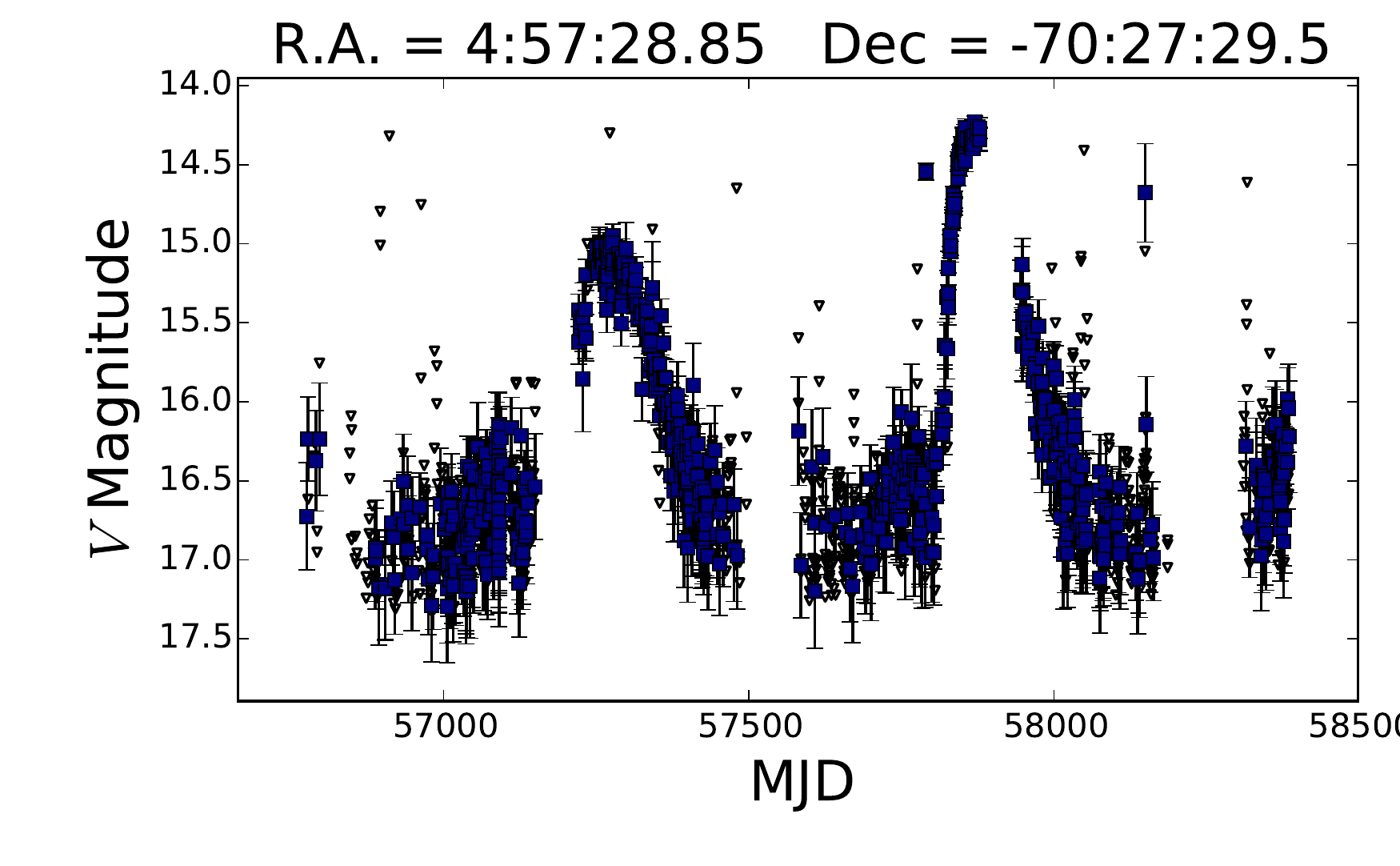}
    \includegraphics[width=0.32\textwidth]{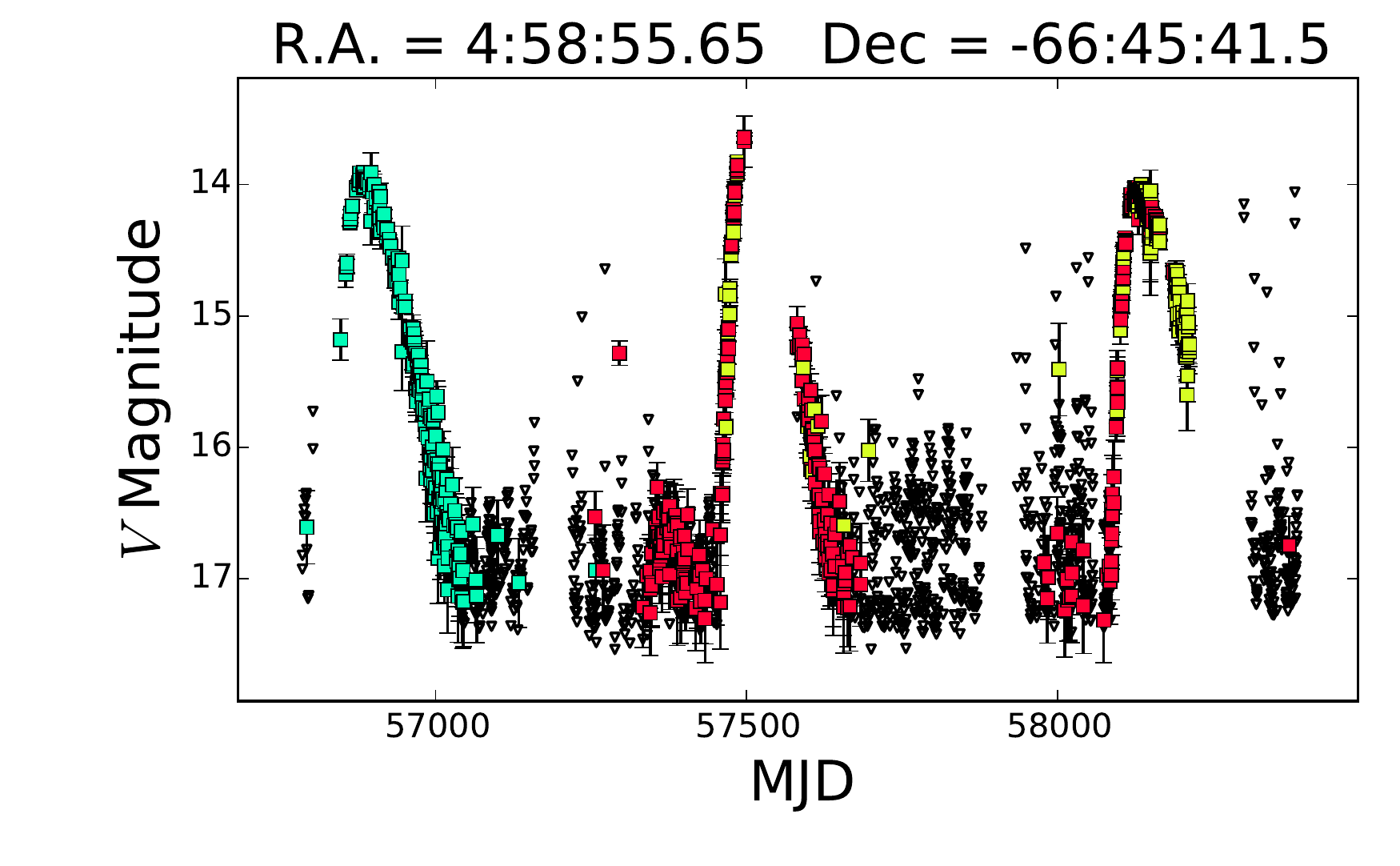}
    \includegraphics[width=0.32\textwidth]{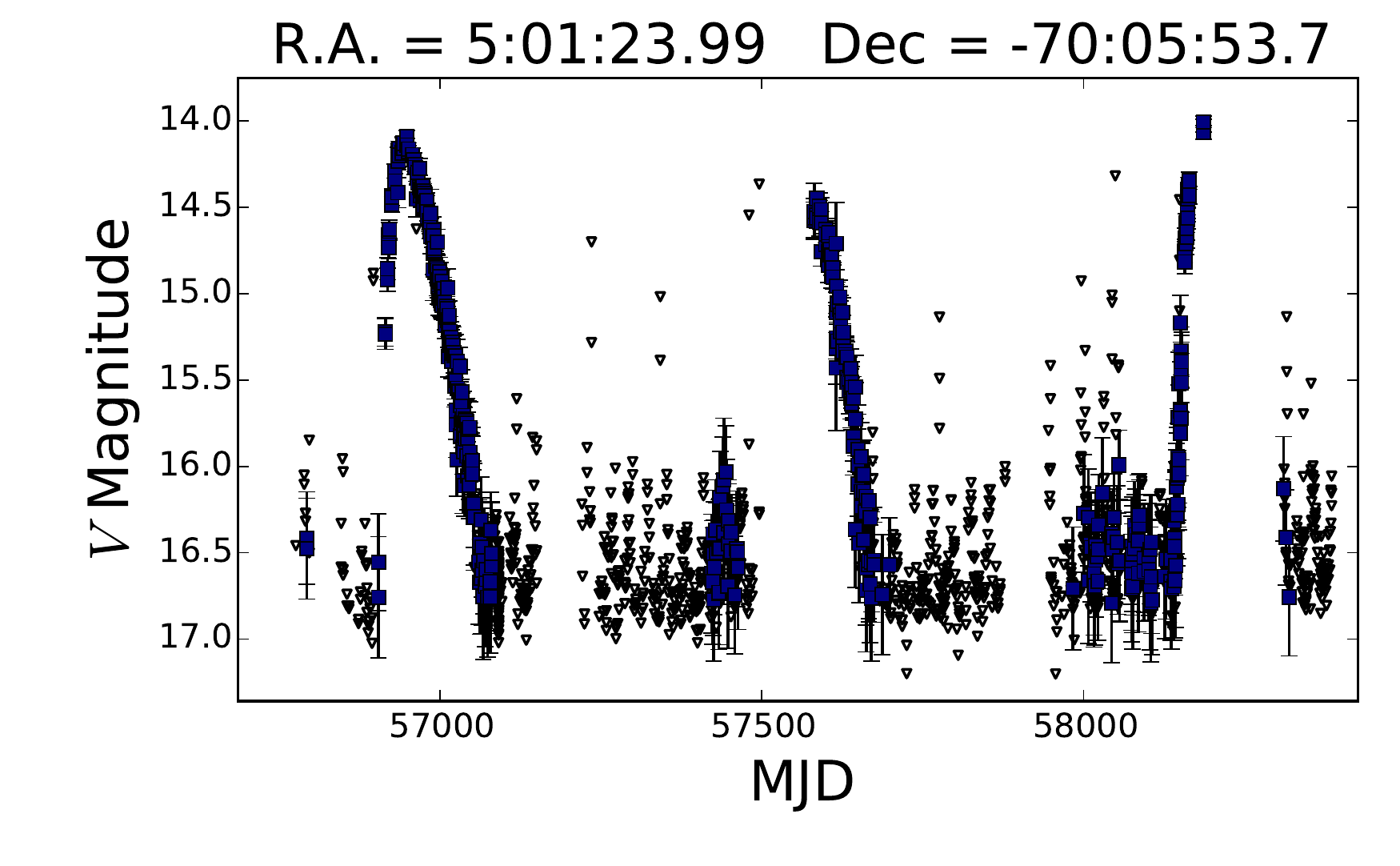}
    \includegraphics[width=0.32\textwidth]{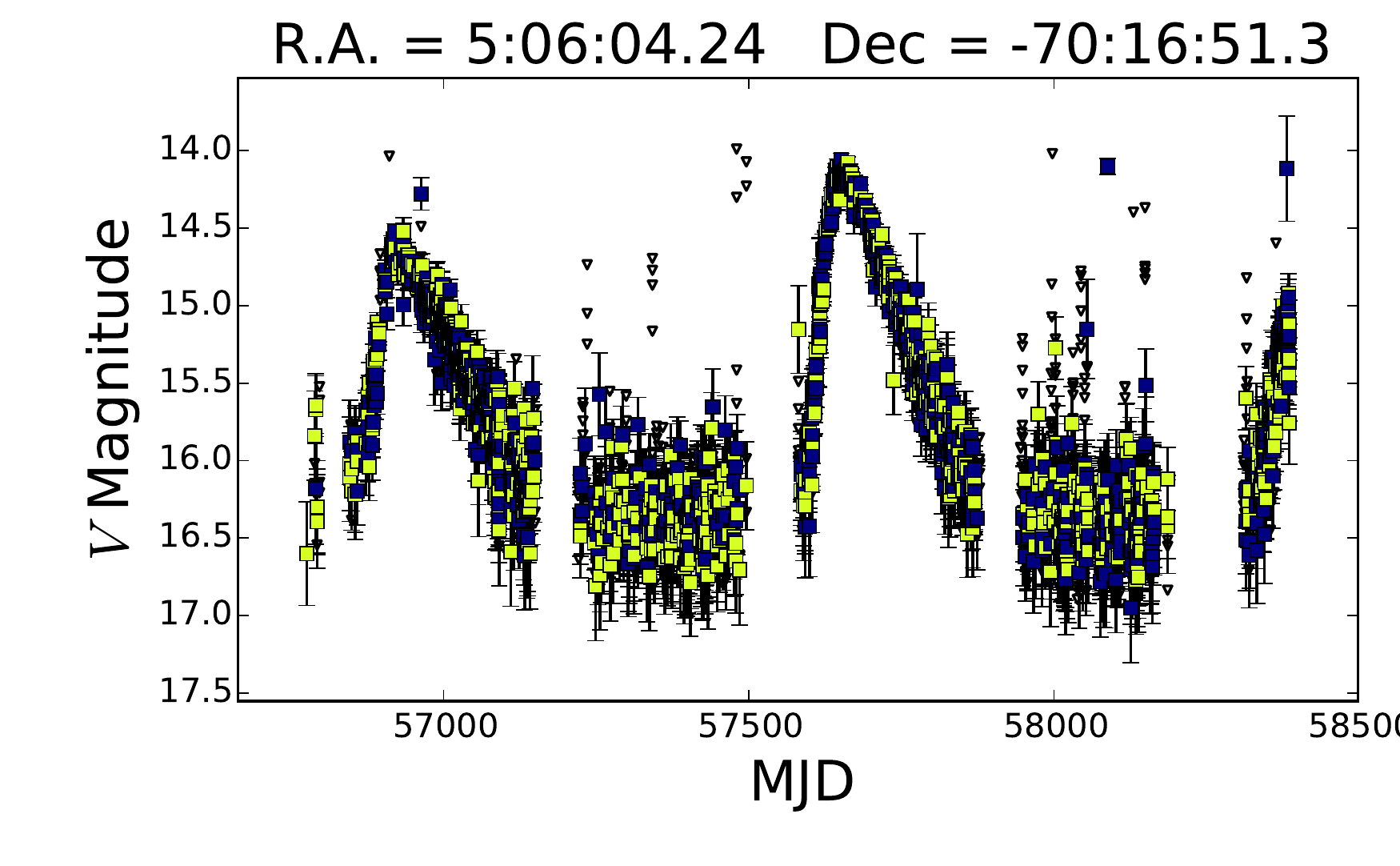}
    \includegraphics[width=0.32\textwidth]{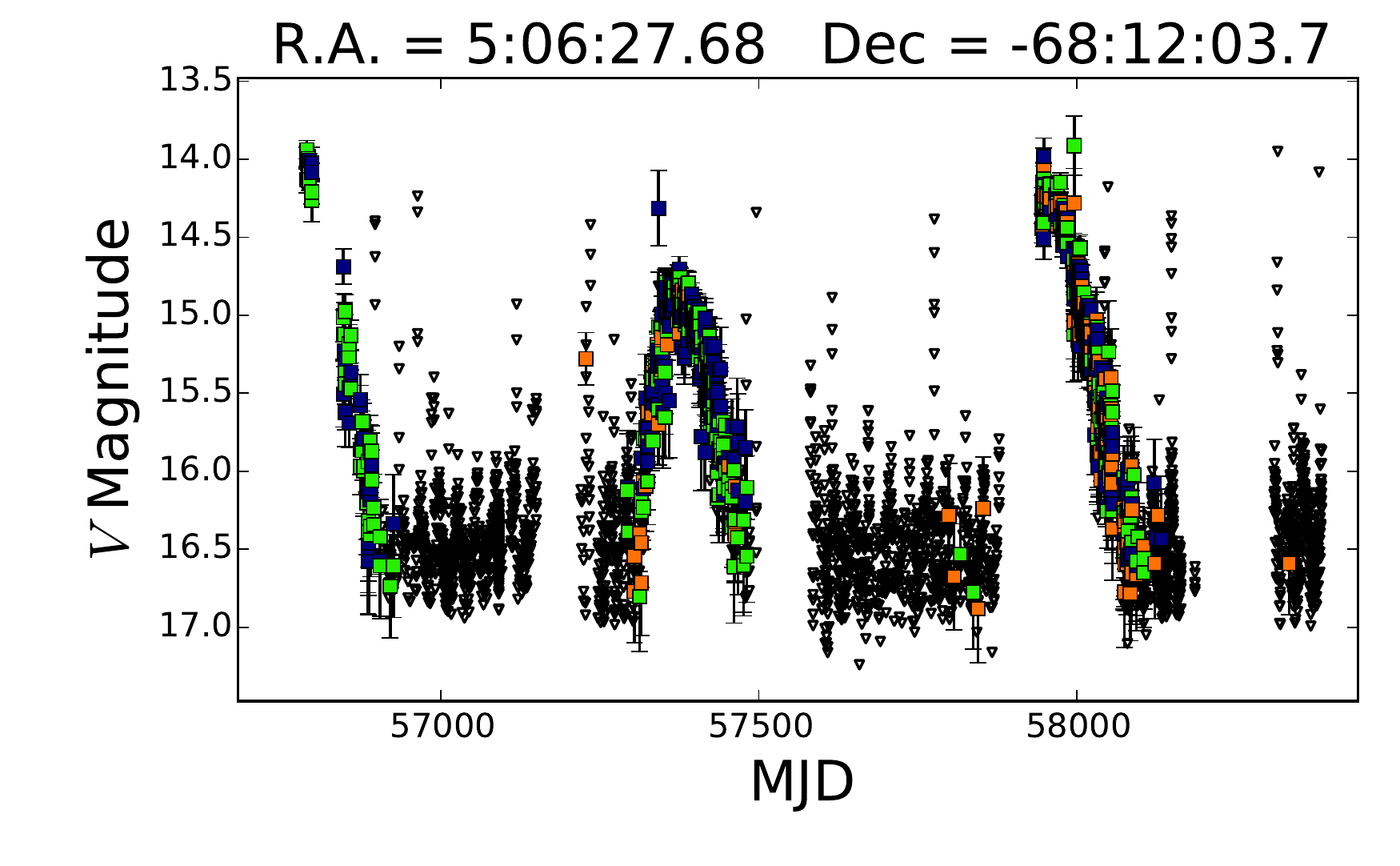}
    \includegraphics[width=0.32\textwidth]{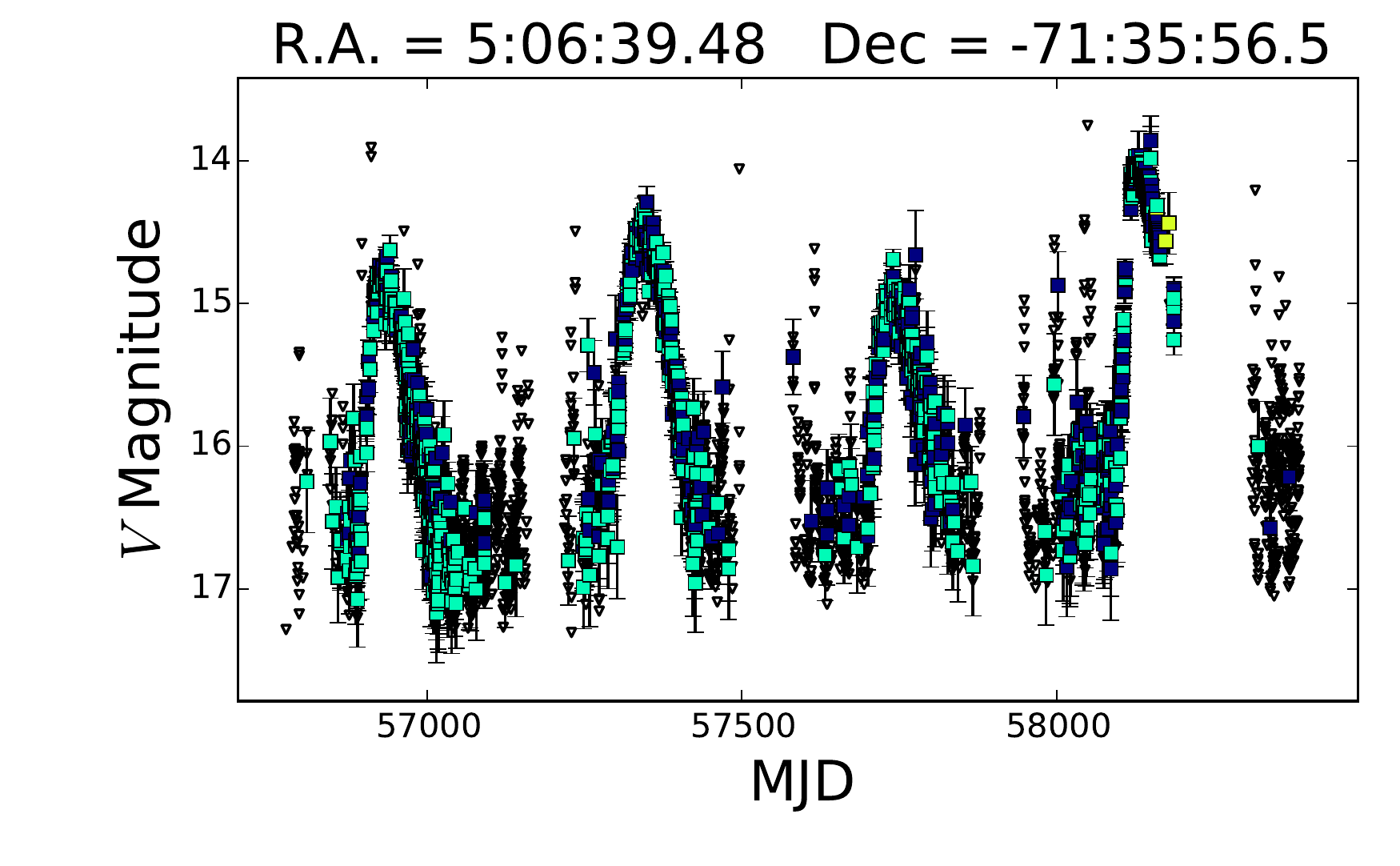}
    \includegraphics[width=0.32\textwidth]{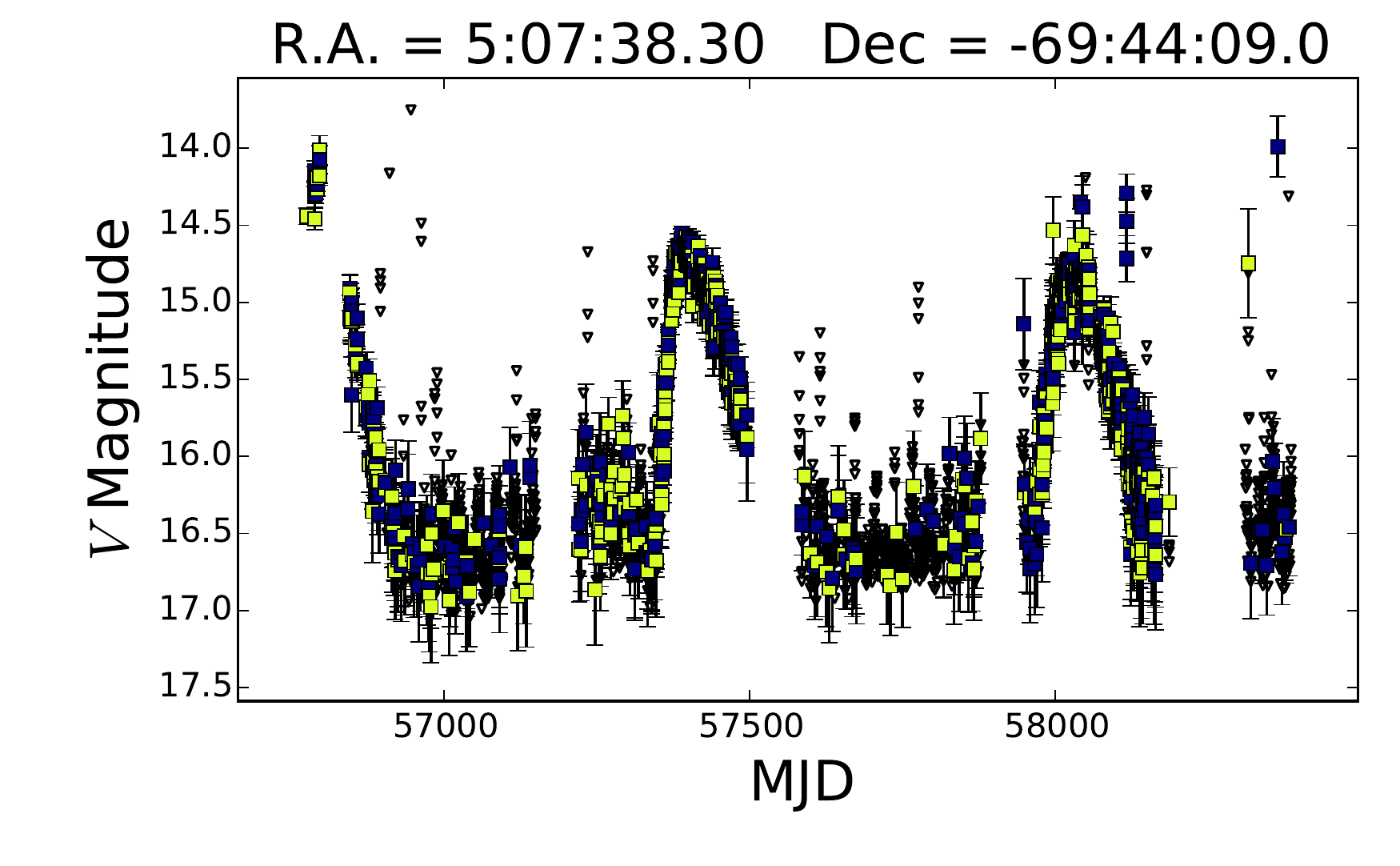}
    \includegraphics[width=0.32\textwidth]{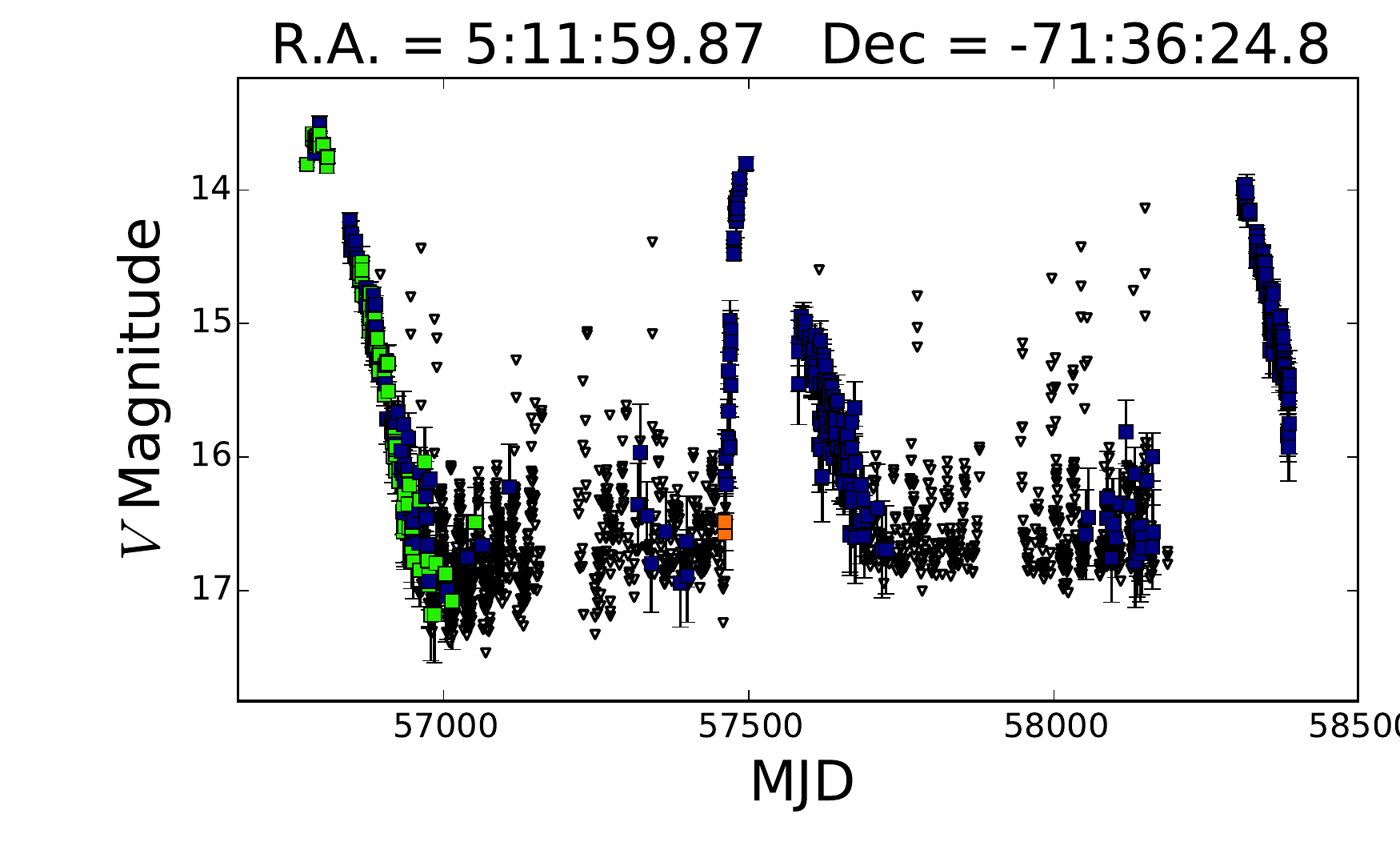}
    \includegraphics[width=0.32\textwidth]{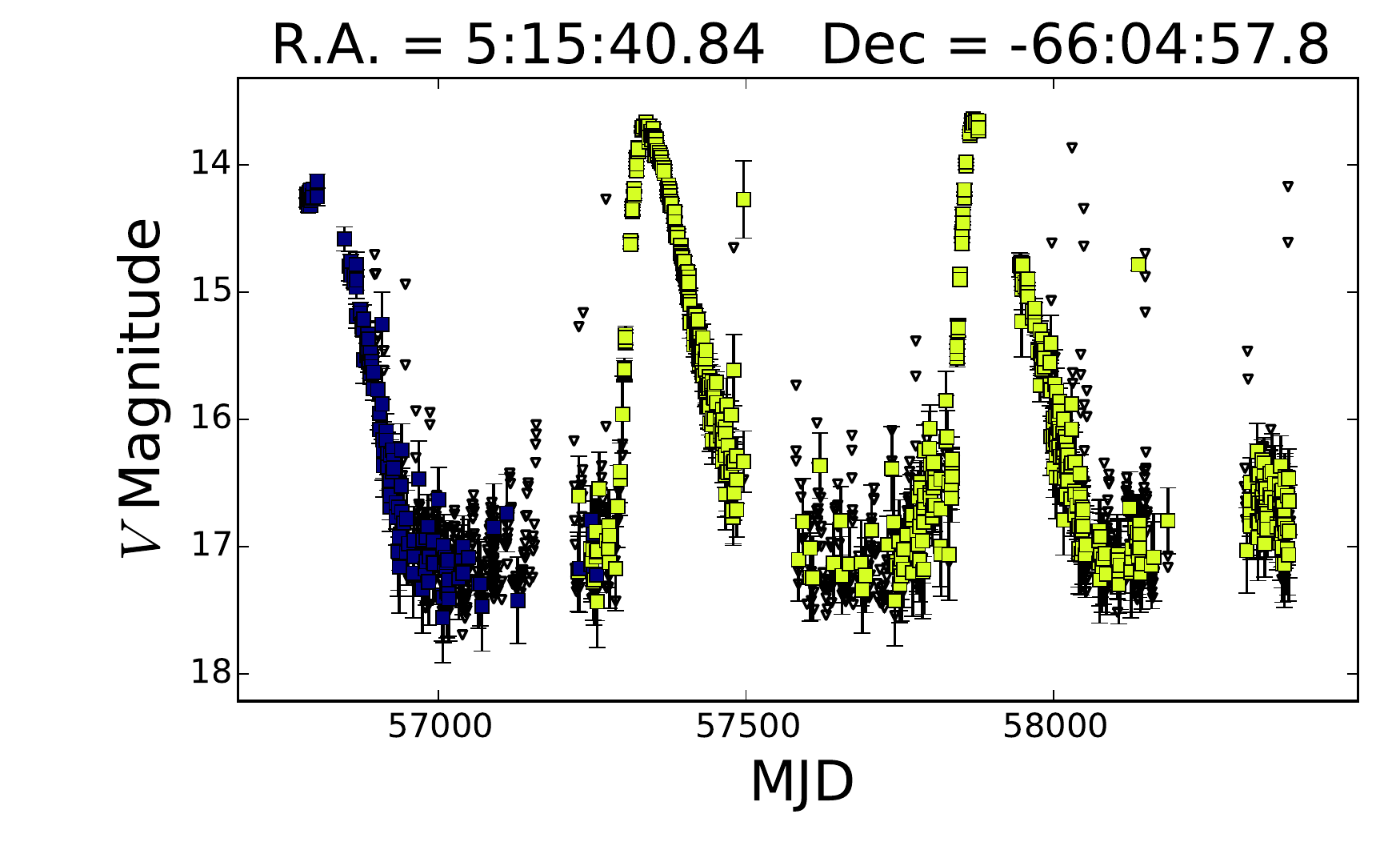}
    \caption{Light curves of high amplitude variables. The properties of these stars are outline in Table \ref{hav_info}.}
    \label{hav_lcs1}
\end{figure*}

\begin{figure*}[htb!]
    \centering
    \includegraphics[width=0.32\textwidth]{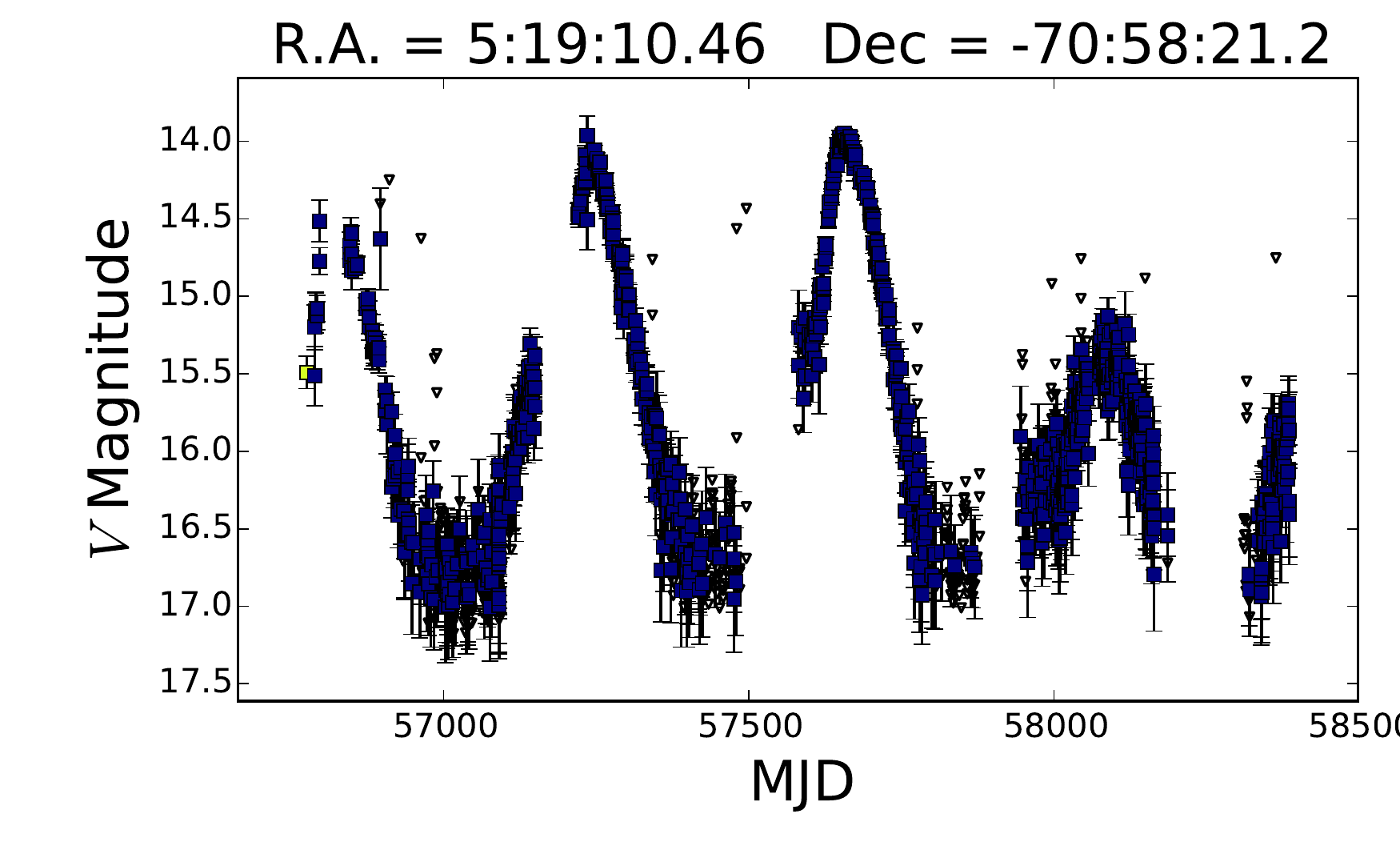}
    \includegraphics[width=0.32\textwidth]{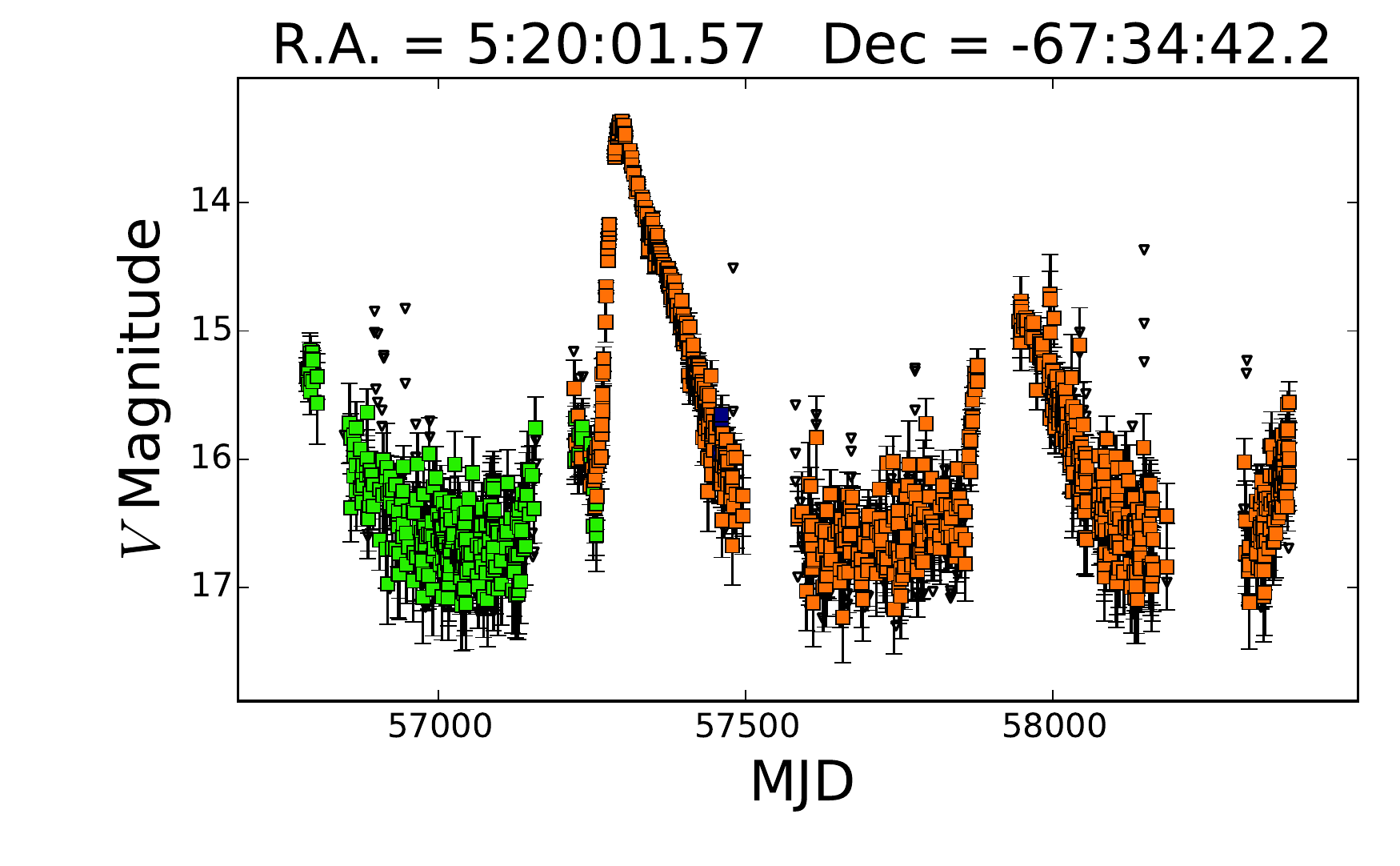}
    \includegraphics[width=0.32\textwidth]{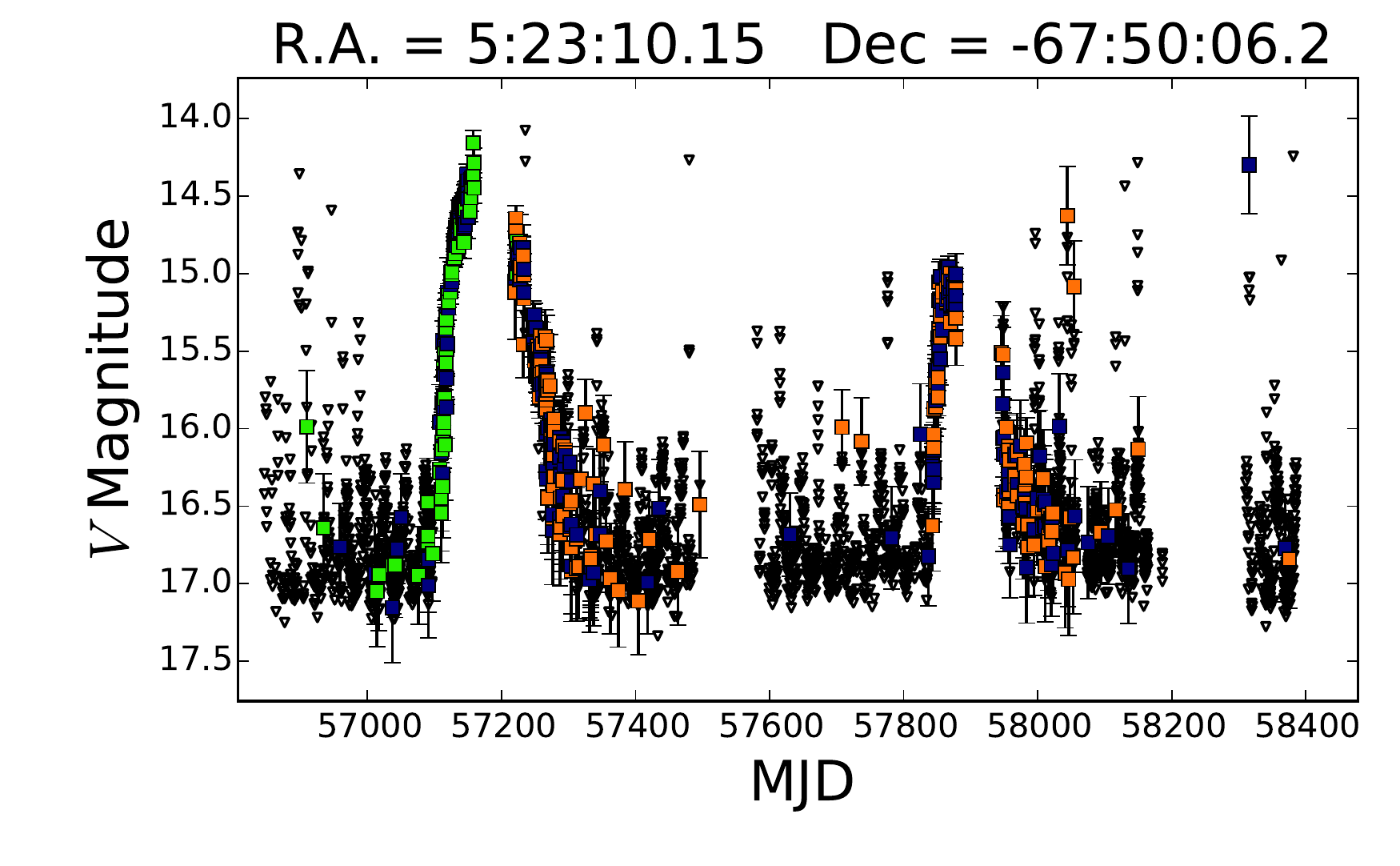}
    \includegraphics[width=0.32\textwidth]{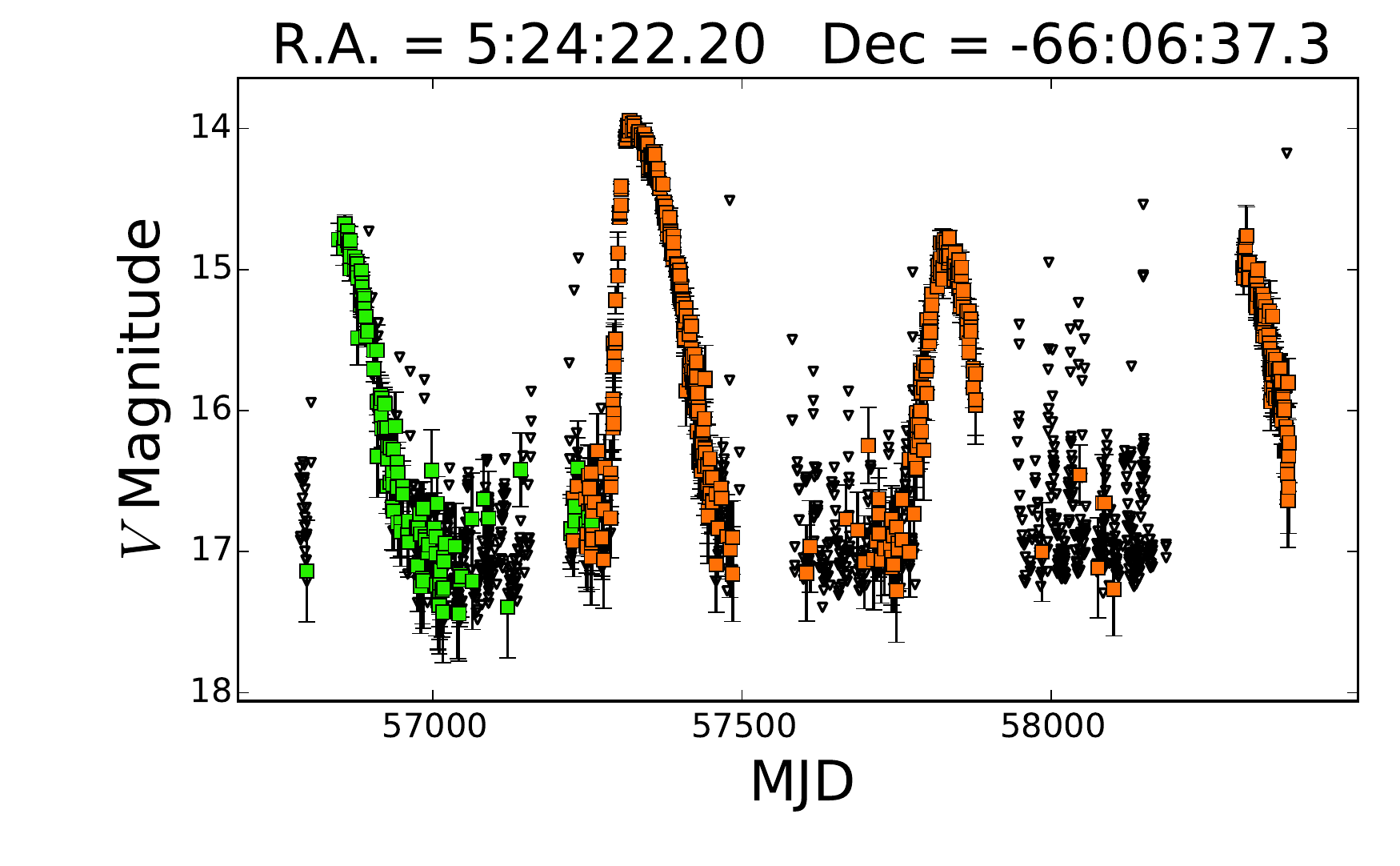}
    \includegraphics[width=0.32\textwidth]{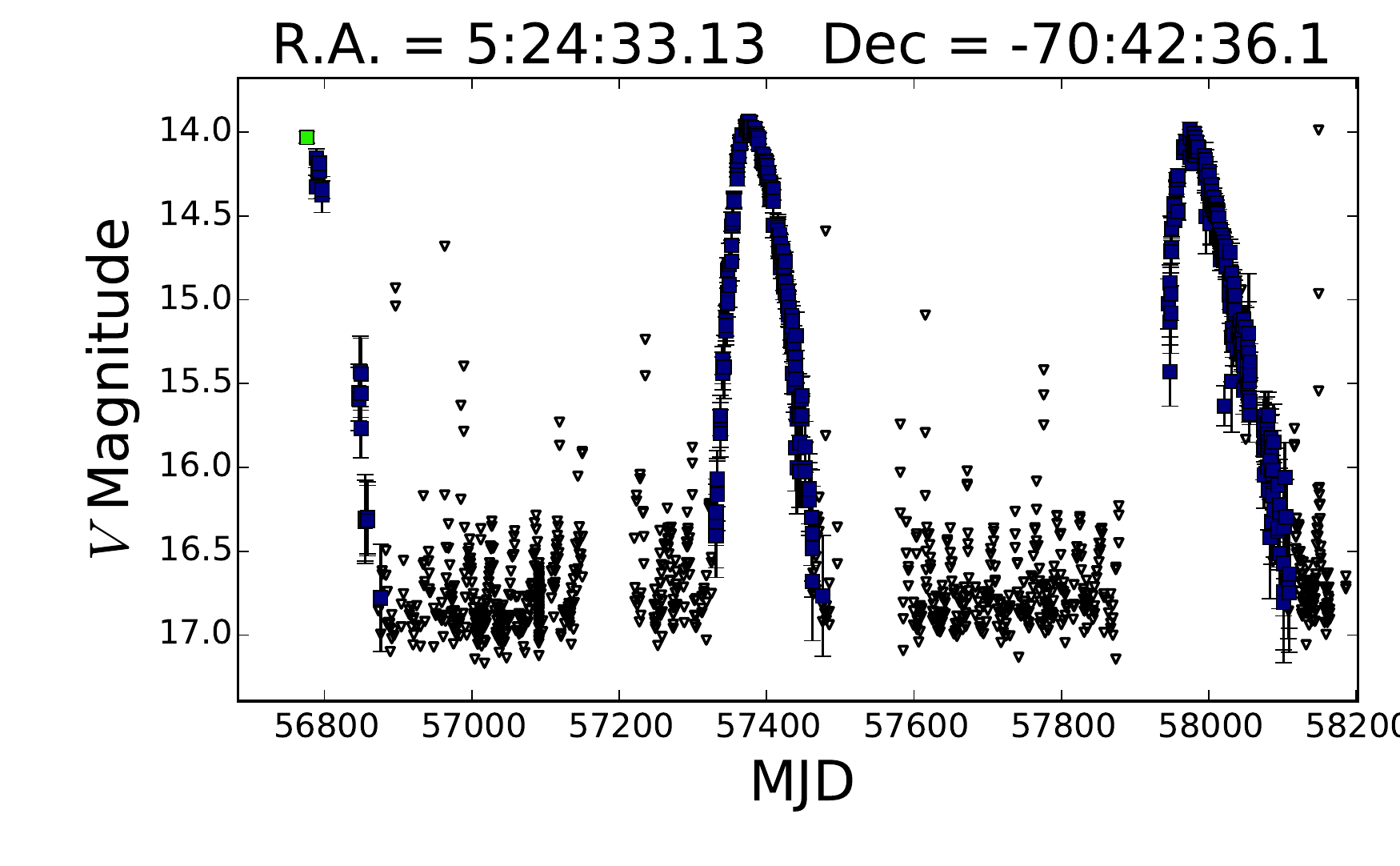}
    \includegraphics[width=0.32\textwidth]{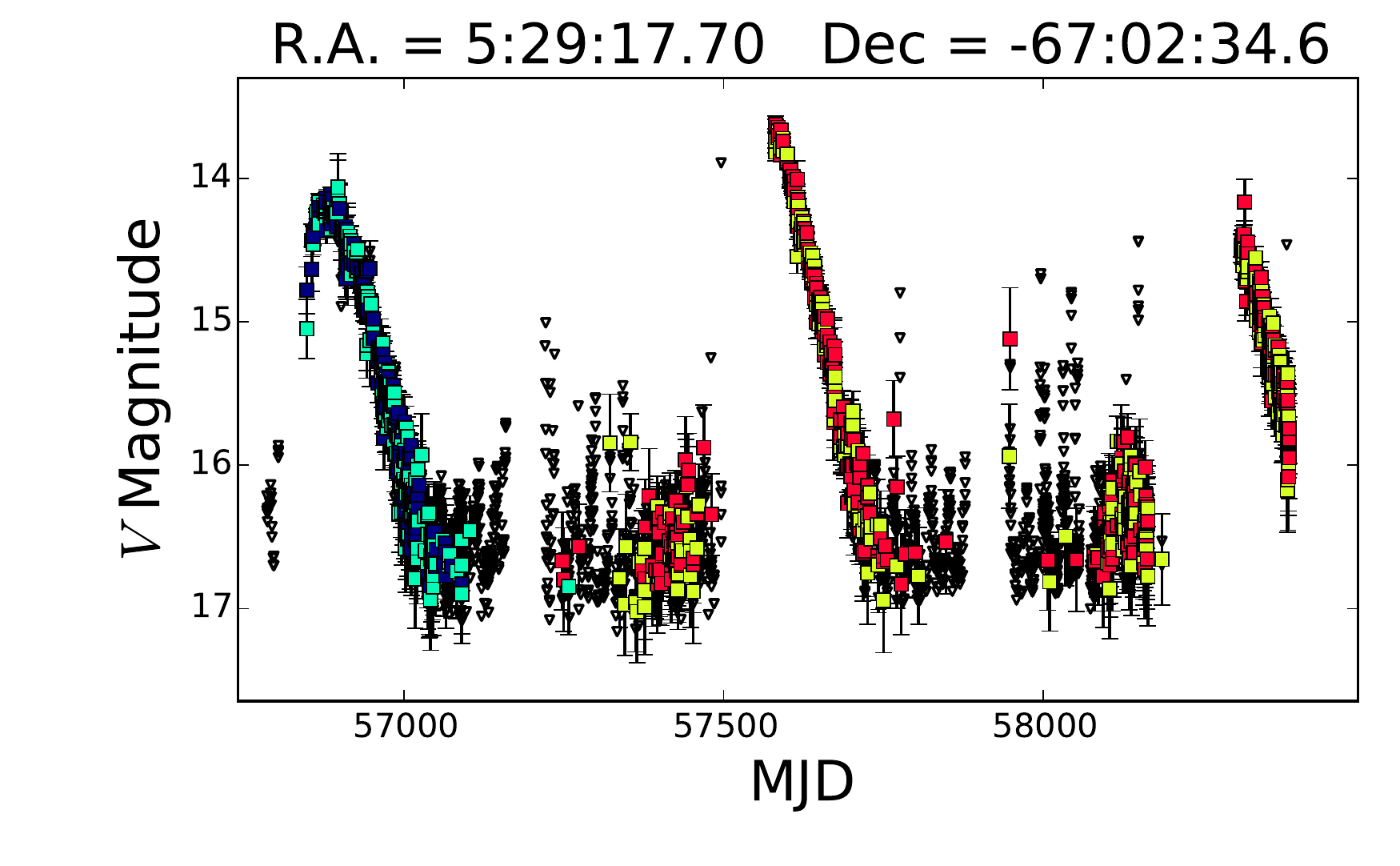}
    \includegraphics[width=0.32\textwidth]{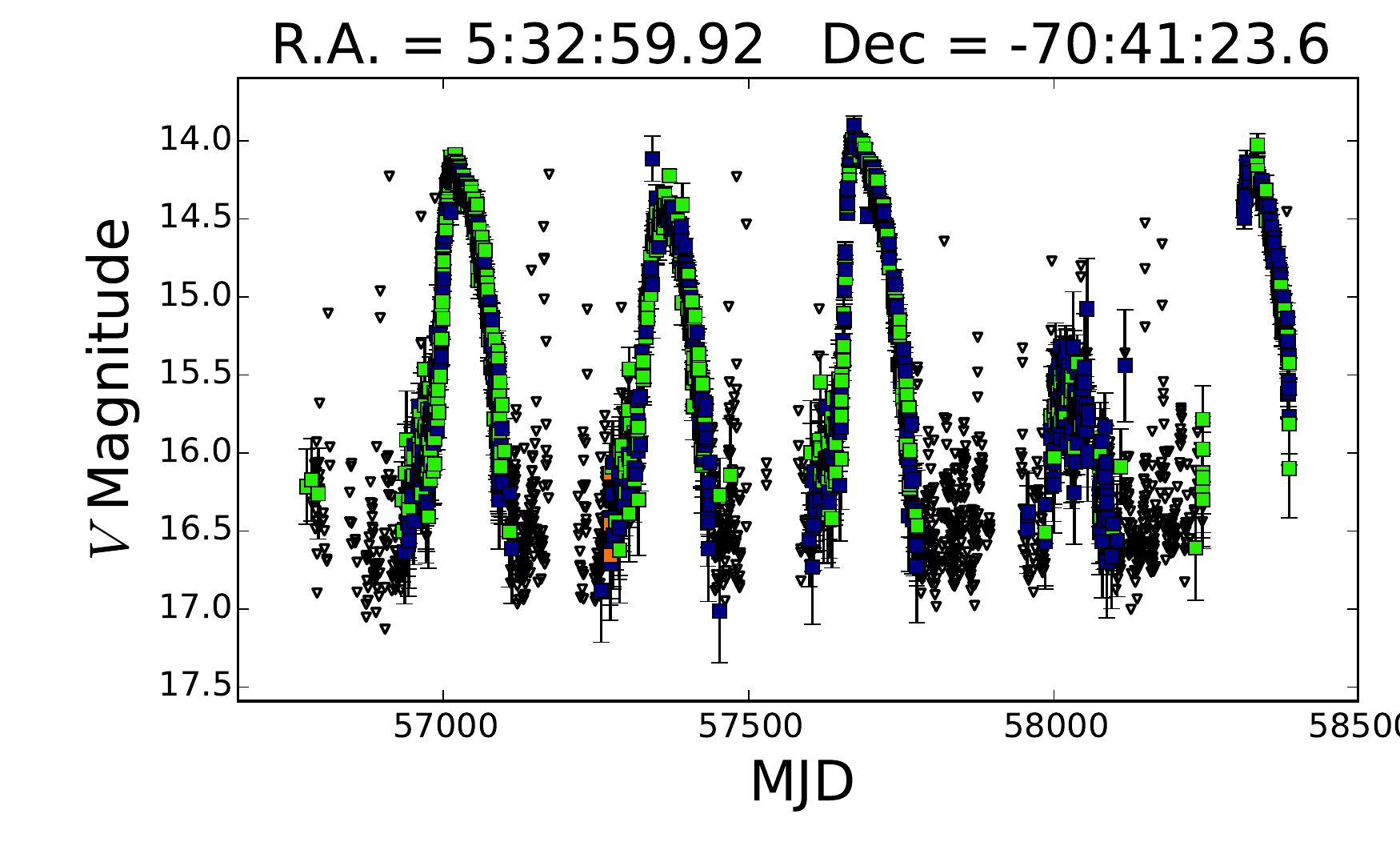}
    \includegraphics[width=0.32\textwidth]{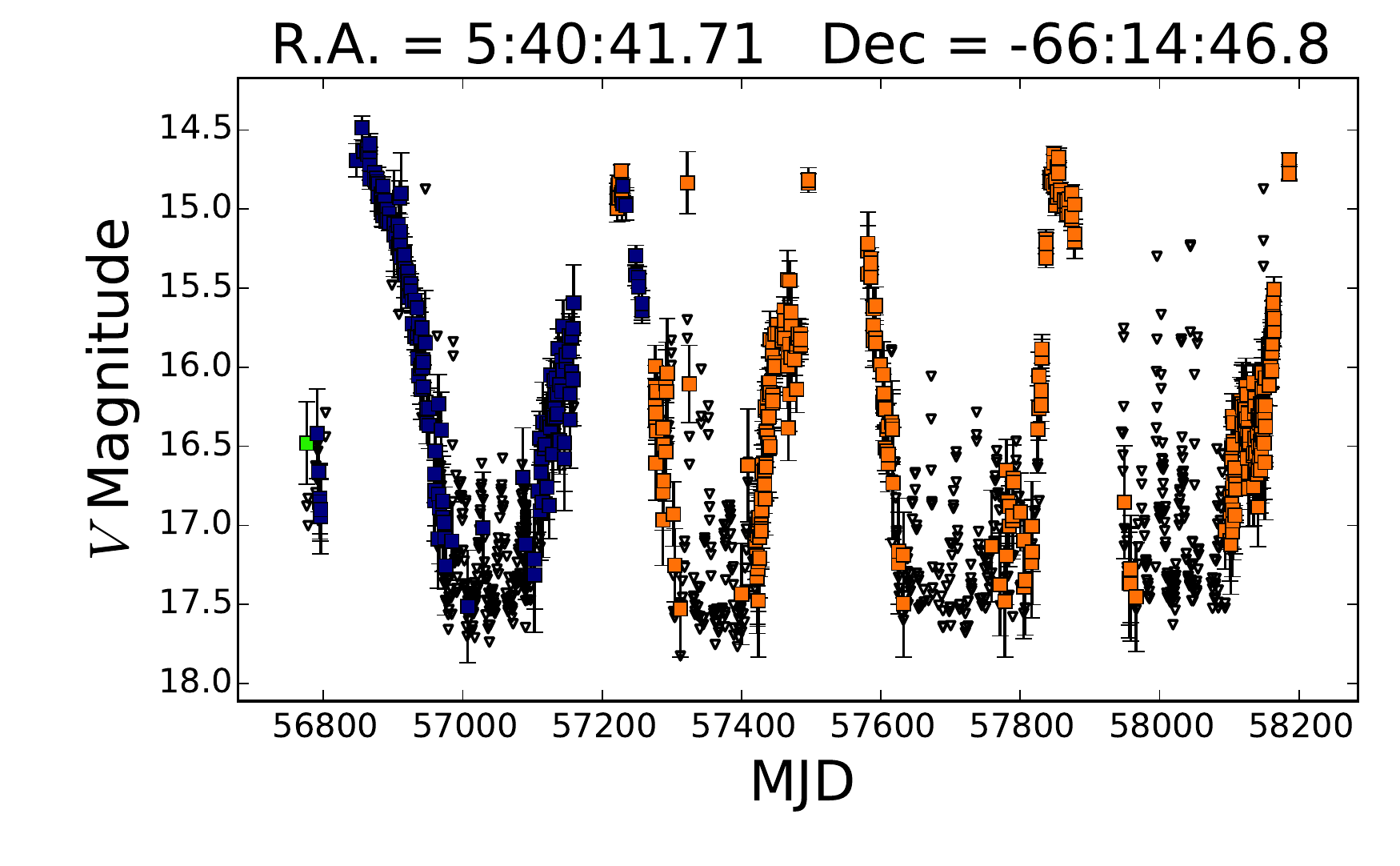}
    \includegraphics[width=0.32\textwidth]{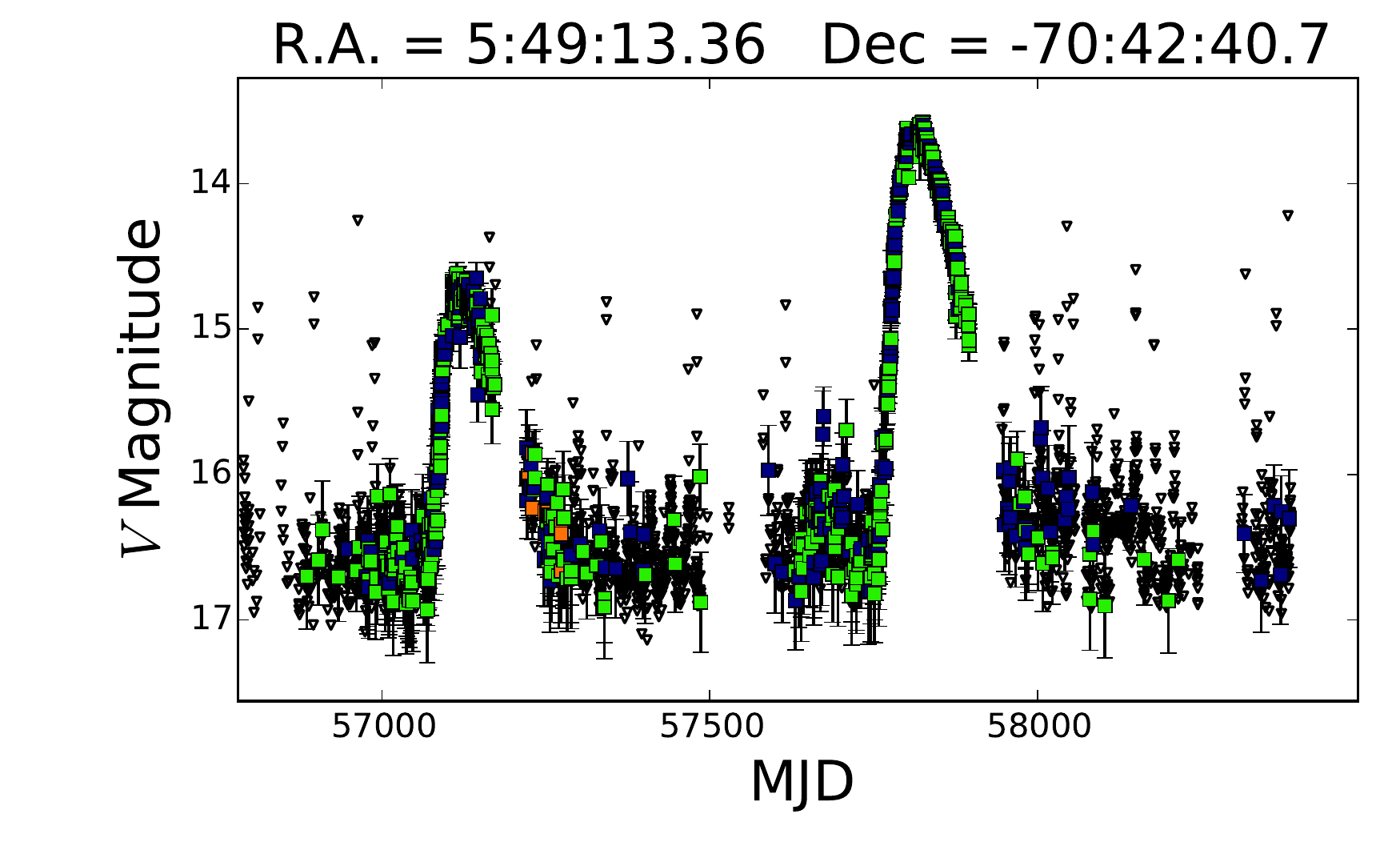}
    \includegraphics[width=0.32\textwidth]{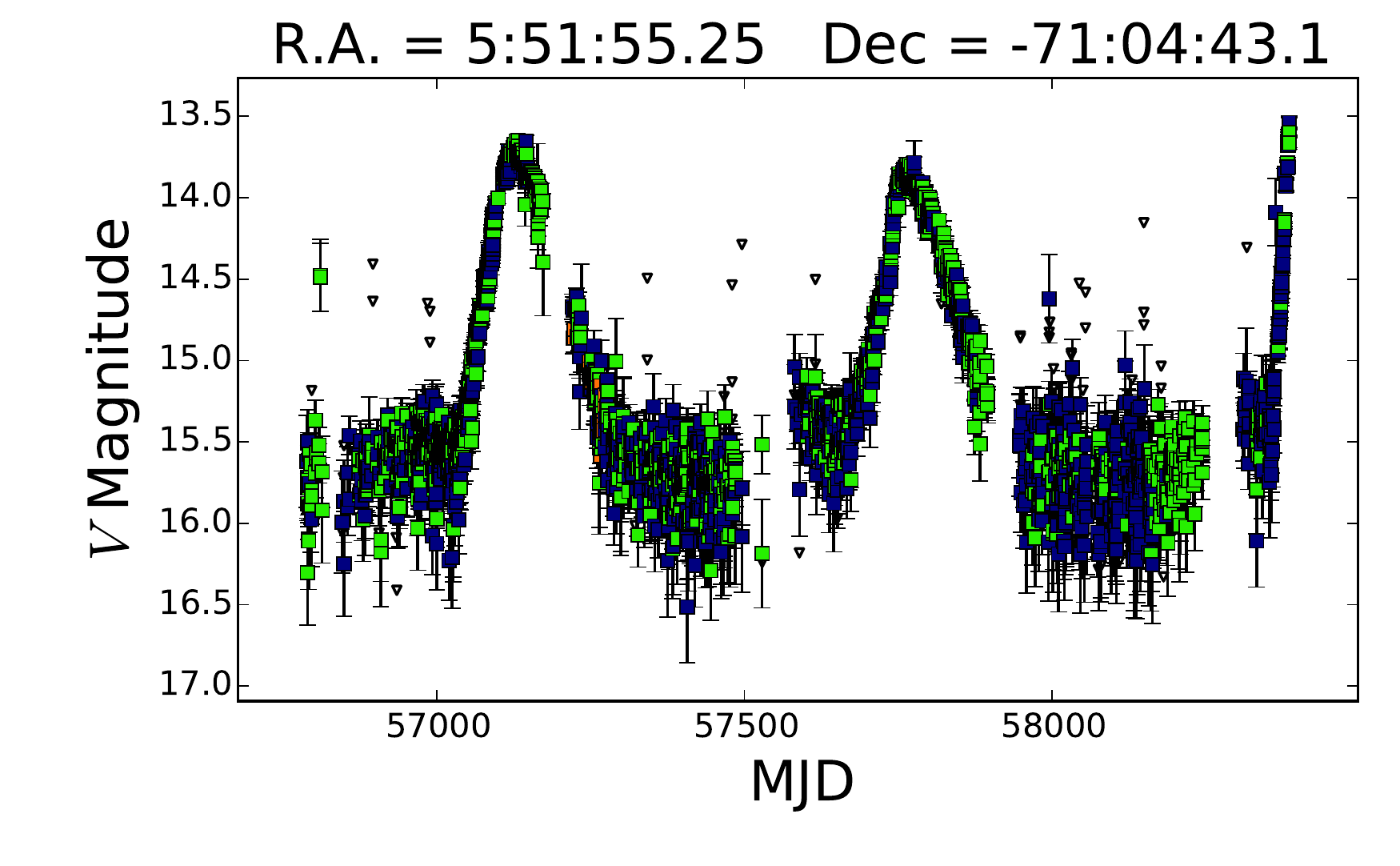}
    \includegraphics[width=0.32\textwidth]{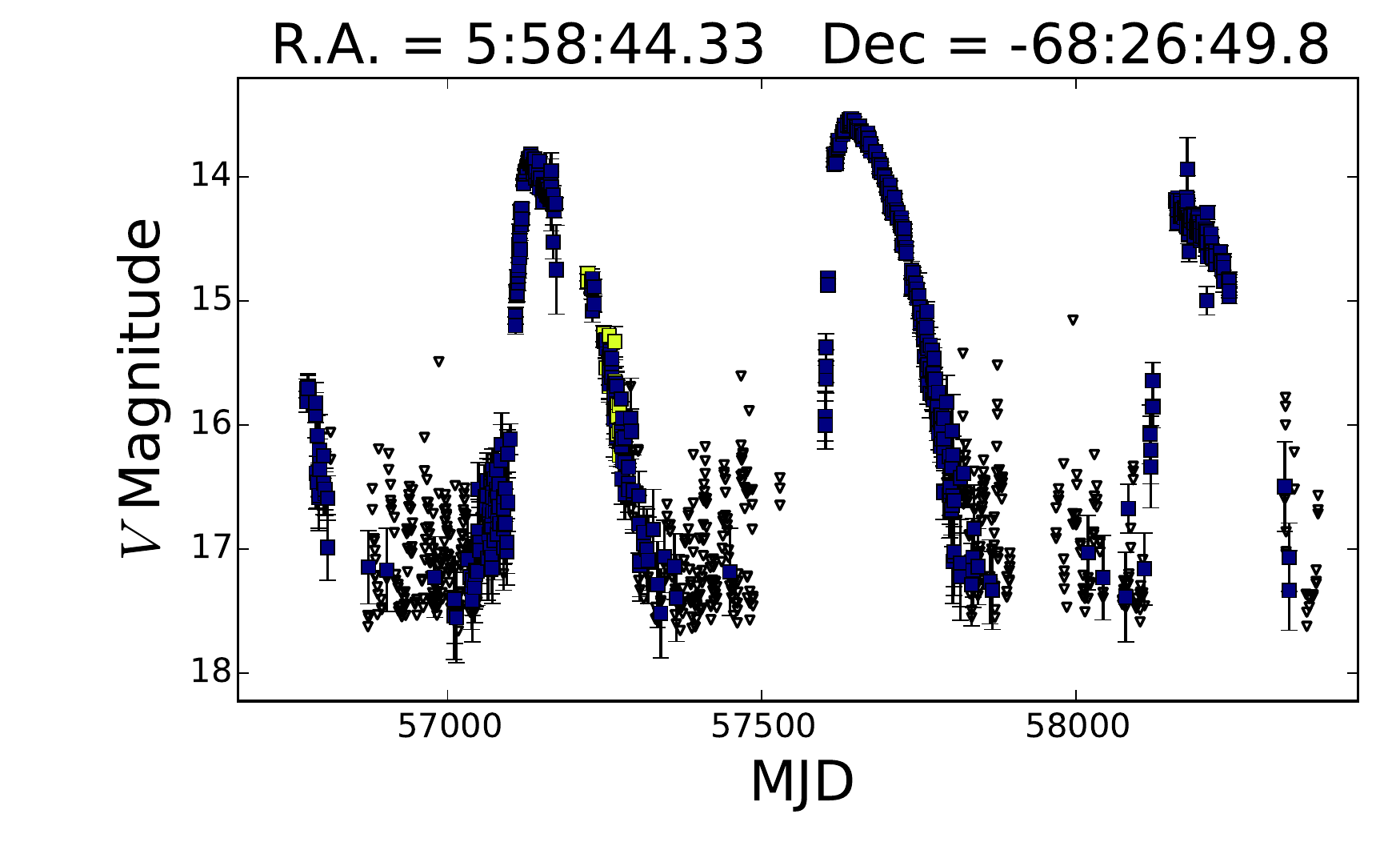}
    \includegraphics[width=0.32\textwidth]{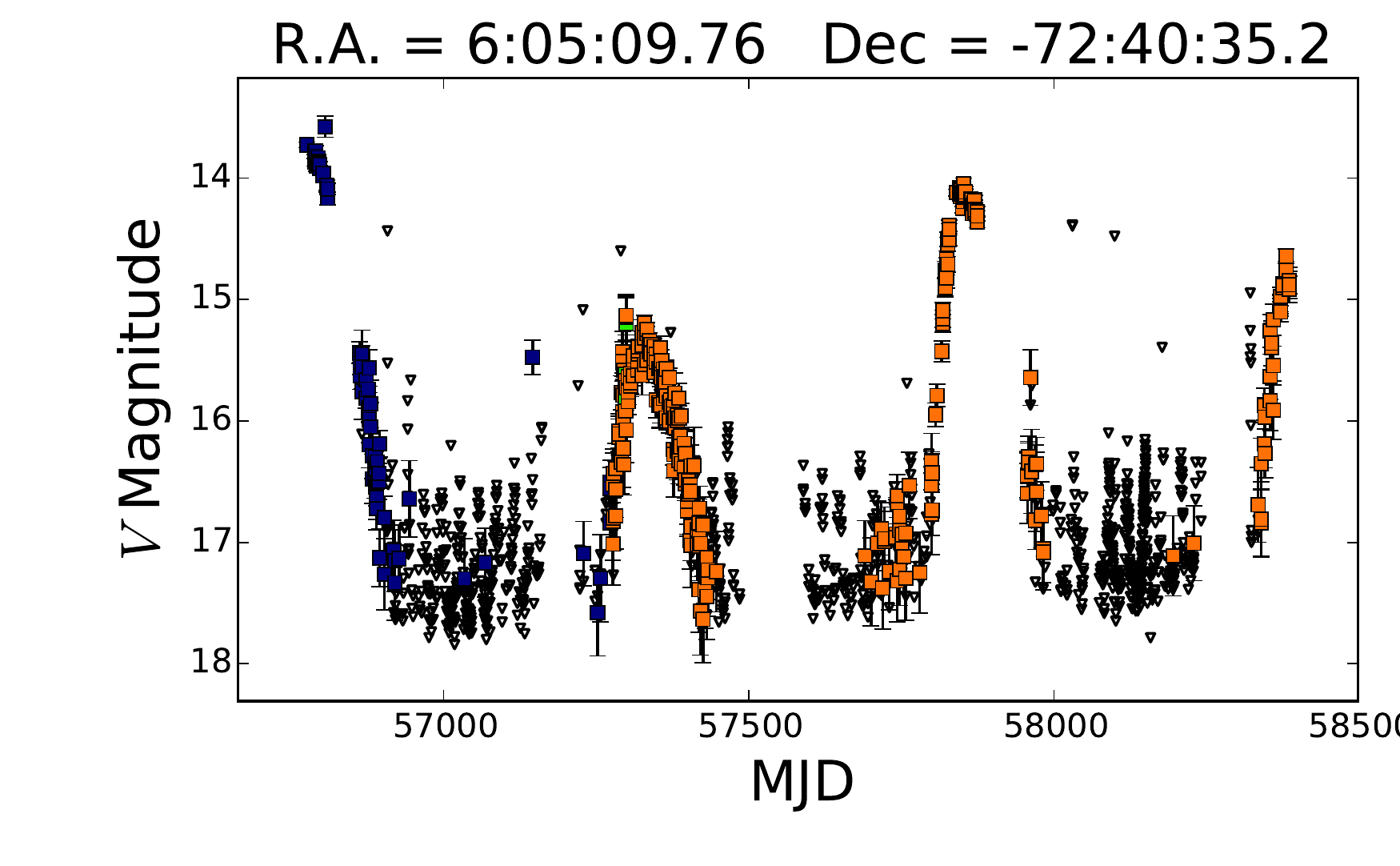}
    \caption{HAV light curves continued.}
    \label{hav_lcs2}
\end{figure*}

\clearpage

\section{Results of MCMC Fitting}\label{mcmc_appendix}

\startlongtable
\begin{deluxetable*}{cc|cccc|ccc}
\renewcommand\thetable{B}
\tabletypesize{\footnotesize}
\tablecolumns{9} 
\tablecaption{Results of SED fitting for the HLOs. The temperature, luminosity, $\tau_{\text{V}}$, and $\chi^{2}$ from the best run (lowest $\chi^{2}$) is shown, as well as the median values of the temperature, luminosity, and $\tau_{\text{V}}$ distributions with statistical errors encompassing 68\% (1-$\sigma$) of the distribution. Median temperatures are rounded to the nearest tenth. We discard the first 1000 runs as a burn-in region, leaving 3000 runs in the distributions. For the best fit values of temperature and luminosity we assume systematic uncertainties of $\pm$ 50K and $\pm 0.05$ dex, respectively. Phases that include ASAS-SN g-band photometry are marked with as * in the Phase column. The phase of HV2112 marked with two asterisks (**) has its posterior distributions shown in a corner plot in Figure~\ref{cornerplot}. \label{mcmc_results}}

\tablehead{ & & \multicolumn{4}{c|}{Best Fit} & \multicolumn{3}{c}{Median}\\\hline
\colhead{Star} & \multicolumn{1}{c|}{Phase} & \colhead{T$_{\text{eff}}$ (K)} & \colhead{log(L/L$_{\odot}$)} & \colhead{$\tau_{\text{V}}$} & \multicolumn{1}{c|}{$\chi^{2}$} & \colhead{T$_{\text{eff}}$ (K)} & \colhead{log(L/L$_{\odot}$)} & \colhead{$\tau_{\text{V}}$}
}
\startdata
   HV2112 & 0.05 & 3659 & 5.00 & 0.06 & 11.8 & 3650$^{+30}_{-30}$ & 4.99$^{+0.01}_{-0.01}$ & 0.11$^{+0.06}_{-0.06}$ \\ 
   '' & 0.25** & 3394 & 4.86 & 0.42 & 7.0 & 3400$^{+20}_{-10}$ & 4.83$^{+0.03}_{-0.01}$ & 0.61$^{+0.13}_{-0.11}$ \\ 
   '' & 0.35 & 3379 & 4.86 & 0.34 & 8.1 & 3380$^{+10}_{-10}$ & 4.86$^{+0.02}_{-0.02}$ & 0.22$^{+0.06}_{-0.04}$ \\ 
   '' & 0.65 & 3350 & 4.67 & 0.48 & 15.7 & 3340$^{+10}_{-20}$ & 4.68$^{+0.02}_{-0.02}$ & 0.31$^{+0.12}_{-0.08}$ \\ 
   '' & 0.74 & 3526 & 4.86 & 0.39 & 13.0 & 3520$^{+20}_{-20}$ & 4.87$^{+0.01}_{-0.01}$ & 0.16$^{+0.13}_{-0.04}$ \\ 
   '' & 0.86* & 3593 & 4.83 & 0.21 & 13.2 & 3590$^{+30}_{-20}$ & 4.83$^{+0.03}_{-0.02}$ & 0.12$^{+0.08}_{-0.06}$ \\ 
    \hline 
   SMC-1 & 0.29 & 3320 & 4.53 & 0.12 & 8.4 & 3310$^{+20}_{-20}$ & 4.53$^{+0.01}_{-0.02}$ & 0.08$^{+0.06}_{-0.05}$ \\
   '' & 0.39* & 3365 & 4.49 & 0.32 & 18.0 & 3360$^{+10}_{-10}$ & 4.50$^{+0.02}_{-0.02}$ & 0.15$^{+0.10}_{-0.06}$ \\
   '' & 0.69 & 3399 & 4.48 & 0.27 & 12.8 & 3410$^{+20}_{-10}$ & 4.47$^{+0.02}_{-0.01}$ & 0.17$^{+0.08}_{-0.03}$ \\
   '' & 0.99 & 3633 & 4.66 & 0.09 & 10.2 & 3640$^{+30}_{-30}$ & 4.66$^{+0.01}_{-0.01}$ & 0.16$^{+0.07}_{-0.10}$ \\
   \hline
   SMC-2 & 0.10 & 3519 & 4.77 & 0.01 & 4.3 & 3530$^{+20}_{-20}$ & 4.75$^{+0.02}_{-0.01}$  & 0.26$^{+0.05}_{-0.22}$ \\
   '' & 0.13 & 3445 & 4.68 & 0.17 & 7.7 & 3450$^{+20}_{-30}$ & 4.68$^{+0.01}_{-0.01}$ & 0.11$^{+0.05}_{-0.05}$ \\
   '' & 0.18 & 3400 & 4.74 & 0.01 & 5.2 & 3420$^{+20}_{-20}$ & 4.73$^{+0.01}_{-0.01}$ & 0.05$^{+0.05}_{-0.03}$ \\
   '' & 0.50* & 3360 & 4.43 & 0.71 & 15.3 & 3350$^{+10}_{-10}$ & 4.43$^{+0.02}_{-0.02}$ & 0.46$^{+0.16}_{-0.08}$ \\
   '' & 0.79 & 3504 & 4.53 & 0.42 & 12.8 & 3490$^{+20}_{-30}$ & 4.54$^{+0.01}_{-0.02}$ & 0.14$^{+0.24}_{-0.06}$ \\
   '' & 0.81 & 3515 & 4.60 & 0.29 & 11.7 & 3520$^{+20}_{-20}$ & 4.59$^{+0.01}_{-0.01}$ & 0.36$^{+0.08}_{-0.06}$ \\
   \hline
   SMC-3 & 0.19 & 3578 & 4.86 & 0.28 & 6.7 & 3570$^{+20}_{-20}$ & 4.87$^{+0.01}_{-0.01}$ & 0.13$^{+0.06}_{-0.04}$ \\
   '' & 0.29 & 3390 & 4.88 & 0.16 & 9.4 & 3390$^{+10}_{-10}$ & 4.89$^{+0.01}_{-0.01}$ & 0.07$^{+0.04}_{-0.04}$ \\
   '' & 0.92 & 3455 & 4.65 & 0.52 & 19.3 & 3450$^{+30}_{-30}$ & 4.66$^{+0.01}_{-0.01}$ & 0.38$^{+0.06}_{-0.07}$ \\
   \hline
   SMC-4 & 0.13* & 3454 & 4.75 & 0.14 & 10.6 & 3460$^{+20}_{-30}$ & 4.74$^{+0.02}_{-0.02}$ & 0.28$^{+0.07}_{-0.04}$ \\
   '' & 0.30 & 3264 & 4.64 & 0.39 & 9.7 & 3250$^{+30}_{-30}$ & 4.64$^{+0.01}_{-0.01}$ & 0.15$^{+0.07}_{-0.05}$ \\
   '' & 0.91 & 3390 & 4.53 & 0.23 & 16.5 & 3390$^{+10}_{-10}$ & 4.55$^{+0.02}_{-0.02}$ & 0.10$^{+0.07}_{-0.06}$ \\
   '' & 0.99 & 3635 & 4.75 & 0.09 & 12.0 & 3640$^{+30}_{-30}$ & 4.75$^{+0.01}_{-0.01}$ & 0.11$^{+0.06}_{-0.04}$ \\
   \hline
   SMC-5 & 0.24 & 3396 & 4.58 & 0.17 & 3.2 & 3400$^{+20}_{-10}$ & 4.56$^{+0.02}_{-0.01}$ & 0.14$^{+0.03}_{-0.04}$ \\
   '' & 0.54 & 3364 & 4.37 & 0.30 & 12.5 & 3360$^{+10}_{-10}$ & 4.38$^{+0.02}_{-0.01}$ & 0.19$^{+0.07}_{-0.09}$ \\
   '' & 0.64* & 3394 & 4.37 & 0.27 & 13.3 & 3390$^{+10}_{-10}$ & 4.38$^{+0.02}_{-0.03}$ & 0.21$^{+0.08}_{-0.05}$ \\
   '' & 0.94 & 3606 & 4.40 & 0.80 & 7.0 & 3620$^{+30}_{-20}$ & 4.40$^{+0.01}_{-0.01}$ & 0.57$^{+0.24}_{-0.12}$ \\
    \hline
   SMC-6 & 0.08* & 3579 & 4.87 & 0.15 & 7.3 & 3580$^{+10}_{-10}$ & 4.88$^{+0.02}_{-0.02}$ & 0.09$^{+0.03}_{-0.05}$ \\
   '' & 0.17 & 3482 & 4.80 & 0.47 & 6.4 & 3490$^{+20}_{-30}$ & 4.79$^{+0.01}_{-0.01}$ & 0.48$^{+0.17}_{-0.16}$ \\
   '' & 0.44 & 3267 & 4.70 & 0.31 & 10.4 & 3250$^{+30}_{-30}$ & 4.72$^{+0.01}_{-0.01}$ & 0.12$^{+0.13}_{-0.05}$ \\
   '' & 0.50 & 3308 & 4.62 & 0.31 & 11.7 & 3300$^{+20}_{-30}$ & 4.64$^{+0.01}_{-0.01}$ & 0.15$^{+0.07}_{-0.07}$ \\
   '' & 0.81 & 3419 & 4.72 & 0.46 & 13.3 & 3440$^{+30}_{-20}$ & 4.71$^{+0.01}_{-0.01}$ & 0.50$^{+0.31}_{-0.11}$ \\
   '' & 0.86 & 3443 & 4.62 & 0.48 & 13.8 & 3430$^{+30}_{-20}$ & 4.63$^{+0.01}_{-0.01}$ & 0.32$^{+0.07}_{-0.09}$ \\
   \hline
   LMC-1 & 0.18 & 3381 & 4.87 & 0.31 & 10.0 & 3380$^{+10}_{-10}$ & 4.87$^{+0.01}_{-0.02}$ & 0.22$^{+0.09}_{-0.06}$ \\
   '' & 0.45 & 3286 & 4.69 & 0.45 & 15.0 & 3270$^{+30}_{-30}$ & 4.71$^{+0.01}_{-0.02}$ & 0.30$^{+0.09}_{-0.08}$ \\
   '' & 0.50* & 3360 & 4.61 & 0.34 & 55.7 & 3350$^{+10}_{-10}$ & 4.62$^{+0.02}_{-0.02}$ & 0.20$^{+0.15}_{-0.07}$ \\
   '' & 0.92 & 3426 & 4.77 & 0.46 & 16.1 & 3440$^{+30}_{-20}$ & 4.76$^{+0.01}_{-0.01}$ & 0.45$^{+0.20}_{-0.19}$ \\
   '' & 0.97 & 3559 & 4.82 & 0.56 & 14.5 & 3530$^{+20}_{-20}$ & 4.82$^{+0.01}_{-0.02}$ & 0.38$^{+0.19}_{-0.16}$ \\
   \hline
   LMC-2 & 0.11 & 3623 & 5.11 & 0.83 & 5.9 & 3630$^{+40}_{-30}$ & 5.11$^{+0.02}_{-0.01}$ & 0.88$^{+0.21}_{-0.37}$ \\
   '' & 0.11* & 3569 & 5.04 & 0.85 & 12.9 & 3570$^{+20}_{-20}$ & 5.04$^{+0.02}_{-0.02}$ & 0.92$^{+0.15}_{-0.16}$ \\
   '' & 0.13 & 3566 & 5.17 & 0.37 & 7.2 & 3560$^{+20}_{-20}$ & 5.18$^{+0.01}_{-0.01}$ & 0.31$^{+0.04}_{-0.05}$ \\
   '' & 0.45 & 3330 & 4.98 & 0.69 & 11.2 & 3320$^{+20}_{-20}$ & 4.99$^{+0.02}_{-0.02}$ & 0.58$^{+0.35}_{-0.14}$ \\
   '' & 0.45 & 3407 & 4.95 & 1.83 & 3.8 & 3390$^{+30}_{-10}$ & 5.00$^{+0.03}_{-0.05}$ & 0.94$^{+0.60}_{-0.37}$ \\
   '' & 0.76 & 3631 & 5.07 & 0.78 & 11.9 & 3620$^{+30}_{-20}$ & 5.08$^{+0.02}_{-0.01}$ & 0.53$^{+0.20}_{-0.10}$ \\
   '' & 0.79 & 3633 & 4.98 & 1.01 & 7.9 & 3620$^{+50}_{-30}$ & 5.00$^{+0.03}_{-0.04}$ & 0.81$^{+0.54}_{-0.29}$ \\
   \hline
   LMC-3 & 0.08 & 3622 & 4.47 & 0.60 & 6.9 & 3630$^{+30}_{-30}$ & 4.47$^{+0.01}_{-0.01}$ & 0.54$^{+0.19}_{-0.18}$ \\
   '' & 0.77* & 3422 & 4.26 & 1.15 & 12.4 & 3420$^{+20}_{-20}$ & 4.26$^{+0.01}_{-0.01}$ & 0.84$^{+0.12}_{-0.21}$ \\
   '' & 0.88 & 3394 & 4.21 & 0.73 & 12.7 & 3390$^{+10}_{-10}$ & 4.23$^{+0.02}_{-0.02}$ & 0.56$^{+0.13}_{-0.09}$ \\
   '' & 0.97 & 3427 & 4.18 & 0.86 & 16.4 & 3420$^{+20}_{-20}$ & 4.19$^{+0.01}_{-0.01}$ & 0.54$^{+0.14}_{-0.05}$ \\
   \hline
   LMC-4 & 0.16* & 3387 & 4.79 & 0.13 & 5.5 & 3390$^{+10}_{-10}$ & 4.80$^{+0.02}_{-0.02}$ & 0.07$^{+0.07}_{-0.05}$ \\
   '' & 0.22 & 3280 & 4.70 & 0.22 & 6.4 & 3260$^{+30}_{-30}$ & 4.72$^{+0.01}_{-0.01}$ & 0.07$^{+0.06}_{-0.05}$ \\
   '' & 0.51 & 3335 & 4.51 & 0.67 & 16.5 & 3320$^{+20}_{-20}$ & 4.52$^{+0.02}_{-0.02}$ & 0.44$^{+0.09}_{-0.09}$ \\
   '' & 0.54 & 3336 & 4.59 & 0.18 & 16.3 & 3330$^{+20}_{-20}$ & 4.59$^{+0.02}_{-0.02}$ & 0.17$^{+0.08}_{-0.05}$ \\
   '' & 0.87 & 3398 & 4.55 & 0.66 & 10.9 & 3400$^{+20}_{-10}$ & 4.54$^{+0.03}_{-0.01}$ & 0.40$^{+0.16}_{-0.10}$ \\
   '' & 0.90 & 3520 & 4.69 & 0.38 & 8.3 & 3510$^{+20}_{-30}$ & 4.70$^{+0.02}_{-0.02}$ & 0.23$^{+0.09}_{-0.13}$ \\
   '' & 0.94 & 3524 & 4.68 & 0.45 & 10.2 & 3510$^{+20}_{-30}$ & 4.69$^{+0.01}_{-0.02}$ & 0.16$^{+0.18}_{-0.05}$ \\
    \hline
   LMC-5 & 0.00* & 3557 & 4.83 & 0.20 & 23.0 & 3550$^{+20}_{-20}$ & 4.85$^{+0.02}_{-0.02}$ & 0.10$^{+0.04}_{-0.04}$ \\
   '' & 0.08 & 3481 & 4.72 & 0.23 & 12.7 & 3460$^{+30}_{-30}$ & 4.74$^{+0.01}_{-0.01}$ & 0.12$^{+0.06}_{-0.04}$ \\
   '' & 0.15 & 3383 & 4.80 & 0.29 & 8.5 & 3380$^{+10}_{-10}$ & 4.81$^{+0.02}_{-0.02}$ & 0.18$^{+0.05}_{-0.09}$ \\
   '' & 0.31 & 3249 & 4.77 & 0.54 & 12.9 & 3230$^{+30}_{-30}$ & 4.78$^{+0.01}_{-0.01}$ & 0.32$^{+0.14}_{-0.08}$ \\
   '' & 0.70 & 3250 & 4.64 & 0.39 & 19.6 & 3240$^{+30}_{-30}$ & 4.65$^{+0.01}_{-0.01}$ & 0.25$^{+0.06}_{-0.08}$ \\
   '' & 0.77 & 3331 & 4.61 & 0.32 & 18.1 & 3320$^{+20}_{-20}$ & 4.62$^{+0.02}_{-0.02}$ & 0.26$^{+0.08}_{-0.08}$ \\
\enddata
\end{deluxetable*}

\begin{figure*}[ht!]
    \centering
    \includegraphics[width=\textwidth]{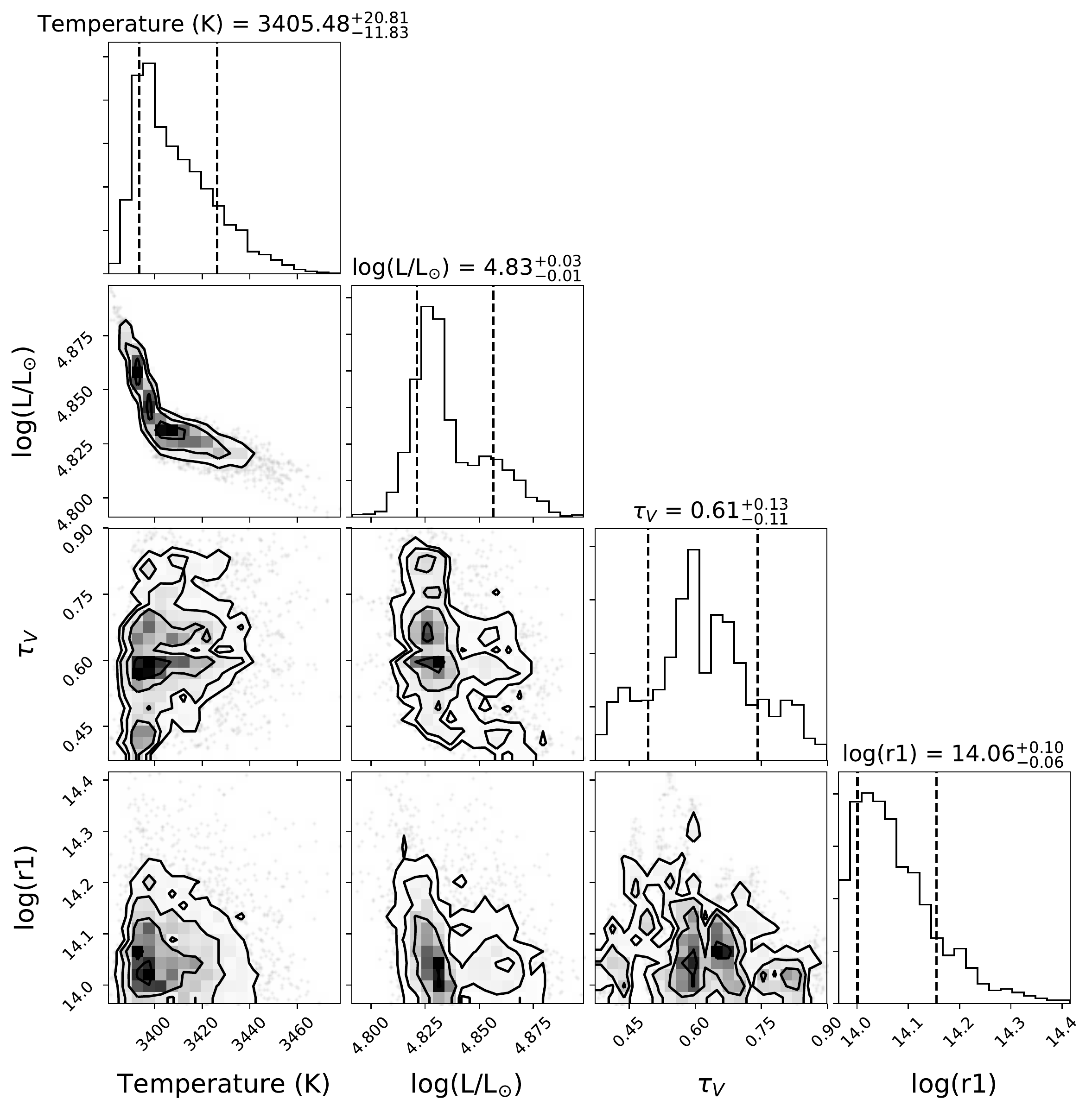}
    \caption{An example corner plot from the MCMC fitting of one phase of HV2112's variability cycle. The phase is indicated in Table \ref{mcmc_results} with a double asterisk (**). The posterior distributions of the HLOs generally look the same; this particular phase of HV2112 was randomly chosen. The text above each 1D histogram shows the median value and 1-$\sigma$ uncertainties for each parameter. There is some degree of degeneracy between the luminosity and temperature parameters, but due to the small magnitudes of the statistical errors resulting from the MCMC fitting, this does not impact our conclusions.}
    \label{cornerplot}
\end{figure*}

\clearpage

\bibliography{biblio}

\end{document}